\documentclass[reqno]{amsart}
\usepackage{ifthen}
\newboolean{full}
\setboolean{full}{true}
\newboolean{tcs}
\makeindex

\newcounter{proofcount}
\newcommand{\addproof}[2]{%
  \iffull{#2}\else%
    \expandafter\gdef\csname proof\roman{proofcount}\endcsname{%
      \subsection{Proof of #1}
      #2
      }%
    \stepcounter{proofcount}
  \fi
}

\def\doproofs{%
  \newcounter{lauf}
  \setcounter{lauf}{1}
  \whiledo{\value{lauf}<\value{proofcount}}{%
    \csname proof\roman{lauf}\endcsname
    \stepcounter{lauf}
    }
}

\usepackage{graphicx,style,rotating,fancybox,array,longtable,amsmath,longtable}

\newlength{\mylen}    

\begin{document}  

%
% Frontmatter
% 
\iftcs
\begin{frontmatter}
\def\email{\ead}
\fi
\title{The Category Theoretic Solution of Recursive Program Schemes}  
\author{Stefan Milius}  
\address{Institute of Theoretical Computer Science, Technical University of  
         Braunschweig, Germany}  
\email{milius@iti.cs.tu-bs.de}
\author{Lawrence S.~Moss}
\address{Department of Mathematics, Indiana University, Bloomington, IN, USA}
\email{lsm@cs.indiana.edu}

\iftcs
\begin{keyword}
\else
\keywords{
\fi
  recursive program scheme, Elgot algebra,
  coalgebra, completely iterative monad, algebraic
  trees, second-order substitution
\iftcs
\end{keyword}
\else
}
\fi

\date{19 March 2007}  

\begin{abstract}
  This paper provides a general account of the notion of recursive program
  schemes, studying  both uninterpreted and interpreted solutions.
  It can be regarded as the category-theoretic version of the classical area of
  algebraic semantics.  The overall assumptions needed are small indeed:
  working only in categories with ``enough final coalgebras'' we show how to
  formulate, solve, and study recursive program schemes.  Our general theory is
  algebraic and so avoids using ordered, or metric structures.  Our work
  generalizes the previous approaches which do use this extra structure by
  isolating the key concepts needed to study   substitution in
  infinite trees, including second-order substitution.  As special cases of our
  interpreted solutions we obtain the usual denotational semantics using
  complete partial orders, and the one using complete metric spaces.  Our
  theory also encompasses implicitly defined objects which are not usually
  taken to be related to recursive program schemes.  For example, the
  classical Cantor two-thirds set falls out as an interpreted solution (in our
  sense) of a recursive program scheme.  
\end{abstract}  
\iftcs
\end{frontmatter}
\else
\maketitle  
\fi
\iftcs
\pagebreak
\enlargethispage{10pt}
\fi
\tableofcontents
\pagebreak

%
% Here begins the paper's text
%
\section{Introduction}
\label{sec:intro}

The theory of {\em recursive program schemes\/}
(see~\cite{courcelle,guessarian,nivat}) is a topic at the heart of
semantics.  One takes a system of equations such as
\begin{equation}\label{eq:algex}
  \begin{array}{rcl}
    \phi(x) & \approx & F(x, \phi(Gx)) \\
    \psi(x) & \approx & F(\phi(Gx), GGx)
  \end{array}
\end{equation}
where $F$ and $G$ are {\em given\/} functions and where $\phi$ and $\psi$ are
defined in terms of them by (\ref{eq:algex}).  The problems are: to give some
sort of semantics to schemes, and to say what it means to {\em solve\/} a
scheme.  Actually, we should distinguish between \emph{interpreted} schemes,
and \emph{uninterpreted} schemes.

An uninterpreted scheme such as~\refeq{eq:algex} above is a purely syntactic
construct; one has no information about the data on which given functions and
recursively defined ones operate on. Hence, the semantics is independent of this data.
In our example the semantics provides two infinite trees 
\begin{equation}\label{eq:soltrees}
\sol{\phi}(x) = 
\vcenter{
  \xy
  \POS (00,00) *+{F} = "A"
     , (-5,-10) *+{x} = "x"
     , (5,-10) *+{F} = "A1"
     , (0,-20) *+{Gx} = "Bx"
     , (10,-20) *+{F} = "A2"
     , (5,-30)  *+{GGx} = "BBx"
     , (15,-30) *+{} = "end"
  %%%
  \ar@{-} "A";"x"
  \ar@{-} "A";"A1"
  \ar@{-} "A1";"Bx" 
  \ar@{-} "A1";"A2"
  \ar@{-} "A2";"BBx"
  \ar@{.} "A2";"end"
  \endxy
}
\qquad\qquad
\sol{\psi}(x) =
\vcenter{
  \xy
  \POS (000,000)  *+{F}   = "A"
     , ( -5,-10)  *+{F}   = "F"
     , (  5,-10)  *+{GGx} = "BBx"
     , (-10,-20)  *+{Gx}  = "Gx"
     , (  0,-20)  *+{F}   = "F2"
     , ( -5,-30)  *+{GGx} = "GGx"
     , (  5,-30)  *+{}    = "end"
  %%%
  \ar@{-} "A";"F"
  \ar@{-} "A";"BBx"
  \ar@{-} "F";"Gx"
  \ar@{-} "F";"F2"
  \ar@{-} "F2";"GGx"
  \ar@{.} "F2";"end"
  \endxy
}
\end{equation}
over the signature of given functions; one tree for each of $\phi$ and
$\psi$. We explain this in more detail in Section~\ref{sec:algsem} below.  

An interpreted scheme is one which comes with an algebra with operations for
all the given operation symbols. The recursive program scheme then defines new
operations on that algebra. Here is the standard example in the subject.
Let $\Sigma$ be the signature of given operation symbols with a constant
$\one$, one unary symbol $\pred$, a binary symbol $*$ and a 
ternary one $\cond$. The interpretation we have in mind is the natural numbers
where $\cond_\Nat(k,n,m)$ returns $n$ if $k$ is $0$ and $m$ otherwise, and all
other operations are as expected. The signature $\Phi$ of the recursively defined
operations consists just of one unary symbol $f$. Consider the recursive
program scheme 
\index{factorial}%
\begin{equation}\label{eq:run}
  f(n) \approx \cond(n, \one, f(\pred(n)) * n))\,.
\end{equation}
Then~\refeq{eq:run} is a recursive program scheme defining the factorial
function. 

This paper presents a generalization of the classical theory based on
\emph{Elgot algebras} and \emph{final coalgebras}.  The point in a nutshell is
that knowing that the infinite trees for a signature are the final coalgebra
for a functor on sets leads to a purely algebraic account of first-order
substitution and \mbox{(co-)}recursion, as shown in~\cite{aamv,moss}.  One does
not need to assume any metric or order to study infinite trees: the finality
principle is sufficient for developing the crucial parts of the theory of
infinite trees.  In this paper we show that this same finality principle allows
us to give an uninterpreted semantics to a scheme: we show how to solve
a scheme such as \refeq{eq:algex} to get a pair of infinite trees,
see~\refeq{eq:soltrees} above.

For our interpreted semantics we work with Elgot algebras, a simple and
fundamental notion introduced in~\cite{amv3}. We show how to give an
interpreted solution to recursive program schemes in arbitrary Elgot algebras.
We believe that our results in this area generalize and extend the previous
work on this topic. Furthermore, we claim that our abstract categorical
approach allows a unified view of several well-known approaches to the
semantics of implicit recursive function definitions. Our method for obtaining
interpreted solutions easily specializes to the usual denotational semantics
using complete partial orders.  As a second application we show how to solve
recursive program schemes in complete metric spaces. For example, 
it follows from our work that there is a
unique contracting $f: [0,1]\to [0,1]$ such that
\begin{equation} 
  f(x) = \frac{1}{4}\left(x + f\left(\frac{1}{2}\sin x\right)\right).
  \label{eq-sin}
\end{equation}
In addition, our work on Elgot algebras points to new connections
to classical subjects. For example, recall that the Cantor set $c$ is the
unique non-empty closed  $c\subseteq [0,1]$ such that
\index{Cantor set}%
\begin{equation} 
  c = \frac{1}{3}c \ \cup\ \left(\frac{2}{3} + \frac{1}{3}c\right)\,,
  \label{eq-cantor}
\end{equation}
where $\frac{1}{3}c$ denotes the set $\{\,\frac{1}{3} x \mid x \in c \,\}$ as
usual. 
The general fact that (\ref{eq-cantor}) has  a
unique solution in the set $C([0,1])$
of  non-empty closed subsets of $[0,1]$ follows
from $C([0,1])$ being an Elgot algebra.

Finally, our theory also encompasses examples of recursive program schemes and
their solutions which are not treated with the classical theory; in this
paper we present recursive definitions of operations satisfying equations like
commutativity. Other examples are recursive program scheme solutions in
non-wellfounded sets (solving $x = \{\,\{\,y \mid y \subseteq x\ 
\textrm{finite}\,\}\,\}$). We will present these applications
 in a future paper~\cite{non-well}.

Our purpose in this paper is to isolate general principles
which imply the existence and uniqueness results 
for uninterpreted solutions of systems such as (\ref{eq:algex})
and interpreted schemes such as (\ref{eq:run}--\ref{eq-cantor}).

\iffull
\subsection{Several Semantics} 
There are several semantics for recursive program schemes, and it
is worth mentioning a bit about them, both to situate our work
and also because we shall be interested in showing that one can
recover different flavors of semantics from our more general approach.

\subsubsection{Operational Semantics}
\index{factorial!classical semantics}
This gives semantics to interpreted schemes only. Solutions are defined by
rewriting. In our factorial example, the semantics of~\refeq{eq:run} would be as
follows: to compute the solution $f_{\Nat}(n)$ for a natural number
$n$, start with the term $f(n)$, replace $f$ by the
right-hand side of the scheme, and then evaluate this term
as much as possible.  If the evaluated
right-hand side still contains one or more symbols $f$, then replace those
symbols again by their right-hand side
and then evaluate the resulting term, and so on. If after finitely many steps
this process returns a natural number,
 we have computed $f_{\Nat}(n)$; otherwise
we declare $f_{\Nat}(n)$ to be undefined.
(Of course there are important and subtle points to be considered
here pertaining to issues like call by name and call by value interpretations
of function symbols, and also about the overall strategy of rewriting.)
In the factorial example,
the described process always stops. But in general this may not be the case.

\subsubsection{Denotational Semantics}
%\index{factorial!denotational semantics}

Again this provides semantics for interpreted schemes only. The algebra that
comes as the interpretation of the given functions is a complete partial order
(cpo) $A$ with continuous operations for all given operation symbols. A system
like~\refeq{eq:algex} gives then rise to a continuous function $R$ on a
function space. In our factorial example from (\ref{eq:run}), one would consider
the natural numbers as a flat cpo and $R$ assigns to a continuous function $f$
on that cpo the function $Rf$ defined by
\begin{equation}\label{eq:R}
  Rf(n) = \left\{
    \begin{array}{c@{\qquad}p{3cm}}
      \bot & if $n = \bot$ \\
      1    & if $n = 0$ \\
      f(n-1) \cdot n & else
    \end{array}
  \right.
\end{equation}
The semantics of the given scheme is then the least fixed point of $R$.
More generally, denotational semantics provides a continuous operation
$\phi_A$ on $A$ for each  recursively defined operation symbol $\phi$ of a given
recursive program scheme.

It is true but by no means obvious that the operational 
and denotational semantics agree in
the appropriate sense.  Thus, there is a matter of
semantic  equivalence to be investigated.  This was one of the
primary starting points of the original theory of recursive
program schemes, see~\cite{courcelle,guessarian,nivat}.
In any case, the equivalence raises a question:
the
very definition of the
denotational semantics seems to require an order.
 But 
 is an order really necessary?
  One of the themes of work in coalgebra
is that for some semantic problems, order-theoretic methods
may be replaced by purely algebraic ones. 
   The reason is that coalgebra
is often about semantics spaces for infinite behaviors
of one type or another, and these are studied using 
universal properties (typically finality) instead of
the extra structure coming from a cpo or complete metric space. 
In broad terms, this is our move as well. 

\comment{The second
general theme is what we might call {\em recovery of
one semantics by another}.
The first instance of this is the recovery of the operational
semantics by the denotational semantics.   
One therefore feels that the denotational semantics is getting
at something deeper.  (At the same time,   subtle points
of operational semantics such as the distinction between
call-by-name and call-by-value
are not easy to address with the
methods of denotational semantics, so there is a loss as well.)
}

\subsubsection{Algebraic Semantics}
\label{sec:algsem}

This considers \emph{uninterpreted} recursive program schemes. These are
schemes where no interpretation is given. It is then an issue in this approach
to make sure that one can recover the denotational and operational semantics
inside the algebraic semantics.  But first, one must say what a solution to
(\ref{eq:algex}) should be. Normally, one considers for a signature $\Sigma$
and a set $X$ of generators the set $T_\Sigma X$ of all finite and infinite
$\Sigma$-trees over $X$.  These are ordered and rooted trees labelled so that
an inner node with $n$ children, $n > 0$, is labelled by an $n$-ary operation
symbol from $\Sigma$, and leaves are labelled by a constant symbol or a
variable of $X$.  Then one defines an appropriate notion of {\em second-order
  substitution\/} whereby $\Sigma$-trees may be substituted for operation
symbols in $\Gamma$-trees for another signature $\Gamma$
(see~\cite{courcelle}). (Recall that the ordinary first-order
substitution\index{substitution!first-order} replaces $\Sigma$-trees for
generators in the leaves of other $\Sigma$-trees; only one signature is
involved.)  In general, a recursive program scheme is given by two signatures:
$\Sigma$ (of given operations) and $\Phi$ (of recursively defined operations),
and a set of formal equations providing for each $n$-ary operation $\phi$ from
$\Phi$ on the left-hand side a $(\Sigma+\Phi)$-tree over $n$ syntactic
variables on the right-hand side of the equation. A solution for such a
recursive program scheme then assigns to each $n$-ary operation symbol $\phi$
from $\Phi$ a $\Sigma$-tree $\sol{\phi}(x_1, \ldots, x_n)$ in $n$ syntactic
variables. This $\Sigma$-tree is required to be equal to the $\Sigma$-tree
obtained by performing the following second-order substitution on the tree from
the right-hand side of the formal equation defining $\phi$: one replaces each
operation symbol $\psi$ from $\Phi$ by the $\Sigma$-tree provided by the
solution.

As an example, consider the signature $\Sigma$ with a binary operation symbol
$F$ and a unary symbol $G$. One might want to define new operations $\phi$ and
$\psi$ recursively as in~\refeq{eq:algex} above.  Notice that the system is
\emph{guarded}\index{guarded!recursive program scheme}, or in 
\emph{Greibach normal form}\index{Greibach normal form}: the right-hand sides
start with a symbol from $\Sigma$. One key opening result of algebraic
semantics is that a unique solution exists for guarded systems among
$\Sigma$-trees. For 
instance, for the above system \refeq{eq:algex} the solution consists of the
$\Sigma$-trees from~\refeq{eq:soltrees}.
%%%%% Here was the picture of the trees!
The standard approach views infinite trees as either the completion of the
finite trees, considered as a metric space, or else as the ideal completion of
the set of finite trees.  Second-order substitution is defined and studied
using one of those methods, and using this one can say what it means to solve a
recursive program scheme.  We shall 
rework the standard theory by considering $\Sigma$-trees as final coalgebras.
 More precisely, let $H_\Sigma$ be the functor on
 sets associated to $\Sigma$; see Example~\ref{exmp-signature-functors}
 below. Then the collection $T_\Sigma X$ of all $\Sigma$-trees over $X$ is
 the carrier of a final coalgebra structure for the functor
 $H_\Sigma(\,\_\,) + X$.
It is the universal property of those final coalgebras for
any set $X$ which allows us to give a semantics to recursive program schemes.
We feel that this move leads to a nicer treatment on a conceptual level.

For example, the solution to~\refeq{eq:algex} is obtained from finality alone,
as follows: the original recursive program scheme gives rise to a \emph{flat
  system of equations}. In this case, this means a function of the form
$e : Z \to H_\Sigma Z +X$,
where $Z$  is a set called the set of \emph{formal variables}%
\index{formal variables!arising from an RPS}  
of the system and $X$ is the set of syntactic variables from the recursive
program scheme. 
In this particular case, we have $X = \{\,x\,\}$, and the set $Z$ of formal
variables is the set $T_{\Sigma+\Phi} \{x\}$ of all  finite and infinite
$(\Sigma + \Phi)$-trees in one variable $x$. Since the formal variables 
are trees, and since trees are involved with the intended solution,
there is much opportunity for confusion.
To minimize it, we shall \underline{underline} the formal variables.

Again, we have  a 
formal  variable $\ul{t}$ for each $(\Sigma +
\Phi)$-tree $t$.  The system $e$ itself 
works as follows: if $t$ is the variable $x$ we have $e(\ul{x}) = x$;
we usually prefer to write this as $\ul{x} \approx x$. 
If $t$ is not $x$, then $e(\ul{t})$ is the result of replacing all appearances of symbols
of $\Phi$ by their right-hand sides in the given scheme.
%In other words, it is $t^*$ from above, considered as an
%element of $H_{\Sigma}(Z) + X$.

\comment{
Before we return to recursive
program scheme solutions, it will be helpful to go into details
on second-order substitution.
Let $Z$ be 
the set  $T_{\Sigma+\Phi} \{x\}$ of all  finite and infinite
$(\Sigma + \Phi)$-trees in one variable $x$.
Second-order substitution according to~\refeq{eq:algex}
 is an operation ${}^*: Z\to H_{\Sigma}(Z) +  \{x\}$ satisfying
the following equations:
$$\begin{array}{rcl@{\quad}rcl}
F(t,u)^* & = & F(t^*,u^*) & G(t)^* & =  & G(t^*) \\
\phi(t)^* & = &
F(t^*, \phi(G(t^*)))& 
\psi(t)^* & = &   F(\phi(G(t^*)), G(G(t^*))) \\
\end{array}
$$
For finite trees, it is obvious that there is a unique 
second-order substitution operation.  For infinite trees,
it is a classical result that we shall derive in great generality.
}

In more technical terms, one performs on $t$ the second-order substitution
which is given by the original recursive program scheme (read as an assignment
from left to right). This second-order subtitution provides a map
$\hatop: \Tsf \{\,x\,\} \to  \Tsf \{\,x\,\}$ satisfying
the following equations:
\begin{displaymath}
\begin{array}{rcl@{\qquad}rcl@{\qquad\qquad}rcl}
  \st{F(t,u)}  & = & F(\st t, \st u) & \st{G(t)} & =  & G(\st t) & \st x = x \\
  \st{\phi(t)} & = & F(\st t, \phi(G(\st t))) & 
  \multicolumn{6}{l}{
    \st{\psi(t)} = F(\phi(G(\st t)), G(G(\st t)))}
\end{array}
\end{displaymath}

In order to arrive in the codomain of $e$ one underlines the
maximum proper subtrees of $\st t$. So from~\refeq{eq:algex} we obtain the
following flat system of equations
\begin{equation}\label{eq:induced}
\begin{array}{rcl@{\quad}rcl}
  \ul x & \approx & x & 
  \ul{\phi(Gx)} & \approx & F(\ul{Gx}, \ul{\phi(GGx)}) \\
  \ul{\phi(x)} & \approx & F(\ul x, \ul{\phi(Gx)}) &
  \ul{F(x, \phi(Gx))} & \approx & F(\ul x, \ul{F(Gx, \phi(GGx))}) \\
  \ul{\psi(x)} & \approx & F(\ul{\phi(Gx)}, \ul{GGx}) &
  \ul{GGx} & \approx & G(\ul{Gx}) \\
  \ul{\phi(\psi(x))} & \approx & \multicolumn{4}{l}{
    F(\ul{F(\phi(Gx), GGx)}, \ul{\phi(G(F(\phi(Gx), GGx)))})} \\
  & \vdots 
\end{array}
\end{equation}
and so on.  Notice that each tree (here written as a term) on the right-hand side
is a $\Sigma$-tree which is either just a syntactic variable or \emph{flat};
i.\,e., one operation symbol from $\Sigma$ applied to formal variables. Notice
also that the formation of $e$ as described above relies on the fact that
the original recursive program scheme is guarded. Now $e$ is, of course, a
coalgebra for $H_\Sigma(\,\_\,) +X$, and it is well-known that the unique
homomorphism from $Z$ to the final coalgebra $T_\Sigma X$ assigns to every
formal variable its tree unfolding. Hence, this homomorphism assigns to the
formal variables $\ul{\phi(x)}$ and $\ul{\psi(x)}$ their uninterpreted solution
in the original recursive program scheme. 

At this point, we have explained {\em how\/} we solve recursive
program schemes in the algebraic semantics.  But much has been
omitted.  For example, we gave no general account of \emph{why}
ours or
any solution method works.  Our work to come does provide the
missing pieces on this, and much more.

\comment{
Incidentally, there is no real reason in the algebraic semantics literature to
work with trees in lieu of \emph{pointed term graphs modulo bisimulation}. In
fact, these provide just a different description of the very same final
coalgebras $T_\Sigma X$. For example, the system~\refeq{eq:algex} has
as its solution for  $\phi$ the pointed term graph
\begin{displaymath}
\xymatrix{
  *+[F-]{F} \ar[d]_0 \ar[r]^1 & F \ar[d]_0 \ar[r]^1 & 
  F \ar[d]_0 \ar[r]^1 & \cdots \\
  x & G \ar[l]_0 & G \ar[l]_0 & \cdots \ar[l]_0
  }
\end{displaymath}
where we just write the head symbols of the terms labelling the nodes. 
}

\subsubsection{Category Theoretic Semantics}

As the title of our paper suggests, our goal is to propose a category-theoretic
semantics of program schemes. The idea is to be even deeper than the algebraic
semantics, and to therefore obtain results that are more general. Our theory is
based on notions from category theory (monads, Eilenberg-Moore algebras) and
coalgebra (finality, solution theorems, Elgot algebras). The overall
assumptions are weak: there must be finite coproducts, and all functors we deal
with must have ``enough final coalgebras''. More precisely, we work in a
category $\A$ with finite coproducts and with functors $H: \A \to \A$ such that
for all objects $X$ a final coalgebra $TX$ for $H(\,\_\,) + X$ exists.  We
shall introduce and study recursive program schemes in this setting.  In
particular, we are able to prove the existence and uniqueness of interpreted
and uninterpreted solutions to schemes. The price we pay for working in such a
general setting is that our theory takes somewhat more effort to build. But
this is not excessive, and perhaps our categorical proofs reveal more
conceptual clarity than the classical ones.

We shall interpret schemes in Elgot algebras. The formal definition of
Elgot algebras appears in Section~\ref{section-Elgot-algebras}.
To be brief, they are algebras for a functor $H$ together with an
operation which assigns `solutions' to `flat equations in $A$'.
This operation must satisfy two easy and well-motivated axioms.
One of the key examples of an Elgot algebra is the algebra
$T_\Sigma X$ of all $\Sigma$-trees over $X$; and indeed, $T_\Sigma X$ is a free
Elgot algebra on $X$. This fact implies all of the structural
properties that are of interest when studying recursion.  But in addition,
there are many other interesting examples of Elgot algebras.  As it happens,
continuous algebras are Elgot algebras.  We shall show that our general results
on solving recursive program schemes in Elgot algebras directly specialize to
{\em least fixed-point recursion\/} in continuous algebras. This yields the
usual application of recursive program scheme solutions for the semantics of
functional programs such as~\refeq{eq:run}. Furthermore, algebras on complete
metric spaces with contracting operations are Elgot algebras, and so our
results specialize to the \emph{unique fixed-point recursion} in completely
metrized algebras provided by Banach's fixed-point theorem. This yields
applications such as the ones in~\refeq{eq-sin} and~\refeq{eq-cantor} above. 

\fi

\medskip
\noindent
{\bf Related Work.} The classical theory of recursive program schemes is
compactly presented by Guessarian~\cite{guessarian}. There one finds results
on uninterpreted solutions of program schemes and interpreted ones in
continuous algebras.

The first categorical accounts of infinite trees as monads of final coalgebras
appear independently at almost at the same time in work of the second
author~\cite{moss}, in Ghani et al.~\cite{glmp,glmp2}, and in Aczel et
al.~\cite{aav}. Furthermore, in~\cite{moss} and in~\cite{aav,aamv} it is shown
how solutions of formal recursive equations can be studied with coalgebraic
methods. In~\cite{aamv,milius} the monad of
final coalgebras is characterized as the free completely iterative monad,
generalizing and extending results of Elgot et al.~\cite{elgot,ebt}. Hence,
from~\cite{aamv,milius} it also follows how to generalize \emph{second-order
  substitution} of infinite trees (see Courcelle~\cite{courcelle}) to final
coalgebras.  The types of recursive equations studied in~\cite{moss,aamv} did
not go as far as program schemes.  It is thus an important test problem for
coalgebra to see if work on solving systems of equations can extend to
(un)interpreted recursive program schemes.  We are pleased that this paper
reports a success in this matter.

Ghani et al.~\cite{glm} obtained a general solution theorem with the aim of
providing a categorical treatment of uninterpreted program scheme
solutions. Part of our proof for the solution theorem for uninterpreted
schemes is essentially the same as their proof of the same
fact. However, the notion of recursive program scheme in~\cite{glm} is
different from ours, and more importantly, our presentation of recursive
program scheme solutions as fixed points with respect to (generalized)
second-order substitution as presented in~\cite{aamv} is new here. 

Complete metric spaces as a basis for the semantics of recursive program
schemes have been studied by authors such as Arnold and Nivat~\cite{an}.
Bloom~\cite{bloom} studied interpreted solutions of recursive program schemes
in so-called contraction theories. The semantics of recursively defined data
types as fixed points of functors on the category of complete metric spaces has
been investigated by Ad\'amek and Reitermann~\cite{are} and by America and
Rutten~\cite{aru}.  We build on this with our treatment of self-similar
objects.  These have also recently been studied in a categorical framework by
Leinster, see~\cite{l1,l2,l3}. The examples in this paper use standard results
on complete metric spaces, see, e.g.~\cite{barnsley}. We are not aware of any
work on recursive program schemes that even mentions connections to examples
like self-similar sets in mathematics, let alone develops the connection.

Finally, some of the results of this paper have appeared in our
extended abstract~\cite{mm}.   However, most of the technical details and all 
of the 
proofs were omitted. This full version explains our theory in much more detail
and provides full proofs of all new results.

\medskip
\noindent{\bf Contents.\/}
The paper is structured as follows:  Section~\ref{section-background-monads}
contains a brief summary of the definitions concerning monads.
It also states the overall assumptions that we make in the 
rest of the paper.  Section~\ref{section-Elgot-algebras} presents
the notions of {\em completely iterative algebra\/}
and {\em Elgot algebra}, following~\cite{amv3}.  
Except for the Section~\ref{section-EASUEAC},
none of the results in this section is new.  But the paper as
a whole would not make much sense to someone unfamiliar
with completely iterative algebras and Elgot algebras.  So we have included
sketches of proofs in this section.
The completely iterative monads of Section~\ref{section-completely-iterative-monads}
are also not original. The properties of them developed
in Section~\ref{sec:main} are needed for the work on
uninterpreted schemes (Section~\ref{sec:sol}) and then interpreted 
schemes (Section~\ref{section:interpreted}).  These are the heart of the paper.
For the convenience of the reader we have included a list of notations and 
an index at the end. 

\section{Background: Iteratable Functors and Monads}
\label{section-background-monads}

This section contains most of the background which we need
and also connects that background with recursive program schemes.
Before reviewing monads in Section~\ref{section-monads} we should mention
the main base categories of interest in this paper, and   our overall
assumptions on  those categories and endofunctors on them.

\subsection{Iteratable Functors}
\label{sec:ass}\index{functor!iteratable $\sim$}

Throughout this paper we assume that a category $\A$ with finite coproducts
(having monomorphic injections) is given. In addition, all endofunctors $H$ on
$\A$ we consider are assumed to be \emph{iteratable} [sic]: for each object
$X$, the functor $H(\,\_\,) + X$ has a final coalgebra.
We denote for an iteratable endofunctor $H$ on $\A$ by 
\begin{displaymath}\label{p:thx}\T H X\end{displaymath}
the final coalgebra  for $\HX$.
We shall later see that $\T{H}$ is the object assignment of a monad.  
  Whenever confusion is unlikely we will drop 
  the parenthetical $(H)$ and simply write
  $T$ for $\T H$. By the Lambek
  Lemma~\cite{lambek}, the structure morphism
  \begin{displaymath}\label{eq:alpha}
    \alpha^H_X: \T{H}X \to H\T{H}X + X
  \end{displaymath}
  of the final coalgebra is an isomorphism, and consequently, $\T{H}X$ is a
  coproduct of $H\T{H}X$ and $X$ with injections 
  \begin{displaymath}\label{p:taueta}
  \begin{array}{rll}
    \eta^H_X & : X \to \T{H}X & 
    \qquad \textrm{``injection of variables''} 
    \\
    \tau^H_X & : H\T{H}X \to \T{H}X & 
    \qquad \textrm{``$\T{H}X$ is an $H$-algebra''}
  \end{array}  
  \end{displaymath}
  Again, the superscripts will be dropped if confusion is unlikely.
  
  Having coproduct injections that are monomorphic is a mere technicality and
  could even be totally avoided, see~\cite{milius2}, Sections~4 and~5. Dropping
  that assumption would mean replacing the notions of ideal monads and their
  homomorphisms in Section~\ref{section-completely-iterative-monads} by two
  slightly more involved notions.  Therefore we decided to keep the additional
  assumption to simplify our presentation.

More serious is the assumption of iteratability.    This is a mild
assumption: experience shows most endofunctors of interest are iteratable.

\index{$\Sigma$-trees (over $X$)}
\begin{exmp}\label{ex:signature}
  We recall that ordinary signatures of function symbols $\Sigma$ (as in
  general algebra) give functors on $\Set$.  This is one of the central
  examples in this paper because it will allow us to recover the classical
  algebraic semantics from our category-theoretic one.  A signature may be
  regarded as a functor $\Sigma: \Nat \to \Set$, where $\Nat$ is the discrete
  category with natural numbers as objects.  For each $n$, write $\Sigma_n$ for
  $\Sigma(n)$; this is the set of function symbols of arity $n$ in $\Sigma$.
  There is no requirement that different numbers should give disjoint sets of
  function symbols of those arities.  Given any signature $\Sigma$ there is an
  associated polynomial endofunctor (henceforth called a \emph{signature
  functor}\index{functor!signature $\sim$}\index{functor!polynomial $\sim$}) 
  \begin{equation}\label{eq:poly}
    H_\Sigma X = \coprod\limits_{i < \omega} \Sigma_i \times X^i
  \end{equation}
  on $\Set$. 
  When we need to refer to elements of $H_\Sigma X$, we shall use the notation
  $(f,\vec{x})$ for a generic element of $H_\Sigma X$;
  $n$ is understood in this notation, $f\in \Sigma_n$, and $\vec{x} = (x_1,
  \ldots, x_n)$ is an $n$-tuple of elements of $X$. Also observe that on
  morphisms $k: X \to Y$ the action of $H_\Sigma$ is given by   
  $H_{\Sigma} k (f,\vec{x}) = (f, \overrightarrow{kx})$, where we write
  $\overrightarrow{kx}$ for $(kx_1, \ldots, kx_n)$.  Notice that if $f$ is a
  constant symbol from $\Sigma_0$, then $\vec{x}$ and $\overrightarrow{kx}$ are
  the unique empty tuple. 
  
  Signature functors $H_\Sigma$ of $\Set$ are iteratable. In fact, consider 
  the set
  \iffull
  \begin{displaymath}\label{p:trees}T_\Sigma X\end{displaymath}
  \else $T_\Sigma X$ \fi 
  of finite and infinite $\Sigma$-labelled
  trees with variables from the set $X$.  More precisely, $T_\Sigma X$ consists
  of ordered and rooted trees
  labelled so that a node with $n$ children, $n > 0$, is labelled by an
  operation symbol from $\Sigma_n$, and leaves are labelled by constant symbols
  or variables (elements of $X + \Sigma_0$).

  This set $T_\Sigma X$ is the carrier of a final coalgebra for
  $H_\Sigma(\,\_\,) + X$. The coalgebra structure is the inverse of the pairing
  of the maps $\tau_X: H_\Sigma T_\Sigma X \to T_\Sigma X$ and $\eta_X : X \to
  T_\Sigma X$. The map $\tau_X: (f, \vec{t}) \mapsto f(\vec{t})$ performs
  ``tree tupling'': it takes an $n$-ary operation symbol $f$ from $\Sigma$ and
  an $n$-tuple $\vec{t}$ of $\Sigma$-trees and returns the $\Sigma$-tree
  obtained by joining all the trees from $\vec{t}$ with a common root node
  labelled by $f$. And $\eta_X$ is the obvious injection which
  regards each variable as a one-point tree.
\label{exmp-signature-functors}
\end{exmp}

So at this point we have seen signature functors on $\Set$ as examples of
iteratable functors.  Unfortunately, we must admit that iteratability is not a
very nice notion with respect to closure properties of functors---for example,
iteratable functors need not be closed under coproducts or composition.  But the
main examples of base categories $\A$ in this paper are $\Set$, $\CPO$ and
$\CMS$. In these categories there are stronger yet much nicer conditions that
ensure iteratability.

\begin{exmp}\label{ex:cats_i}
  Accessible endofunctors of $\Set$. Let $\lambda$ be a regular cardinal.
  An endofunctor $H$ of the category $\Set$ of sets and functions is called
  \emph{$\lambda$-accessible}\index{functor!accessible $\sim$} it it preserves $\lambda$-filtered colimits. It
  is shown in~\cite{ap}, Proposition~5.2, that $\lambda$-accessible functors
  are precisely those endofunctors where for each set $X$ any element $x \in
  HX$ lies in the image of $Hm: HY \to HX$ for some subset $Y \subto X$ of
  cardinality less than $\lambda$.  As usual, we call an endofunctor $H$
  \emph{finitary} if it is $\omega$-accessible and we call $H$
  \emph{accessible} if it is $\lambda$-accessible for some regular cardinal
  $\lambda$.  Any accessible endofunctor is iteratable, see~\cite{barr}.  In
  particular, signature functors are finitary, whence iteratable (as we already
  know).
\end{exmp}

\begin{exmp}\label{ex:cats_ii}\index{cpo|see{complete partial
  order}}\index{complete partial order}\index{functor!locally continuous $\sim$}
%\index{locally continuous functor}
\index{map!continuous $\sim$}
  Locally continuous endofunctors. $\CPO$ is the category of $\omega$-complete
  partial orders (partially ordered sets with joins of all increasing
  $\omega$-chains, but not necessarily with a least element $\bot$). Morphisms
  of $\CPO$ are the continuous functions (preserving joins of all
  $\omega$-chains). Hence the morphisms are monotone (preserve the order).
  Notice that coproducts in $\CPO$ are disjoint unions with elements of
  different summands incomparable. $\CPO$ is understood to be enriched in the
  obvious way. For given cpo's $X$ and $Y$, the hom-set $\CPO(X,Y)$ is a cpo in
  the pointwise order. An endofunctor $H$ on $\CPO$ is \emph{locally
    continuous} if it preserves the extra structure just noted---that is, each
  derived function $\CPO(X,Y) \to \CPO(HX, HY)$ is continuous.  Observe that
  not all locally continuous functors need be iteratable.  For a counterexample
  consider the endofunctor assigning to a cpo $X$ the powerset of the set of
  order components of $X$. This is a locally continuous endofunctor but it does
  not have a final coalgebra. However, any accessible endofunctor $H$ on $\CPO$
  has a final coalgebra, see~\cite{barr}, and moreover, $H$ is iteratable.
\end{exmp}

\begin{exmp}\label{ex:cats_iii}\index{complete metric space}
  Contracting endofunctors of complete metric spaces.  $\CMS$ is the category
  of complete metric spaces with distances measured in the interval $[0,1]$
  together with maps $f: X \to Y$ such that $d_Y(fx,fy)
  \leq d_X(x,y)$ for all $x,y \in X$.  
These maps are called \emph{non-expanding}\index{map!non-expanding $\sim$}.
A stronger condition is that $f$ be
  \emph{$\eps$-contracting}\index{map!contracting
  $\sim$}: for some $\eps < 1$ we have that $d_Y(fx,fy) \leq
  \eps\cdot d_X(x,y)$ for $x, y \in X$. Again, $\CMS$ is understood to be
  enriched. For complete metric spaces $(X, d_X)$ and $(Y, d_Y)$ the hom-set
  $\CMS(X,Y)$ is a complete metric space with the metric given by
  \begin{displaymath}
    d_{X,Y}(f, g) = \sup\limits_{x \in X} d_Y(f(x), g(x))\,.
  \end{displaymath}
    
  Recall that a functor $H$ on $\CMS$ is called
  $\eps$-\emph{contracting}\index{functor!contracting $\sim$} if
  there exists a constant $\eps < 1$ such that each derived function $\CMS(X,
  Y) \to \CMS(HX, HY)$ is an $\eps$-contracting map; that is,
  \iftcs $\else\begin{displaymath}\fi
    d_{HX,HY}(Hf,Hg) \leq\eps\cdot d_{X,Y}(f,g)%
    \iftcs\/$ \else\end{displaymath}\fi
  for all non-expanding maps
  $f,g: X \to Y$.  Contracting functors on $\CMS$ are iteratable,
  see~\cite{are}.
\end{exmp}

\begin{construction}\label{constr:coalg}
  Let $H$ be an endofunctor of $\Set$. We recall here that a final coalgebra
  $T$ for $H$  can (if it exists) be constructed as a limit of an
  \mbox{(op-)chain}. Let us define by transfinite recursion the following chain
  indexed by all ordinal numbers:

  $
  \begin{array}{l@{\quad}rcl@{\qquad}rcl}
    \textrm{Initial step:} & 
    T_0 & = & 1, & t_{1,0} & \equiv & \xymatrix@1{H1 \ar[r]^{!} &  1}, 
    \\
    \rule{0mm}{15pt}
    \textrm{Isolated step:} & 
    T_{\beta+1} & = & HT_\beta, & t_{\beta+1,\gamma+1} & \equiv & 
    \xymatrix@1{
      HT_\beta \ar[r]^-{Ht_{\beta,\gamma}} & HT_\gamma
      }
    \\
    \rule{0mm}{15pt}
    \textrm{Limit step:} & 
    T_\lambda & = & \lim\limits_{\beta < \lambda} T_\beta &
    \multicolumn{3}{l}{%
      \textrm{with limit cone $t_{\lambda,\beta}: T_\lambda \to T_\beta$, $\quad \beta <  \lambda$,}%
      }
  \end{array}
  $

  where the connecting map $t_{\lambda+1,\lambda}$ is uniquely determined by
  the commutativity of the squares
  \begin{displaymath}
    \vcenter{
      \xymatrix@C+1pc{
        T_{\beta + 1} = HT_\beta
        \ar[d]_{t_{\beta+1,\beta}}
        &
        HT_\lambda = T_{\lambda+1}
        \ar[l]_-{H t_{\lambda,\beta}}
        \ar@{.>}[d]^{t_{\lambda+1,\lambda}}
        \\
        T_\beta
        &
        T_\lambda
        \ar[l]^-{t_{\lambda,\beta}}
        }
      }
      \qquad \beta < \lambda\,.
  \end{displaymath}
  This chain it said to \emph{converge} if $t_{\beta+1, \beta}$ is an
  isomorphism for some ordinal number $\beta$. 
\end{construction}

It has been proved by Ad\'amek and Koubek~\cite{ak} that an endofunctor $H$ has
a final coalgebra iff the above chain converges; moreover, if $\beta$ is an
ordinal number such that $t_{\beta+1,\beta}$ is invertible, then $t_{\beta+1,\beta}^{-1}: T_\beta \to HT_\beta$
is a final coalgebra for $H$. For many set endofunctors one can give a bound
for the number of steps it will take until the above final coalgebra chain
$T_\beta$ converges. The following result has been established by
Worrell~\cite{w}. 
\begin{thm}
  For a $\lambda$-accessible endofunctor $H$ of $\Set$ the final coalgebra
  chain $T_\beta$ converges after $\lambda \o 2$ steps, and $(T_{\lambda\o 2},
  t_{\lambda\o 2 + 1, \lambda \o 2}^{-1})$ is a final coalgebra for $H$. 
\end{thm}
In particular, for a finitary endofunctor a final coalgebra is obtained after
$\omega + \omega$ steps.  For some functors one can further improve on this
bound. For endofunctors preserving limits of countable op-chains the final
coalgebra chain converges after countably many steps so that $(T_\omega,
t_{\omega +1, \omega}^{-1})$ is a final coalgebra. For example, each signature
functor $H_\Sigma$ of $\Set$ preserves limits of op-chains. 

We mention additional examples of iteratable endofunctors $H$ with their
final coalgebras $TX$ on the categories of interest. 
\begin{examples}\label{ex:finitary} 
   \iftcs\else\par\noindent\fi%
  \begin{myenumerate}
  \myitem A functor $H: \Set\to \Set$ is finitary (it preserves
    filtered colimits) iff it is a quotient of some polynomial functor
    $H_\Sigma$, see~\cite{at}, III.4.3. The latter means that we have a natural
    transformation $\eps: H_\Sigma \to H$ with epimorphic components $\eps_X$.
    These components are fully described by their kernel equivalence whose
    pairs can be presented in the form of so-called basic equations
    \begin{displaymath}\sigma(x_1, \ldots, x_n) = \rho(y_1, \ldots, y_m)\end{displaymath}
    for $\sigma \in \Sigma_n$, $\rho \in \Sigma_m$ and $(\sigma, \vec{x}),
    (\rho, \vec{y}) \in H_\Sigma X$ for some set $X$ including all $x_i$ and
    $y_j$. Ad\'amek and Milius~\cite{am} have proved that the final coalgebra
    $TX$ for $H(\,\_\,) + X$ is given by the quotient $T_\Sigma
    X/\mathord{\sim_X}$ where $\sim_X$ is the following congruence: for every
    $\Sigma$-tree $t$ denote by $\cut_n t$ the finite tree obtained by cutting
    $t$ at level $n$ and labelling all leaves at that level by some symbol
    $\bot$ not from $\Sigma$. Then we have $s \sim_X t$ for two $\Sigma$-trees
    $s$ and $t$ iff for all $n < \omega$, $\cut_n s$ can be obtained from
    $\cut_n t$ by finitely many applications of the basic equations describing
    the kernel of $\eps_X$.  For example, the functor $H$ which assigns to a
    set $X$ the set $\{\,\{\,x,y\,\}\mid x,y \in X\,\}$ of unordered pairs of
    $X$ is a quotient of $H_\Sigma X = X \times X$ expressing one binary
    operation $b$ where $\eps_X$ is presented by commutativity of $b$; that is, 
    by the basic equation $b(x,y) = b(y,x)$.  And $TX$ is the coalgebra of all
    unordered binary trees with leaves labelled in the set $X$.

  \myitem Consider the finite power set functor $\Pfin: \Set \to \Set$. Under
    the Anti-Foundation Axiom (AFA)~\cite{aczel:88}, its final coalgebra is the set $H\!F_1$ of
    hereditarily finite sets; see~\cite{vc}. Analogously, the final coalgebra
    for $\Pfin(\,\_\,) + X$ is the set $H\!F_1(X)$ of hereditarily finite sets
    generated from the set $X$. Even without AFA, the final
    coalgebra for $\Pfin$ may be described as in Worrell~\cite{w}; it is the
    coalgebra formed by all strongly extensional trees.  These are unordered trees
    with the property
    that for every node the subtrees defined by any two different children
    are not bisimilar.  Analogously, the final coalgebra for $\Pfin(\,\_\,) +
    X$ is the coalgebra of all strongly extensional trees where some leaves
    have a label from the set $X$.
    
  \myitem The (unbounded) power set functor $\mathcal{P}: \Set \to \Set$ does
    not have a final coalgebra, whence it is not iteratable.  However, moving
    to the category of classes and functions between them, the power set
    functor turns out to be iteratable, see e.\,g.~\cite{amv_class}.  Indeed,
    some of the machinery that comes from iteratable functors turns out to have
    a surprisingly set-theoretic interpretation when specialized to this
    setting; see~\cite{moss_uniform}.
  \end{myenumerate}
\end{examples}

\begin{exmp}\label{ex:lift_cpo}\index{lifting|see{lifted
    functor}}\index{functor!lifted $\sim$!on $\CPO$}
  In our applications, the key point is that certain $\Set$ endofunctors
  lift to (iteratable) endofunctors on $\CPO$. And we need to know that those
  liftings are locally continuous. In fact, let $H$ be an iteratable $\Set$
  functor with a locally continuous lifting $H'$ on $\CPO$.  Thus 
    $H'$ is a functor
  and the  forgetful functor $U: \CPO \to \Set$ gives a
  commutative square
  \begin{displaymath}
    \xymatrix{
      \CPO 
      \ar[r]^-{H'}
      \ar[d]_U
      &
      \CPO 
      \ar[d]^U 
      \\
      \Set 
      \ar[r]_-H
      &
      \Set
      }
  \end{displaymath}
  Then $H'$ is iteratable, and moreover, the final coalgebra functor
  $\T{H'}$ is a lifting of $\T{H}$:
  \begin{equation}\label{eq:cpolift}
    \vcenter{
      \xymatrix{
        \CPO
        \ar[r]^-{\T{H'}}
        \ar[d]_U
        &
        \CPO
        \ar[d]^{U}
        \\
        \Set
        \ar[r]_-{\T{H}}
        &
        \Set
        }
      }
  \end{equation}
  To see this, first recall that for any set $X$ the final coalgebra $\T{H}X$
  is obtained from the final coalgebra chain $T_\beta$ of $H(\,\_\,) + X$, see
  Construction~\ref{constr:coalg}. In fact, $\T{H}X$ is the coalgebra $(T_\gamma,
  t^{-1}_{\gamma+1,\gamma})$ for the smallest ordinal number $\gamma$ for which
  $t_{\gamma+1,\gamma}$ 
  is invertible. As $\CPO$ is a complete category we can define for any
  endofunctor $H'$ of $\CPO$ a final coalgebra chain $T_\beta'$ in precisely the
  same way as in~\ref{constr:coalg}. Since the forgetful functor $U$
  preserves limits, it follows that for a cpo $X$ the final coalgebra chain
  of $H'(\,\_\,) + X$ has the $T_\beta$ as underlying sets. However, in $\CPO$
  the continuous map $t_{\gamma+1,\gamma}$ might not be invertible. But since the chain
  of underlying sets converges at index $\gamma$ we know that for all ordinal
  numbers $\delta$ the connecting maps $t_{\gamma+\delta,\gamma}:
  T_{\gamma+\delta} \to T_\gamma$ are monomorphisms of $\CPO$. Moreover, all
  cpos $T_{\gamma+\delta}$ have (up to 
  isomorphism) the same underlying set $T_\gamma$ and therefore the partial orders
  on the $T_{\gamma+\delta}$, $\delta \geq 0$, form a decreasing chain of
  subsets of $T_\gamma \times T_\gamma$. This implies that the final coalgebra
  chain has to converge at some index $\gamma + \delta$, $\delta \leq
  \card(T_\gamma \times T_\gamma)$. By standard arguments it follows that the
  cpo $T_{\gamma+\delta}$ is the final coalgebra for $H'(\,\_\,) + 
  X$. Thus, we may choose $\T{H'}X = T_\gamma$ equipped with the cpo structure
  given by the cpo $T_{\gamma+\delta}$, whence the square~\refeq{eq:cpolift}
  commutes as desired.
  
  For example, every signature functor $H_\Sigma$ has a locally continuous
  and iteratable lifting $H'$. This lift is the functor
  \begin{displaymath}
    H' X = \coprod\limits_{i< \omega} \Sigma_i \times X^i
  \end{displaymath}
  on $\CPO$.  Here each $\Sigma_n$ is a discretely ordered set,
  $+$ is the coproduct of $\CPO$ (a lift of the coproduct of $\Set$),
  and $\times$ the usual product.   It should be noted that even if $X$ has a
  least element, $H' X$ almost never has one. Finally, $\T{H'}X$
  is the $\Sigma$-tree algebra $T_\Sigma X$ with the order induced by the
  order of the cpo $X$---we describe this order in more detail later in
  Example~\ref{ex:intsolcpo}(i).
\end{exmp}

\begin{exmp}\label{ex:lift_cms}\index{functor!lifted $\sim$!on $\CMS$}
  Let $H: \Set\to \Set$  have  a contracting lifting $H'$ on
  $\CMS$;
  so we have
  \begin{displaymath}
    \xymatrix{
      \CMS
      \ar[r]^-{H'}
      \ar[d]_U
      & 
      \CMS
      \ar[d]^U
      \\
      \Set
      \ar[r]_-H
      & 
      \Set
      }
  \end{displaymath}
  for $U: \CMS \to \Set$ the forgetful functor.  Then $H$ is
  iteratable and $U \o \T{H'} = \T{H} \o U$. In fact, this follows from the
  results of~\cite{are}, since $U$ preserves limits. Any signature functor
  $H$ on $\Set$ has a contracting lifting to $\CMS$.  For $HX = X^n$, define
  $H'(X,d) = (X, \frac{1}{2} d_\mathrm{max})$ (where $d_\mathrm{max}$ is the
  maximum metric).  This is a contracting functor with $\eps = \frac{1}{2}$.
  And coproducts of $\frac{1}{2}$-contracting liftings are
  $\frac{1}{2}$-contracting liftings of coproducts. The final coalgebra
  $\T{H'}X$ is the $\Sigma$-tree algebra $T_\Sigma X$ equipped with a
  suitable complete metric. We will provide details of this metric later in
  Remark~\ref{rem:treemetric}.  
\end{exmp}

\begin{rem}
  We comment on the way we formulated the category $\CPO$. Recall from our
  first mention of it in Example~\ref{ex:cats_ii} that we did not insist that
  objects in $\CPO$ have a least element. This might have seemed mysterious at
  the time, but now we can explain the reason for this. Requiring a least
  element and working with strict maps implies that coproducts would
  \emph{not} be based on the disjoint union of posets. So the lifting property
  that we saw in Example~\ref{ex:lift_cpo} above would fail. Since this
  property is important for all of our work with $\CPO$, it dictated the
  formulation.
\end{rem}

\iffull
\subsection{Monads}
\label{section-monads}\index{monad}

A {\em monad\/} on a category $\A$ is a triple $(T,\mu,\eta)$  consisting
of a functor 
$T : \A \to \A$, and natural transformations
 $\mu : TT \to T$  and $\eta : \Id \to T$, satisfying  
the {\em unit laws\/} $\mu\o T\eta = \mu \o \eta T = \id$, and 
the {\em associative law\/}   $\mu \o T\mu = \mu \o \mu T$:
\begin{displaymath}
\xymatrix{
  T \ar[r]^{T\eta} \ar@{=}[rd] & TT \ar[d]^\mu & 
  T \ar[l]_{\eta T} \ar@{=}[ld] \\
  & T
}
\qquad 
\xymatrix{ TTT \ar[r]^{T\mu} \ar[d]_{\mu T}  & TT \ar[d]^{\mu} \\
TT \ar[r]_{\mu} & T \\
}
\end{displaymath}
For a functor $H: \A \to \A$ and a natural transformation $\alpha: F \to G$,
 we
use the usual notations $H\alpha$ and $\alpha H$ to denote the natural
transformations with components $H(\alpha_X)$ and $\alpha_{HX}$, respectively. 
%Looking at our statement $\mu \o \eta T = T$ the reader can see
%that  we usually use the same notation for a functor and for
%the evident natural transformation from that functor to itself.  
%We do the same with objects, writing the name of the object for the identity
%morphism on it. 
Also it is customary to write just $T$ for the monad in lieu of
the triple, and we will follow this convention.

Let $(S,\eta,\mu)$ and $(T,\eta',\mu')$ be monads
on the same category $\A$.  A {\em morphism of monads\/}%
\index{monad!morphism of $\sim$} $\phi$ from $S$
to $T$ is a natural transformation $\map{\phi}{S}{T}$ such that 
$\phi\o \eta = \eta'$, and $\phi\o \mu = \mu'\o (\phi * \phi)$:
\begin{displaymath}
\xymatrix 
{\Id_{\A} \ar[dr]_{\eta'}
\ar[r]^{\eta} & S \ar[d]^{\phi} \\
 & T 
}
\qquad
\xymatrix  
{SS \ar[r]^{\mu} \ar[d]_{\phi * \phi} & S  \ar[d]^{\phi}\\
TT \ar[r]_{\mu'} & T}
\qquad
\end{displaymath}
This operation $*$ is the {\em parallel composition\/}\index{parallel composition} of natural
transformations.  In general, if $\alpha : F\to G$ and $\beta: H \to K$ are
natural transformations, $\alpha * \beta: FH \to GK$ is $\alpha{K}\o F\beta =
G\beta\o\alpha{H}$.
\begin{displaymath}\label{p:parallel}
  \xymatrix{
    FH \ar[r]^{F\beta} \ar[d]_{\alpha{H}} \ar[rd]|{\alpha*\beta}
    & 
    FK \ar[d]^{\alpha{K}}
    \\
    GH \ar[r]_{G\beta} 
    & 
    GK
    }
\end{displaymath}
From naturality one easily infers the \emph{double interchange
  law}\index{double interchange law} which
states that for $\alpha$ and $\beta$ as above and $\alpha': G \to G'$ and
$\beta': K \to K'$, we have
\begin{equation}\label{eq:dil}
  (\alpha' * \beta') \o (\alpha * \beta) = (\alpha' \o \alpha) * (\beta' \o
  \beta) 
\end{equation}

We will denote by 
\begin{displaymath}\label{p:MA}\MA\end{displaymath}
the category of monads on $\A$ and their morphisms.
\fi

\begin{exmp}\label{ex:monad}
  Let $H_\Sigma$ be a signature functor on $\Set$.  We already know how to
  define an object assignment $T_\Sigma$.  In fact, $T_\Sigma$ is a functor.
  We also have a natural transformation $\eta : \Id \to T_\Sigma$ which for
  each set $X$ regards the elements of $X$ as elements of $T_\Sigma X$.  We
  additionally define a natural transformation $\mu_X: T_\Sigma(T_\Sigma X)\to
  T_\Sigma X$, the operation which takes trees over trees over $X$ into trees
  over $X$ in the obvious way.  In this way, we have a monad
  $(T_\Sigma,\mu,\eta)$ on $\Set$.
\end{exmp}   

\begin{exmp}\label{ex:monad2}
  More generally, let $H$ be iteratable on a category $\A$.
  It has been shown in  previous work~\cite{aamv,milius2} that the object
  assignment $T$ (assigning to each object $X$ the final coalgebra for
  $\HX$) gives rise to a monad on $\A$. In fact, in loc.~cit.~this monad is
  characterized by a universal property---it is the free \emph{completely
  iterative monad} on $H$. We will recall the notion of completely iterative
  monad and the result characterising $T$ in detail later in
  Sections~\ref{sec:dramatis} and~\ref{section-completely-iterative-monads}
  below. 
\end{exmp}

\comment{
\begin{exmp} \label{ex:monad:cpobottom}
Coproducts in $\CPO_\bot$ are disjoint unions of partial orders
 with the $\bot$ elements identified.
We also require a version of Example~\ref{ex:monad} for   $\CPO_\bot$.
Again, a signature is a functor $\Sigma: \Nat \to \Set$.
(That is, we do not need an order on the sets $\Sigma_n$.)
The difference is that now we associate to $\Sigma$ the following 
functor $H'_{\Sigma}$ on $\CPO_\bot$: Given a $\CPO_\bot$ $X$,
the set underlying  $H'_{\Sigma}X$ is $H_{\Sigma}UX\cup \{\bot\}$,
where $H$ is as in Example~\ref{ex:monad} and
 $\bot$ is a new element.  The order is the obvious one:
$\bot$ is below everything, and $(f,\vec{x}) \leq (g,\vec{y})$
iff $f =g $ and $\vec{x} \leq \vec{y}$ in the pointwise
order on $X^n$.  The
functor $H'_{\Sigma}$ acts on morphisms in the obvious (strict) way: extend 
the action on sets by preserving $\bot$.

There is some additional relevant structure.   First, $H'_{\Sigma}$ is
\emph{locally  continuous}: the action of $H'$ on the hom-sets
$\CPO_{\bot}(X,Y)$ preserves joins of chains.
 Second, let $\Sigma_\bot$  be 
$\Sigma$ with a new constant element $\bot$.  Then  $H'_\Sigma$ is a 
\emph{lifting} of
 $H_{\Sigma_{\bot}}$ in the sense that the square
    \begin{displaymath}
    \xymatrix{
      \CPO_\bot 
      \ar[r]^-{H'_\Sigma}
      \ar[d]_U
      & 
      \CPO_\bot
      \ar[d]^U
      \\
      \Set
      \ar[r]_-{H_{\Sigma_\bot}}
      & 
      \Set
      }
    \end{displaymath}
commutes,  where $U:
    \CPO_\bot \to \Set$  is  the forgetful functor.
\end{exmp}

We aim lift of the monad structure $(T_\Sigma,\eta,\mu)$ to $\CPO_\bot$.
However, the discrepancy between the coproduct structures on the two categories
complicates results of this type.  We shall use not only the forgetful functor
$U:  \CPO_\bot \to \Set$ but also  the obvious restriction
 $U_0 : \CPO_\bot \to \Set$ taking
a cpo $X$ to $UX\setminus \{\bot\}$.
The strictness of maps in $\CPO_\bot$ also gives a natural transformation
$\iota : U_0 \to U$.

\begin{prop}
Let $\Sigma$ be a signature, let $\Sigma_{\bot}$ be $\Sigma$ with a 
new constant $\bot$, and let
$H, H_\bot : \Set \to \Set$ be the associated functors.
Let $(T_\bot,\alpha,\eta,\mu)$  be such that
$(T_\bot,\eta, \mu)$ is a monad, and for every set
$(T_\bot X,\alpha_X)$ is a final $\HX$-coalgebra.
Then there is a tuple  $(T',\alpha',\eta',\mu')$  such that
\begin{myenumerate}
\myitem  $(T',\eta',\mu')$ is a monad on $\CPO_\bot$.
\myitem $U\o T' = T_{\bot} \o U_0$.
\myitem  $U\alpha' = \alpha U_0$.
\myitem $\iota T' \o U_0\eta' = \eta {U_0}$.
\myitem for
all cpo's $X$, $(T'X, \alpha')$ is a final $\HprimeX$ coalgebra.
\end{myenumerate}
\label{prop-lifting}
\end{prop}

\begin{proof}  
For a cpo $X$, we write $U_0 X$ for the set $UX\setminus\{\bot\}$.
Note that we have dropped the bottom element; we recover this
when we pass to the set $H_{\bot}(U_0 X)$.

Let $T' X$ be the set $T_\bot U_0 X$.
We equip $T' X$ with  an order relation described in the following
way: $t\leq u$ if $u$ may be obtained from $t$ by (i) replacing
nodes of $t$ labeled $\bot$   with arbitrary trees, and (ii) replacing
 node labels in $X$ with $\leq$-larger labels.
  It is not hard to show that every chain 
in $\leq$ has a least upper bound: define the structure of this
least upper bound going down the tree.
  We then set $T'X$ be the resulting cpo.

One checks that $\alpha_{U_0 X}$  is continuous and then takes
$\alpha'_X$ to  be this map $\alpha_{U_0 X}$. 

Let $\eta'_X$ be defined in the obvious way: for $x\neq \bot$,
$\eta'_X(x)$ is the one-point tree determined by $X$ in the second summand of
$\HbotX$, and as for $\bot$, $\eta'_X(\bot)$ comes from the first 
summand via the constant symbol $\bot$ of $\Sigma_\bot$.s  
It is straightforward to verify that $\iota T' \o U_0\eta' = \eta {U_0}$.

Now we check the finality of $T'X$.  Suppose that $e : Y \to H'Y + X$
is an $H'$-coalgebra.  Then \begin{displaymath}\can\o Ue : UY \to H_{\bot} Y + U_0 X,\end{displaymath}
where 
\begin{displaymath}\can : U(H'Y + X) \to H_{\bot}UY + U_0 X \end{displaymath}
sends the bottom of $H'Y + X$ to $\bot\in H_{\bot} UY$.
By finality, let $f: UY \to T_\bot U_0 X$ be such that 
\begin{equation}
\alpha_{U_0 X} \o f = (H_\bot f + U_0 X)\o \can\o Ue
\label{eq-f-for-continuity}
\end{equation}
We shall check below that $f$ is continuous so that for some $f': Y \to T'X$,
$f = Uf'$.   
Then we have \begin{displaymath}(H_\bot f + U_0 X)\o \can\o Ue 
 = U((H'f' + X)\o e)\end{displaymath} so that 
\begin{displaymath} \alpha'_X \o  f' = (H'f' + X) \o e.\end{displaymath}
Therefore $f'$   will be a coalgebra morphism, as desired.

For the uniqueness, if $g : Y \to T'X$ is a coalgebra morphism,
then $\alpha'_X\o g = (H'g + X)\o e$.  Applying $U$ to both sides
gives \begin{displaymath}\alpha_{U_0 X} \o Ug = U(H'g + X) \o Ue = (H_\bot Ug + U_0 X)\o \can \circ Ue.\end{displaymath} 
  Thus by finality 
of $T_{\bot}U_0 X$, $Ug = f =  Uf'$.  So $g = f'$.

We omit all of the verifications of the continuity of the maps,
including especially the continuity of $f$ from (\ref{eq-f-for-continuity}).
\comment{
We conclude with the verification
of the  continuity of $f$ as in (\ref{eq-f-for-continuity}).
The other continuity arguments in this proof are similar.
It is easy to check that $f$ is monotone by using our definition
of the order $\leq$. Indeed, if $y\leq z$, then $f(y)$ and $f(z)$ 
are trees with all of the same labels from $\Sigma$; the labels from
$X$ are related in the appropriate ways.  So we check that $f$
commutes with joins of chains.

Consider the functor $C :\CPO_\bot \to \CPO_\bot$ taking $X$
to the set of (possibly empty) chains $s\subseteq X$,
taken to be a cpo under inclusion.   The action on
morphisms $h: X\to Y$ is $Ch(s) = h[s]$.  There is a natural
transformation $\gamma : C \to \Id$ given by $\gamma_X(s) = \bigsqcup s$.
The naturality here amounts to the definition of continuity
for $\CPO$s defining the category.  By monotonicity, we have $f^* : U(CY) \to U(C T_\bot U_0 X)$
again by $f^*(s) = f[s]$. 
The continuity of  
$f: UY \to T_\bot U_0 X$  is equivalent to the assertion that
$\gamma_{T_\bot U_0 X} \o f^* = f\o U\gamma_Y$.
}\end{proof}
}

\subsection{Recursive Program Schemes}
\label{section-rpsfirst}\index{recursive program scheme!explanation of}
We shall now explain how to capture recursive program schemes in a
category-theoretic way. In order to do this we use the fact that there is a
bijective correspondence between natural transformations from a signature
$\Sigma$ and from its signature functor $H_\Sigma$. More precisely, every
signature $\Sigma$ can be regarded as a functor $\Sigma: \Nat \to \Set$, where 
$\Nat$ is the discrete category with natural numbers as objects. Let $J: \Nat \to
\Set$ be the functor which maps a natural number $n$ to the set $\{\,0,\ldots,
n-1\,\}$. Then for every endofunctor $G$ of $\Set$ there is a bijective
correspondence 
\begin{equation}\label{eq:bij}
  \frac{\Sigma \to G \o J}{H_\Sigma \to G}\,,
\end{equation}
and this bijective correspondence is natural in $\Sigma$ and $G$.  It is easy
to prove this directly. In fact, for any natural transformation $\alpha: \Sigma
\to G \o J$ (that is, a family of maps $\alpha_n : \Sigma_n \to G\{\,0,\ldots,
n-1\,\}$), we obtain a natural transformation $\beta: H_\Sigma \to G$ as
follows. The component $\beta_X$ maps an element $(f,\vec{x})$ of $H_\Sigma X$
to $Gs(\alpha_n(f))$, where we consider the $n$-tuple $\vec{x}$ as a function
$s: Jn \to X$. Conversely, given $\beta: H_\Sigma \to G$ define $\alpha$ by
$\alpha_n(f) = \beta_{Jn}(f, i_n)$, where $i_n$ is the $n$-tuple $(0,\ldots,
n-1)$. One may verify that the two constructions yield natural
transformations, are inverses, and are natural in $\Sigma$ and $G$.

The above bijective correspondence~\refeq{eq:bij} also follows from a
well-known adjunction between the category of signatures and the category of
endofunctors of $\Set$. For two categories $\A$ and $\cat{B}$ we denote by 
\begin{displaymath}\label{p:fun}
\fun{\A}{\cat{B}}
\end{displaymath}
the category of functors from $\A$ to $\cat{B}$. Recall that the functor $(\,\_\,) \o J: [\Set, \Set] \to [\Nat,
\Set]$ of restriction to $\Nat$ has a left-adjoint $\Lan{\,\_\,}$, i.\,e.~the
functor assigning to a signature its left Kan extension along $J$. Since $\Nat$
is a discrete category, the usual coend formula for computing $\Lan{\Sigma}$,
see e.\,g.~\cite{ml}, Theorem X.4.1, specializes to the coproduct 
\begin{displaymath}
  \Lan{\Sigma} X = \coprod\limits_{n \in \Nat} \Set(Jn, X) \copower \Sigma(n)\,,
\end{displaymath}
where $M \copower Z = \coprod_{m \in M} Z$ denotes the copower of $Z$, which is
isomorphic to $Z \times M$. Since $\Set(Jn, X)$ is isomorphic to $X^n$ we see
that the above formula is isomorphic to the coproduct in~\refeq{eq:poly}. This
implies that $\Lan{\Sigma}$ is (naturally isomorphic to) the signature functor
$H_\Sigma$. Hence, by virtue of the adjunction $\Lan{\,\_\,} \dashv (\,\_\,) \o
J$, we obtain~\refeq{eq:bij} as desired.

\comment{ % Old Text about Kan extensions
In order to do this we use a well-known adjunction
between the category of signatures and the category of endofunctors of
$\Set$. For two categories $\A$ and $\cat{B}$ we denote by 
\begin{displaymath}
\fun{\A}{\cat{B}}
\end{displaymath}
the category of functors from $\A$ to $\cat{B}$. 

Let $\Sigma$ be a signature; that is, a functor $\Sigma: \Nat \to \Set$,  where
$\Nat$ the discrete category with natural numbers as objects.  Let $J: \Nat \to
\Set$ be the functor which maps a natural number $n$ to the set $\{\,0,\ldots,
n-1\,\}$.  Recall that the functor $(\,\_\,) \o J: [\Set, \Set] \to [\Nat,
\Set]$ of restriction to $\Nat$ has a left-adjoint $\Lan{\,\_\,}$, i.\,e.~the
functor assigning to a signature its left Kan extension along $J$. Since $\Nat$
is a discrete category, the usual coend formula for computing $\Lan{\Sigma}$,
see e.\,g.~\cite{ml}, Theorem X.4.1, specializes to the coproduct 
\begin{displaymath}
  \Lan{\Sigma} X = \coprod\limits_{n \in \Nat} \Set(Jn, X) \copower \Sigma(n)\,,
\end{displaymath}
where $M \copower Z = \coprod_{m \in M} Z$ denotes the copower of $Z$, which is
isomorphic to $Z \times M$. Since $\Set(Jn, X)$ is isomorphic to $X^n$ we see
that the above formula is isomorphic to the coproduct in~\refeq{eq:poly}. This
implies that $\Lan{\Sigma}$ is (naturally isomorphic to) the signature functor
$H_\Sigma$. Hence, by virtue of the adjunction $\Lan{\,\_\,} \dashv (\,\_\,) \o
J$, there is for any signature $\Sigma$ and any endofunctor $G$ of $\Set$ a
bijection between natural transformations $\Sigma \to G \o J$ and natural
transformation $H_\Sigma \to G$, and this bijection is natural in $\Sigma$ and
$G$. In fact, for any natural transformation $\alpha: \Sigma \to G \o J$
(that is, a family of maps $\alpha_n : \Sigma_n \to G\{\,0,\ldots, n-1\,\}$), we
obtain a natural transformation $\beta: H_\Sigma \to G$ as follows. The
component $\beta_X$ maps an element $(f,\vec{x})$ of $H_\Sigma X$ to
$Gs(\alpha_n(f))$, where we consider the $n$-tuple $\vec{x}$ as a function $s:
Jn \to X$. Conversely, given $\beta: H_\Sigma \to G$ define $\alpha$ by
$\alpha_n(f) = \beta_{Jn}(f, i_n)$, where $i_n$ is the $n$-tuple $(0,\ldots,
n-1)$. It is easy to see that the two constructions yield natural
transformations, are   inverses, and  are natural in
$\Sigma$ and $G$.
}

We shall now use the above bijective correspondence~\refeq{eq:bij} to express recursive
program schemes as natural transformations. 
\comment{
Let $J: \Nat \to \Set$ be $n\mapsto\{1,2, \ldots, n\}$.  (There is nothing
special about this particular $J$: we could have also used
the map $n\mapsto\{0,1, \ldots, n-1\}$ which is more common in set theory
(since it turns out to be the identity on $n$). One could also begin with a countably infinite set $X$ of syntactic variables
and use $n\mapsto\{x_1,x_2, \ldots, x_n\}$.  The particular
definition of $J$  is immaterial,
but one must be fixed.) The point is that 
$H_\Sigma(X)$ is the disjoint union
of the sets $\Sigma_n\times X^{Jn}$, where here the notation $X^{Jn}$ means 
the set of functions from $Jn$ to $X$.  So we also use the notation $(f,s)$ for
a generic element of $H_\Sigma(X)$: $n$ is understood, and $s: Jn \to X$.
In this notation, $H_\Sigma k (f,s) = (f,k\o s)$.

 We shall use a general fact:  if $G : \Set\to \Set$ is any 
functor, then natural transformations $\alpha:  \Sigma \to  G\o J$ correspond
to natural transformations $\beta: H_\Sigma \to G$.
Given $\alpha$, we define $\beta = \beta(\alpha)$ by specifying
its value on an arbitrary set $A$; for this we consider some $n\in \Nat$,
some $f\in \Sigma_n$, and some $s: Jn\to A$.  We set
$\beta_A(f,s) = (Gs \o \alpha_n)(f)$.
To check the naturality, let $k: A \to B$.   For $n$, $f$, and $s$ as above,
\begin{displaymath}   Gk\o\beta_A(f,s) = Gk((Gs \o \alpha_n)(f)) = 
(G(k\o s)  \o \alpha_n)(f)   = \beta_B(f,k\o s) = \beta_B(H_\Sigma k(f,s)).
\end{displaymath}
In the other direction, given $\beta$, we define $\alpha = \alpha(\beta)$
by $\alpha_n(f) = \beta_{Jn}(f,i_n)$, where $i_n$ is the identity on $Jn$.
The naturality is easy to check.   This gives the associations between
the two types of natural transformations.  To check that they are 
inverses, note  first that
$\alpha(\beta(\alpha))_n(f) = \beta(\alpha)_{Jn}(f,i_n) = (Gi_n \o \alpha_n)(f) = \alpha_n(f)$.
Second, we use the naturality of $\beta$ to see that
\begin{displaymath}\beta(\alpha(\beta))_A(f,s) =
 (Gs \o \alpha(\beta)_n)(f)  = Gs(  \beta_{Jn}(f,i_n)) = 
 \beta_B(H_\Sigma s(f,i_n))  =  \beta_B(f,s).\end{displaymath} 
It will be advantageous in this paper to turn 
 $\alpha:  \Sigma \to  G\o J$   into its associated $\beta: H_\Sigma \to G$.
} %
Suppose we have a signature $\Sigma$ of given
operation symbols. Let $\Phi$ be a signature of new operation symbols.
Classically a \emph{recursive program scheme} (or shortly, \emph{RPS}) gives
for each operation symbol $f \in \Phi_n$ a term $t^f$ 
over $\Sigma + \Phi$ in $n$ variables.  We thus have a system of formal equations
\begin{equation}\label{eq:crps}
  f(x_1, \ldots, x_n) \approx t^f(x_1, \ldots, x_n), 
  \iffull\qquad\else\ \fi 
  f \in \Phi_n, 
  \iffull\qquad\else\ \fi
  n \in \Nat\,.
\end{equation}
Recall~\refeq{eq:algex} and~\refeq{eq:run} for concrete examples. 

Now observe that the names of the variables in~\refeq{eq:crps} do not matter.
More precisely, an RPS can be presented as a family of functions 
\begin{displaymath}
  \Phi_n \to \Tsf \{\,0,\ldots, n-1\,\}, \quad \textrm{with} \quad 
  f \mapsto t^f(0,\ldots, n-1)\,.
\end{displaymath}
So regarding $\Phi$ as a functor from $\Nat$ to $\Set$, any RPS as
in~\refeq{eq:crps} gives rise to a natural transformation
\begin{equation}\label{eq:crps2}
  \Phi \to \Tsf \o J\,.
\end{equation}
The formulation in~\refeq{eq:crps2} insures that in each equation of an RPS
such as~\refeq{eq:crps}, if the symbol on the left side is $n$-ary, then the only
variables that can appear on the right are the $n$ elements of $\{\,0, \ldots,
n-1\,\}$.  Notice as well that our formulation extends the classical notion of
RPS. Since we used $\Tsf$ in the codomain of~\refeq{eq:crps2} 
we have allowed infinite trees instead of just finite terms on the right-hand
sides. And we will further generalize this notion of RPS.

The natural transformation in~\refeq{eq:crps2} corresponds to a unique
natural transformation
\begin{equation}
H_\Phi \to \Tsf \, .  
\label{eq-overallpoint}
\end{equation}
as explained above. The point is that the formulation
in~\refeq{eq-overallpoint} is more useful to us than the one
in~\refeq{eq:crps2} because~\refeq{eq-overallpoint} involves a natural
transformations between endofunctors on one and the same category.  Notice that
for the signature functors $H_\Sigma$ and $H_\Phi$ we have $H_{\Sigma + \Phi} =
H_\Sigma + H_\Phi$ and hence $\Tsf = \T{H_\Sigma + H_\Phi}$. With this in mind,
we can rewrite~\refeq{eq-overallpoint}, and we see that recursive program
schemes correspond to natural transformations of the following form:
\begin{displaymath}H_\Phi \to \T{H_\Sigma + H_\Phi}\,.\end{displaymath}
This explains the work we have done so far.
  
To summarize: we abstract away from signatures and sets and study the
uninterpreted and the interpreted semantics of recursive program schemes
considered as natural transformations of the form 
\iffull 
\begin{displaymath}\label{p:rpsfirst}
\else
$
\fi \Var \to \T{H+\Var}, 
\iffull 
\end{displaymath} 
\else
$
\fi
where $H$, $\Var$ and $H+\Var$ are iteratable endofunctors on the category
$\A$. Now recall from our discussion in the Introduction that to say what a
solution of such a recursive program scheme is we first 
need to have a notion of (generalized) second-order substitution
(see~\cite{courcelle} for the classical notion---we explain this notion in
Example~\ref{ex:2nd}). It turns out that the 
universal property of the free completely iterative monads $\T{H}$, see
Theorem~\ref{thm:freecim}, readily yields this desired generalization. And this
is the reason we are interested in monads in this paper.

\begin{exmp}\label{ex:rps}\index{factorial!recursive program scheme for $\sim$}
  We mention explicitly how the two recursive program schemes
in~\refeq{eq:algex}
  and~\refeq{eq:run} give rise to natural transformations because we later
  often come back to these running examples. 
  Let $\Sigma$ be the signature that contains a unary operation symbol $G$ and
  a binary one $F$---so we have $\Sigma_1 = \{\,G\,\}$, $\Sigma_2 = \{\,F\,\}$
  and $\Sigma_n = \emptyset$ else. The signature $\Phi$ of recursively defined
  operations consists of two unary symbols $\phi$ and $\psi$. Consider the
  recursive program scheme~\refeq{eq:algex} as a natural transformation $r:
  \Phi \to \Tsf \o J$ with the components given by
  \begin{displaymath}
    r_1: \phi \mapsto F(0, \phi(G0)) \qquad \psi \mapsto F(\phi(G0), GG0)
  \end{displaymath}
  (we write trees as terms above) and where $r_n$, $n \neq 1$, is the empty
  map. The bijective correspondence~\refeq{eq:bij} yields a natural
  transformation $H_\Phi \to \T{H_\Sigma + H_\Phi}$, where $H_\Phi X = X + X$
  expresses the recursively defined operations and where $H_\Sigma X = X +
  (X\times X)$ expresses the givens $F$ and $G$. 
  Similarly, the RPS~\refeq{eq:run} defining the
  factorial function with the signature $\Sigma$ of givens and the signature
  $\Phi$ containing only the unary operation symbol $f$ gives rise to a natural
  transformation $H_\Phi \to \T{H_\Sigma + H_\Phi}$, where $H_\Sigma X = 1 + X
  + X \times X + X \times X \times X$ and $H_\Phi X = X$.
\end{exmp}

\subsection{Eilenberg-Moore Algebras}
\label{sec:Eilenberg-Moore}\index{algebra!Eilenberg-Moore
  $\sim$}\index{algebra!for a monad}
Recall that if $(F, G, \eta, \epsilon)$ is an adjunction, we get the associated
monad on the domain of $F$ by taking $T = GF$, $\mu = G\epsilon F$, and $\eta$
from the adjunction.  We also need a converse of this result.  Given a monad
$T$ on $\A$, the {\em Eilenberg-Moore\/} category $\A^T$ of $T$ has as objects
the (monadic) {\em $T$-algebras}: these are morphisms $\alpha: TA \to A$ such
that the diagrams below commute:  
\begin{displaymath}
  \vcenter{
    \xymatrix{
      A 
      \ar[r]^-{\eta_A} 
      \ar@{=}[rd]
      & 
      TA 
      \ar[d]^\alpha \\
      & 
      A
      }
    }
  \qquad\qquad
  \vcenter{
    \xymatrix{
      TTA
      \ar[r]^-{\mu_A}
      \ar[d]_{T\alpha}
      &
      TA 
      \ar[d]^\alpha
      \\
      TA 
      \ar[r]_-\alpha
      & 
      A\,.
      }
    }
\end{displaymath}
A morphism from $\alpha: TA \to A$ to $\beta : TB \to B$ is a morphism $h: A
\to B$ in $\A$ such that the square
\begin{displaymath}
  \xymatrix{
    TA
    \ar[r]^-\alpha
    \ar[d]_{Th}
    &
    A
    \ar[d]^h
    \\
    TB
    \ar[r]_-\beta
    &
    B
    }
\end{displaymath}
commutes. We sometimes write $T$-algebras
using the notation of pairs, as in $(A,\alpha)$, and often simply abbreviate to
$A$.

The relation between this construction and monads is that for any monad $T$,
there is an adjunction $(F^T,U^T,\eta,\epsilon)$ from $\A$ to $\A^{T}$ to which
$T$ is associated.  Here $F^T$ is the functor taking $A$ to the free
$T$-algebra $\mu_A: TTA \to TA$; $U^T: \A^T \to \A$ is the forgetful functor
taking the $T$-algebra $\alpha: TA \to A$ to its carrier $A$; and in the same
notation, $\epsilon_{(A, \alpha)}$ is $\alpha$ itself, taken to be a morphism
of $T$-algebras, see~\cite{ml}, Section~VI.2.

\comment{
\begin{prop}
Let $T = (T,\mu^T,\eta^T)$ and $S = (S, \mu^S,\eta^S)$ be monads on 
a category $\A$, and let $\phi: T\to S$ be a monad morphism.
If $(A,\alpha)$ is an $S$-algebra, then $(A,\alpha\o \phi_A)$ is a $T$-algebra.
\label{prop-EM-monad}
\end{prop}

\begin{proof}  Let $\beta = \alpha\o \phi_A$.  
First, $\beta\o \eta^T_A =\alpha\o \eta^S_A=    A$.
To show that $\beta\o T \beta= \beta \o \mu^T_A$, consider the diagram below:
\begin{displaymath}\xymatrix{
TTA \ar[rr]^{\mu^T_A}   \ar[dr]^{(\phi * \phi)_A} 
\ar[d]_{T\phi_A} & & TA \ar[d]^{\phi_A}  \\
 TSA \ar[r]^{\phi_{SA}}  \ar[d]_{T\alpha}
 & SSA \ar[r]^{\mu^S_A}  \ar[d]_{S\alpha} & SA \ar[d]^{\alpha} \\
TA \ar[r]_{\phi_A}   &  SA \ar[r]_{\alpha} &  A  \\
}
\end{displaymath}
All parts commute: the triangle by the definition of $\phi * \phi$,
the  region to its right because $\phi$ is a monad morphism,
the lower left square by naturality of $\phi$, and the lower right square
because $(A,\alpha)$ is an $S$-algebra.    The outside of the figure
shows that  $\beta\o T \beta= \beta \o \mu^T_A$.
\end{proof}  
}  % end comment

\section{Completely Iterative Algebras and Complete Elgot algebras}
\label{section-Elgot-algebras}

For interpreted solutions of recursive program schemes we need a suitable
notion of an algebra which can serve as interpretation of the givens. By a
``suitable algebra'' we mean, of course, one in which recursive program schemes
have a \emph{solution}. For example, for the recursive program scheme
in~\refeq{eq:algex}, we are interested in those $\Sigma$-algebras $A$, where
$\Sigma$ is the signature that consists of a binary symbol $F$ and a unary one
$G$, in which we can obtain two new operations $\phi_A$, $\psi_A$ on $A$ so
that the formal equations of~\refeq{eq:algex} become valid identities in $A$.
In the classical theory one works with continuous algebras---algebras
carried by a cpo such that all operations are continuous maps. Alternatively,
one can work with algebras carried by a complete metric space such that all
operations are contracting maps. In both of these approaches one imposes extra
structure on the algebra in a way that makes it possible to obtain the
semantics of a recursive definition as a join (or limit, respectively) of
finite approximations.

\comment{
Another possible approach is the axiomatic one. For example, a possible class
of suitable algebras are the iteration theory algebras of~\cite{be-iteration},
Chapter~7. An iteration theory algebra is a $\Sigma$-algebra is equipped with
a choice of solutions of a class of very simple recursive equations. And this
choice is required to satisfy certain axioms. }

The two types of algebras
mentioned above share
a crucial feature that allows for the solution of recursive program schemes:
these algebras induce an \emph{evaluation} of all $\Sigma$-trees. More
precisely, we consider a $\Sigma$-algebra $A$ with a canonical map $T_\Sigma A
\to A$, providing for each $\Sigma$-tree over $A$ its evaluation in $A$. It
seems to us that in order to be able to obtain solutions of recursive program
schemes in a $\Sigma$-algebra, the minimal requirement is the existence of such
an evaluation map turning $A$ into an Eilenberg-Moore algebra for the monad
$T_\Sigma$ (see Example~\ref{ex:monad}). More generally, we work here with
\emph{complete Elgot algebras} for an iteratable endofunctor $H$, which turn
out to be precisely the Eilenberg-Moore algebras for the monad $\T H$,
see~\cite{amv3}. An important subclass of all complete Elgot algebras are
\emph{completely iterative algebras}~\cite{milius2}. One of our main results
(Theorem~\ref{thm:intsol}(ii)) states that recursive program schemes have unique
solutions in completely iterative algebras. 

We have already seen in the introduction that a recursive program scheme gives
rise to a flat system of equations (see~\refeq{eq:induced}). So in order to
solve recursive program schemes it is sufficient to solve flat systems of
equations. Completely iterative algebras are defined as algebras in which every
flat system of equations has a unique solution. And an Elgot algebra comes with
a choice of a solution for every flat system of equations, and this choice
satisfies two easy and natural axioms. 

Let us begin by explaining the notion of completely iterative algebra with an
example. Let $\Sigma$ be a signature, and let $Y$ be any set. We think of $Y$
as a set of \emph{parameters}\index{parameters} 
for which we may later substitute $\Sigma$-trees.
This is in contrast to the constant symbols from $\Sigma$ which we think of as being
fixed once and for all. Consider the $\Sigma$-algebra $T_\Sigma Y$ of all
(finite and infinite) $\Sigma$-trees over $Y$. Let $X$ be a set  
(whose elements may be considered to be \emph{formal
variables}\index{formal variables!of a flat system}) which is disjoint from $Y$. A \emph{flat system} of
recursive equations is a systems of formal equations
\begin{equation}\label{eq:system}
  x \approx t_x\,, \qquad \textrm{for every $x \in X$,}
\end{equation}
where either $t_x \in T_\Sigma Y$ is a $\Sigma$-tree with no formal variables or
else $t_x = \sigma(x_1, \ldots, x_n)$, $\sigma \in \Sigma_n$, $x_1, \ldots, x_n
\in X$.  In this 
last case, $t_x$ is a flat $\Sigma$-tree,  a $\Sigma$-tree of height at most
$1$.  

We have already begun in this paper to use the standard
practice of using $\approx$  in a system
to denote  formal equations (recursive specifications of functions or other
objects).  We use  $=$  to denote 
actual  identity (see just below for an example). 
Flat systems have a unique \emph{solution} in $T_\Sigma Y$: there
exists a unique tuple $\sol{x}$, $x \in X$, of trees in $T_\Sigma Y$ such
that the identities
\begin{displaymath}
\sol{x} = t_x [\,z := \sol{z}\,]_{z \in X}
\end{displaymath}
hold. 
\iffull
For example, let $\Sigma$ consist of a binary operation symbol $*$ and a
unary one $s$. The following flat system of equations
  \begin{displaymath}
  x_0 \approx 
  \vcenter{
    \xy
    \POS   (0,0)   *+{*} = "A"
    ,     (-5,-10) *+{x_1} = "B"
    ,     (5,-10)  *+{x_2} = "C"
    \ar@{-} "A";"B"
    \ar@{-} "A";"C"
    \endxy
    }
  \qquad
  x_1 \approx
  \vcenter{
    \xy
    \POS    (0,0)   *+{s} = "A"
    ,       (0,-10) *+{x_0} = "B"
    \ar@{-} "A";"B"
    \endxy
    }
  \qquad
  x_2 \approx 
  \vcenter{
    \xy
    \POS    (0,0) *+{*} = "A"
    ,     (-5,-10) *+{y_0} = "B"
    ,     (5,-10)  *+{y_1} = "C"
    \ar@{-} "A";"B"
    \ar@{-} "A";"C"
    \endxy
    }
  \end{displaymath}
  with formal variables $X = \{\,x_1, x_2, x_3\,\}$ and parameters $Y = \{\,y_0,
  y_1\,\}$ has as its unique solution the following trees in $T_\Sigma Y$:
  \begin{displaymath}
  \sol{x_0} = 
  \vcenter{
    \xy
    \POS    (0,0)    *+{*} = "A"
    ,      (-5,-10)  *+{s} = "B"
    ,      (5,-10)   *+{*} = "A1"
    ,     (-10,-20)  *+{*}  = "C"
    ,     (0,-20)    *+{y_0} = "A11"
    ,     (10,-20)   *+{y_1} = "A12"
    ,     (-15,-30)  *+{s}  = "D"
    ,     (-5,-30)   *+{*} = "C1"
    ,     (-20,-40)           ="E"
    ,     (-10,-40)  *+{y_0} = "C11"
    ,     (0,-40)    *+{y_1} = "C12"
    %%%
    \ar@{-} "A";"B"
    \ar@{-} "A";"A1"
    \ar@{-} "A1";"A11"
    \ar@{-} "A1";"A12"
    \ar@{-} "B";"C"
    \ar@{-} "C";"D"
    \ar@{-} "C";"C1"
    \ar@{-} "C1";"C11"
    \ar@{-} "C1";"C12"
    \ar@{.} "D";"E"
%    \ar@{-} "D";"D1"
    \endxy
    }
  \qquad
  \sol{x_1} = 
  \vcenter{
    \xy
    \POS     (00,00)    *+{s} = "A"
    ,        (00,-17)   *+{\sol{x_0}} = "x"
    ,        (00,-10)   = "U"
    ,        (-5,-20)   = "L"
    ,        ( 5,-20)   = "R"
    \ar@{-} "A";"U"
    \ar@{-} "U";"L"
    \ar@{-} "U";"R"
    \ar@{-} "L";"R"
    \endxy
    }
  \qquad
  \sol{x_2} = 
  \vcenter{
    \xy
    \POS    (0,0) *+{*} = "A"
    ,     (-5,-10) *+{y_0} = "B"
    ,     (5,-10)  *+{y_1} = "C"
    \ar@{-} "A";"B"
    \ar@{-} "A";"C"
    \endxy
    }
  \end{displaymath}
  \else
  For example, let $\Sigma$ consist of a binary operation symbol, $*$. The flat
  system 
  \begin{displaymath}x_1 \approx x_1 * x_2 \qquad x_2 \approx t\end{displaymath}
  with variables $X = \{\,x_1, x_2\,\}$, and where $t \in T_\Sigma Y$, has as
  its unique solution $\sol{x_1} = \cdots * t) * t) *t)$ and $\sol{x_2} = t$. 
  \fi
  Observe that to give a flat system of equations is the same as to give a
  mapping
  $%\begin{displaymath}
  e: X \to H_\Sigma X + T_\Sigma Y %\,,\end{displaymath}
  $.
  And a solution is the same as a mapping $\sol{e}: X \to T_\Sigma Y$ such that
  the following square 
  \begin{displaymath}
    \vcenter{
      \xymatrix@C+2.5pc{
        X \ar[d]_e \ar[r]^-{\sol{e}} & T_\Sigma Y \\
        H_\Sigma X + T_\Sigma Y \ar[r]_-{H_\Sigma \sol{e} + T_\Sigma Y} 
        & 
        H_\Sigma T_\Sigma Y + T_\Sigma Y \ar[u]_{[\tau, T_\Sigma Y]}
        }
      }
  \end{displaymath}
  commutes, where $\tau$ denotes the tree-tupling map. (We adopt the convention
  of denoting in a commutative diagram the identity on an object by that object
  itself.)
%Recall also that the system which we started with has a unique solution.

\subsection{Completely Iterative Algebras}

\iffull
The example above suggests the following definition, 
originally studied in~\cite{milius2}. 
\fi

\begin{defn}\label{dfn:cia}\index{algebra!completely iterative $\sim$}\index{cia|see{completely iterative algebra}}\index{equation morphism!flat $\sim$}
  Let $H: \A \to \A$ be an endofunctor. By a  \emph{flat equation morphism} in
  an object $A$ (\emph{of parameters}) we mean a morphism 
  \begin{displaymath}e: X \to HX + A\,.\end{displaymath}
  Let $a:  HA \to A$ be an $H$-algebra. 
  We say that $\sol{e}: X \to A$ is a \emph{solution} of $e$
  in $A$ if the square below commutes:
  \index{solution!of a flat equation morphism}
  \begin{equation}\label{diag:ciasol}
    \vcenter{
      \xymatrix@C+1pc{
        X \ar[d]_e \ar[r]^-{\sol{e}} & A \\
        HX + A \ar[r]_-{H\sol{e} + A} & HA + A \ar[u]_{[a, A]}
        }
      }
  \end{equation}

  We call $A$ a \emph{completely iterative algebra} (or \emph{cia}, for short)
  if every flat equation morphism in $A$ has a unique solution in $A$.
\end{defn}

Observe that we have no restriction on our objects of variables.  (That
is, in the case of $\Set$, we do not require that the set of variables be
finite.)  Imposing this restriction weakens the notion to what~\cite{amv} calls
an {\em iterative algebra}.  It will be essential in this paper to consider
equation morphisms whose domain is not finite.  

\begin{rem}
  We shall see in Section~\ref{section:interpreted} how recursive program
  schemes as discussed in Section~\ref{section-rpsfirst} can be reduced to flat
  equation morphisms in an algebra $A$. Then, an interpreted solution of a
  recursive program scheme in $A$ can be obtained by solving the corresponding
  flat equation morphism in $A$. In fact, in Theorem~\ref{thm:intsol} we will
  prove that every guarded recursive program scheme has a unique solution in
  any completely iterative algebra $A$. 
\end{rem}

\begin{examples}\label{ex:cia} %
  \iftcs\else\mbox{ }\fi%
  \begin{myenumerate}
  \myitem Let $\Pfin$ be the finite power set functor on $\Set$,
    and assume the Anti-Foundation Axiom.  Let $H\!F_1$ be the set of
    hereditarily finite sets.  Let $\tau$ be the inclusion of $\Pfin(H\!F_1)$
    into $H\!F_1$.  This map $\tau$ turns $H\!F_1$ into a cia with respect to
    $\Pfin$.
    \label{finitely-branching-sets}
  
  \myitem Consider the subalgebra $H\!F_{1/2}$ of sets whose transitive closure
    is finite, then the {\em complete\/} iterativity is lost.  Only finite
    systems can be solved in this setting. For more on this example and
    the last, see Section~18.1 of~\cite{vc}.

    \myitem Final coalgebras. In~\cite{milius2} it is proved that for a final
    $H$-coalgebra $\alpha: T \to HT$ the inverse $\tau: HT \to T$ of the
    structure map yields a cia for $H$. Analogously, for every object $Y$ of
    $\A$ a final coalgebra $TY$ for $H(\,\_\,) + Y$ yields a cia for $H$, see
    Theorem~\ref{theorem-finalcoalgebra-Elgot} below.  This generalizes the
    first example and the examples $T_\Sigma Y$ of all $\Sigma$-trees over a
    set $Y$.

    \index{algebra!completely iterative $\sim$ on a metric space}
  \myitem Let $H$ be a contracting endofunctor on the category $\CMS$ of
    complete metric spaces, see Example~\ref{ex:cats_iii}.  Then any non-empty
    $H$-algebra $(A, a)$ is completely iterative.  In fact, given any flat
    equation morphism $e:X \to HX + A$ in $\CMS$, it is not difficult to prove
    that the assignment $s \mapsto a \o (Hs + A) \o e$ is a contracting
    function of $\CMS(X,A)$, see~\cite{amv3}. Then, by Banach's Fixed Point
    Theorem, there exists a unique fixed point of that contracting function,
    viz.~a unique solution $\sol e$ of $e$. Notice that $\sol e$ is obtained as
    a limit of a Cauchy sequence. In fact, choose some element $\bot \in A$
    and define the Cauchy sequence $(\sol{e}_n)_{n\in \Nat}$ in $\CMS(X, A)$ by
    recursion as follows: let $\sol{e}_0 = \mathsf{const}_\bot$, and given
    $\sol{e}_n$ define $\sol{e}_{n+1}$ by the commutativity of the square
    \begin{equation}\label{diag:indsol}
      \vcenter{
      \xymatrix@C+1pc{
        X \ar[d]_e \ar[r]^{\sol{e}_{n+1}} & A \\
        HX + A \ar[r]_-{H \sol{e}_n + A} & HA + A \ar[u]_{[\alpha, A]}  
        }}
    \end{equation}
    
  \myitem Non-empty compact subsets form cias. Let $(X, d)$ be a complete metric
    space. Consider the set $C(X)$ of all non-empty compact subspaces of $X$
    together with the so-called Hausdorff metric $h$; for two compact subsets
    $A$ and $B$ of $X$ the distance $h(A, B)$ is the smallest number $r$ such
    that $B$ can be covered by open balls of radius $r$ around each point of
    $A$, and vice versa, $A$ can be covered by such open balls around each
    point of $B$. In symbols, $h(A, B) = \max\{\,d(A \rightarrow B), d(B
    \rightarrow A)\,\}$, where $d(A\rightarrow B) = \max_{a \in A} \min_{b\in
      B} d(a,b)$. It is well-known that $(C(X), h)$ forms a complete metric
    space; see, e.g.~\cite{barnsley}. Furthermore, if $f_i: X \to X$, $i = 1,
    \ldots, n$, are contractions of the space $X$ with contraction factors
    $c_i$, $i = 1, \ldots, n$, then it is easy to show that the map
    \begin{displaymath}\label{p:C}
    \alpha_X: C(X)^n \to C(X) \qquad (A_i)_{i = 1, \ldots, n} \mapsto
    \bigcup\limits_{i = 1}^n f_i[A_i]
    \end{displaymath}
    is a contraction with contraction factor $c = \max_i c_i$ (the product
    $C(X)^n$ is, of course, equipped with the maximum metric). In other words,
    given the $f_i$, we obtain on $C(X)$ the structure $\alpha_X$ of an
    $H$-algebra for the contracting endofunctor $H(X,d) = (X^n, c\cdot
    d_\mathrm{max})$. Thus, if $X$ is not empty and thus has a compact subset,
    then $(C(X), \alpha_X)$ is a cia for $H$.
    
     \index{Cantor set}%
     As an illustration we show that the Cantor ``middle third'' set $c$ may be
     obtained via the cia structure on a certain space. Recall that $c$ is the
     unique non-empty closed subset of the interval $I = [0,1]$ which satisfies
     $c = \frac{1}{3}c \cup (\frac{2}{3} + \frac{1}{3}c)$.  (We write
     $\frac{1}{3}c$ to mean $\{\,\frac{x}{3} \mid x\in c\,\}$, of course.)  So
     let $(X, d)$ be the Euclidean interval $I = [0,1]$ and consider the
     $\frac{1}{3}$-contracting functions $f(x) = \frac{1}{3}x$ and $g(x) =
     \frac{1}{3} x + \frac{2}{3}$ on $I$.  Then $\alpha_I: C(I)\times C(I) \to
     C(I)$ with $\alpha_I(A,B) = f[A] \cup g[B]$ gives the structure of a cia
     on $C(I)$ for the functor $H(X, d) = (X \times X, \frac{1}{3}\cdot
     d_\mathrm{max})$, which is a lifting of the signature functor $H_\Sigma X
     = X \times X$ of $\Set$ expressing one binary operation symbol $\alpha$.
     Now consider the formal equation
     \begin{displaymath}
       x \approx \alpha(x, x)\,.
     \end{displaymath}
     It gives rise to a flat equation morphism $e: 1 \to H1 + C(I)$ which maps
     the element of the trivial one point space $1$ to the element of $H1 = 1$.
     The unique solution $\sol{e}: 1 \to C(I)$ picks a non-empty closed set $c$
     satisfying $c = \alpha(c,c) = f[c] \cup g[c]$.  Hence $\sol e$ picks the
     Cantor set.
     
   \myitem Continuing with our last point, for each non-empty closed $t\in
     C(I)$, there is a unique $c(t)$ with $c(t) = \alpha_I(c(t),t)$. The
     argument is just as above. But the work we have done does \emph{not} show
     that $c$ is a continuous function of $t$.  For this, we would have to
     study a recursive program scheme $\phi(x) \approx \alpha(\phi(x), x)$ and
     solve this in $(C(I),\alpha_I)$ in the appropriate sense.  Our work later
     in the paper does exactly this, and it follows that the solution to
     $\phi(x) \approx \alpha(\phi(x), x)$ in the given algebra is the
     \emph{continuous} function $t\mapsto c(t)$.
   
  \myitem Suppose that $H: \Set \to \Set$ has a lifting to a contracting
    endofunctor $H'$ on $\CMS$, see Example~\ref{ex:lift_cms}. 
    Let $\alpha:HA\to A$ be an $H$-algebra. If there exists a complete metric,
    say $d$, on $A$ such that $\alpha$ is a nonexpanding map $H'(A,d) \to
    (A,d)$, then $A$ is a completely iterative algebra: to every equation
    morphism $e:X\to HX+A$ assign the unique solution of $e:(X,d_0)\to
    H'(X,d_0)+(A,d)$, where $d_0$ is the discrete metric on $X$, the one given by
    $d(x,x')=1$ iff $x\not= x'$.

  \end{myenumerate}
\end{examples}

\subsection{Complete Elgot Algebras}

In many settings, one studies a fixed point operation on a structure like a
complete partial order. And in such settings, one typically does not have {\em
  unique\/} fixed points. So completely iterative algebras are 
  {\em not\/} the unifying
concept capturing precisely what is needed to solve recursive program schemes.
Instead, we shall need a weakening of the notion of a cia.

\begin{rem}\label{rem:after}
  We will need two operations in the statement of Definition~\ref{dfn:elgotalg}
  below.  The first operation formalizes the renaming of parameters in a flat
  equation morphism. More precisely, for a flat equation morphism $e: X \to HX
  + A$ and a morphism $h: A \to B$, we define
  \begin{displaymath}
    h \after e \equiv
    \xymatrix@1{
      X \ar[r]^-e & HX + A \ar[rr]^-{HX + h} & & HX + B\,.
      }
  \end{displaymath}
  The second operation allows us to combine two flat equation morphisms where
  the parameters of the first are the variables of the second into one
  ``simultaneous'' flat equation morphism. More precisely, given two flat
  equation morphisms $e: X \to HX + Y$ and $f: Y \to HY + A$ we define
  \begin{displaymath}
    \smalldiag
    f \plus e \equiv
    \xymatrix@1@C+1pc{
      X + Y \ar[r]^-{[e, \inr]} & HX + Y \ar[r]^-{HX + f} &
      HX + HY + A \ar[r]^-{\can + A} & H(X+Y) + A\,,
      }  
  \end{displaymath}
  where $\can$ is the canonical morphism $[H\inl, H\inr]$, and where $\inl$ and
  $\inr$ denote injections of a binary coproduct. 
\end{rem}

\begin{defn}\label{dfn:elgotalg}\index{algebra!Elgot $\sim$}
  A complete \emph{Elgot algebra} is a triple $(A, a, \funsol)$, where $(A,a)$
  is an $H$-algebra, and $\funsol$ assigns to every flat equation morphism $e$
  with parameters in $A$ a solution $\sol{e}: X \to A$ such that the following
  two laws hold:

  \smallskip\noindent
  {\bf Functoriality.}\index{Functoriality} 
  Solutions respect renaming of variables. Let $e$ and $f$
  be two flat equation morphisms with parameters in $A$, and let $h$
  be a \emph{morphism of equations}\index{morphism of equations} between them,
  that is, the square 
  \iffull
  \begin{displaymath}
  \xymatrix{
    X \ar[d]_h \ar[r]^-e & HX + A \ar[d]^{Hh + A} \\
    Y \ar[r]_-f & HY + A
  }
  \end{displaymath}
  \else
  $f\o h = (Hh + A) \o e$,
  \fi
  commutes. Then we have 
  \iffull
  \begin{displaymath}\sol{e} = \sol{f} \o h\,.\end{displaymath} 
  \else
  $\sol{e} = \sol{f} \o h\,.$
  \fi
   
  \smallskip\noindent {\bf Compositionality.}\index{Compositionality}
  Simultaneous recursion may be
  performed sequentially.  For all flat equation morphisms $e: X \to HX + Y$
  and $f: Y \to HY + A$, the solution of the combined equation morphism $f
  \plus e$ is obtained by first solving $f$ and then solving $e$ ``plugging
  in'' $\sol f$ for the parameters: 
  \iffull
  \begin{displaymath}\sol{(\sol{f} \after e)} = \sol{(f \plus e)} \o \inl\,.\end{displaymath}
  \else
  $\sol{(\sol{f} \after e)} = \sol{(f \plus e)} \o \inl$.
  \fi
\end{defn}

\begin{defn}
  A \emph{homomorphism} $h$ from an Elgot algebra $(A, a, \funsol)$ to an Elgot
  algebra $(B, b, \fundsol)$ 
(for the same functor $H$) 
is a morphism $h: A \to B$ that preserves
  solutions.  That is, for every flat equation morphism $e : X \to HX+ A$ we have
  a commutative triangle
  \begin{displaymath}
    \xymatrix{
      &
      X
      \ar[ld]_{\sol e}
      \ar[rd]^{\dsol{(h \after e)}}
      \\
      A 
      \ar[rr]_-h
      &
      &
      B\,.
      }
  \end{displaymath}
\end{defn}

\begin{prop}\label{prop:sol-pres-hom}\hspace{-3pt}{\rm\cite{amv3}}
  Every homomorphism $h: (A, a, \funsol) \to (B, b, \fundsol)$ of
  Elgot algebras is a homomorphism of $H$-algebras: the square
  \begin{displaymath}
  \xymatrix{
    HA \ar[r]^-a \ar[d]_{Hh} & A \ar[d]^h \\
    HB \ar[r]_-b & B
    }
  \end{displaymath}
  commutes. The converse is false in general. If, however,  $A$ and $B$ are
  cias, then 
  any $H$-algebra morphism is a homomorphism of Elgot algebras.
\end{prop}

\begin{proof}[Sketch of Proof]
We show that the square commutes, omitting some of the details.
First, consider the flat equation morphism
\begin{displaymath}
e_A \equiv
\xymatrix@1@C+1pc{
  HA + A \ar[r]^-{H \inr + A} & H(HA + A) + A\,.
  }
\end{displaymath}
Its solution is $\sol{e}_A = [a, A]: HA + A \to A$,
 as one easily establishes
using the commutativity of Diagram~\ref{diag:ciasol} for
$\sol{e}_A$. Similarly, we have $e_B: HB + B \to H(HB + B) + B$, and
$\dsol{e}_B = [b, B]$. Now consider $h$ as in our proposition. Then the
equation $\dsol{(h \after e_A)} = h \o \sol{e}_A$ holds. Furthermore, $Hh + h:
HA + A \to HB + B$ is a morphism of equations from $h \after e_A$ to $e_B$.
Thus Functoriality yields $\dsol{(h\after e_A)} = \dsol{e}_B \o (Hh + h)$. Now
one readily performs the computation
\begin{displaymath}
[h\o a, h] = 
h \o \sol{e}_A = 
\dsol{(h \after e_A)} =
\dsol{e}_B \o (Hh + h) =
[b\o Hh, h]\,.
\end{displaymath}
The desired equation $h \o a = b \o Hh$ is the left-hand component of the above
equation.
\comment{%
First, consider $e = \inr: A \to HA + A$.  The solution of this flat equation 
morphism is $A$, the identity on $A$.  Second, consider
$f = \inr\o a : HA \to HHA + A$.     Note that $(Ha + A)\o f = a = e\o a$.
So by Functoriality, $\sol{f} = a \o\sol{e}= a$.  
Next, consider $g_A= e\plus f$.  By the Compositionality, 
 $ \sol{g_A} \o \inl = \sol{(\sol{e} \after f)} =  \sol{(A\after f)} = a$.
 Indeed, $\sol{g_A} = [a,A]$.
Now consider $h: A\to B$ as in our proposition.
Let $e' : A \to HA + B$ be $\inr\o h$.  
This is a flat equation morphism in $B$, 
and by Functoriality, $\sol{e'}  = h$.  Let $g'_A = e'\plus f$.
Using the Compositionality, we see that $\sol{g'_A} =  h\o[a, A] = [h\o a, h]$.
There is a flat equation morphism $g_B: HB+B\to H(HB+B)+B$,
defined just as $g_A$ was.  A calculation with Functoriality
shows that $  [h\o a, h]=\sol{g'_A} = \sol{g_B} \o (Hh + h) = [b,B]\o (Hh+ h) = [b\o Hh, h]$.
So we have $h\o a = b\o Hh$, as desired.
}%
\end{proof}

\begin{rem}\iftcs\else\mbox{ }\fi%
  \begin{myenumerate}
  \myitem In~\cite{amv3} there is also a notion of a non-complete Elgot
    algebra. Since we will only be interested in using complete Elgot algebras
    we will henceforth understand by an Elgot algebra a complete one. 
  \myitem The axioms of Elgot algebras are inspired by the axioms of iteration
    theories of Bloom and \'Esik~\cite{be-iteration}. In fact, the two laws
    above resemble the functorial dagger implication and
    the left pairing identity (also known as Beki\'{c}-Scott 
    law) from~\cite{be-iteration}.
    
    One justification for the above axioms is that Elgot algebras (for some
    functor $H$) turn out to be the Eilenberg-Moore category of the monad $\T
    H$, see Subsection~\ref{section-EM-T}. As a result any Elgot algebra $A$ can
    equivalently be presented by a morphism $TA \to A$ satisfying the two usual
    axioms of Eilenberg-Moore algebras. In particular, for a signature functor
    $H_\Sigma$ on $\Set$ we see that if a $\Sigma$-algebra $A$ is an Elgot
    algebra, then there exists a canonical map $T_\Sigma A \to A$ which we may
    think of as an evaluation of all $\Sigma$-trees in $A$.
      
  \myitem Notice that flat equation morphisms are precisely the coalgebras for
    the functor $H(\,\_\,) + A$. Thus, the Functoriality above states that
    $\funsol$ is a functor from the category of coalgebras for $H(\,\_\,) + A$
    to the comma category $\A/A$.

\comment{  
  \myitem In the presence of functoriality the right-hand component of $\sol{(f
      \plus e)}$  is just $\sol{f}$. In fact $\inr: Y \to X + Y$ is a
    coalgebra homomorphism from $f$ to $f \after e$. Thus, in an Elgot algebra 
    we have the equation
    \begin{displaymath}
      \sol{(f\plus e)} = [\sol{(\sol f \after e)}, \sol f]: X + Y \to A\,.
    \end{displaymath}
}
  \end{myenumerate}
\end{rem}

\begin{prop}\label{prop:cia}
  For any endofunctor $H$, every completely iterative algebra is an Elgot
  algebra. 
\end{prop}
It is proved in~\cite{amv3} that for a cia, the assignment of the
unique solution to any flat equation morphism satisfies the Functoriality and
the Compositionality laws.

\begin{examples}\label{ex:elgot} 
  We present some further examples of Elgot algebras. None is in general a cia.
  \mbox{ }
  \begin{myenumerate}
  \myitem Continuous algebras. \index{algebra!continuous $\sim$}%
  Let $H$ be a locally continuous endofunctor on
    $\CPO$, see Example~\ref{ex:cats_ii}.  
    It is shown in~\cite{amv3} that any $H$-algebra $(A, a)$ with a least
    element $\bot$ is an Elgot algebra when to a flat equation morphism $e: X
    \to HX + A$ the least solution $\sol{e}$ is assigned. More precisely,
    define $\sol{e}$ as the join of the following increasing $\omega$-chain in
    $\CPO(X, A)$: $\sol{e}_0$ is the constant function $\bot$; and given
    $\sol{e}_n$ let $\sol{e}_{n+1} = [a, A]\o (H\sol{e}_n +A)\o e$, so
    that Diagram~\refeq{diag:indsol} commutes. \label{part-continuous-algebras}
    
    \myitem Suppose that $H: \Set \to \Set$ is a functor with a locally
    continuous lifting $H': \CPO \to \CPO$.  An $H$-algebra $\alpha: HA \to A$
    is called \emph{$\CPO$-enrichable}\index{algebra!$\CPO$-enrichable $\sim$}
    if there exists a complete partial order $\less$ on $A$ such that $A$
    becomes a continuous algebra $\alpha: H'(A, \less) \to (A, \less)$ with a
    least element. Any $\CPO$-enrichable algebra $A$ is an Elgot algebra for
    $H$: to every flat equation morphism $e: X \to HX + A$, let $\leq$ be the
    discrete order on $X$, consider $\ext{e}: (X, \leq) \to H'(X, \leq) + (A,
    \less)$ defined in the obvious way, and assign $U\sol{\ext e}: X \to A$,
    where $\sol{\ext e}$ is from part~(ii), and $U : \CPO \to \Set$ is the
    forgetful functor.

    \newcommand{\seg}{\mbox{seg}}
    \comment{
  \myitem Let $\mathcal{P}$ be the power set functor on $\Set$.  Let $A$ be any
    set, and let $f: \mathcal{P}(A)\to A$ be any function such that
  %% $f(s)\in s$ for $s\neq \emptyset$, and also such that
    (a) for all $S\subseteq \mathcal{P}(A)$, $f(\bigcup S) = f(\{f(s) : s\in
    S\})$; and (b) there is some $a_0\in A$ such that $f(s) = a_0$ whenever
    $a_0 \in s$.  (One way to get such a function is to take a wellorder $< $
    of $A$.  Let $f(\emptyset)$ be arbitrary; for $s\neq \emptyset$, let $f(s)$
    be the $<$-least element of $s$; and let $a_0$ be the $<$-least element of
    $A$.)  We define an Elgot algebra structure on $(A,f)$ as follows.  Given a
    flat equation morphism $e: X \to \mathcal{P} X + A$, write $x \rightarrow
    x'$
  %% $x'$ is \emph{$e$-reachable from $x$}
    if $x' \in e(x) \in \mathcal{P}(X)$.  Let $\rightarrow^+$ be the transitive
    closure of $\rightarrow$, and let $\rightarrow^* = (\rightarrow \cup
    \rightarrow^+)$.  If $\rightarrow^+$ contains an infinite sequence
    beginning with $x$, let $\sol e(x) = a_0$.  Otherwise, let $\sol e(x) =
    f(s)$, where
    \begin{displaymath}
    s = \{a\in A : \mbox{for some $x'$ such that $x\rightarrow^*x'$,
      $e(x') = a$}\}.
    \end{displaymath}
  % $ is the set of $a\in A$ which
  %are
  % ``reachable from $x$'' in finitely many steps.
    Then $(A, f, \funsol)$ is an Elgot algebra for $\mathcal{P}$.  But this is
    not usually a cia: When $f$ is derived from a wellorder, we have $f(\{ a\})
    = a$ for all $a\in A$.  In general, if $f(\{ a\}) = a$, then $a$ is a
    solution to $x\approx \{ x \}$.
    }
\myitem Every complete lattice $A$ is an Elgot algebra for the endofunctor $HX = X
  \times X$ of $\Set$. In fact, taking binary joins yields an $H$-algebra
  structure $\vee: A \times A \to A$.  Furthermore, observe that the
  algebra $TA$ of all binary trees over $A$ has an evaluation $\alpha: TA
  \to A$ mapping every binary tree in $TA$ to the join of its leaf labels. For
  any flat equation morphism $e: X \to X \times X + A$ form the flat
  equation morphism $\eta_A \after e: X \to X \times X + TA$. Then take its
  unique solution $\sol{(\eta_A \after e)}: X \to TA$, and let $e^* = \alpha \o
  \sol{(\eta_A \after e)}$. One may check that $(A, a, \starop)$ is an Elgot
  algebra for $H$, see~\cite{amv3}, Example~3.9.  Notice that this is usually
  not a cia since the formal equation $x \approx x \vee x$ has in general many
  different solutions in a complete lattice.

\comment{
  \myitem Operational semantics. Let $H_\Sigma$ be the polynomial functor on sets
    associated to the signature $\Sigma$. Let $(A,a)$ be any
    $\Sigma$-algebra.  Let $\uparrow$ be a new element, and let $A^{\uparrow}$
    be the $\Sigma$-algebra with carrier $A \cup \{\,\uparrow\,\}$, with $a $
    extended {\em strictly}.  Given a flat equation morphism $e : X \to HX +
    A^{\uparrow}$, we define a solution by rewriting.  That is, we define a
    finite or infinite sequence $\sol{e}_n$ of maps from $X$ to finite
    $\Sigma$-trees (or terms) over $A$.  The sequence is constructed by
    successively substituting for each variable $x\in X$ the corresponding term
    $e(x)$.  If for some $x\in X$ and $n\in \Nat$, $\sol{e}_n(x)$ contains no
    variables, evaluate it in $A$ pointwise and declare this to be
    $\sol{e}(x)$.  If for $x$ no $n$ exists like this, we declare $\sol{e}(x) =
    \mathord{\uparrow}$.
    
    With these definitions, it can be verified that we have an Elgot algebra.
    Moreover, it is well-known that the solution coincides with the
    denotational solution, where we add a fresh bottom element to make a flat
    cpo, extend the original algebra strictly again, and then iterate in the
    usual way to find the least fixed point, see (iii) above.
    For more discussion of this (Elgot algebra) structure, see, e.g.,
    Nivat~\cite{nivat}. 
}
  \end{myenumerate}
\end{examples}

\subsection{Computation Tree Elgot Algebras}
\label{section-EASUEAC}
\index{algebra!computation tree Elgot $\sim$}

In this section we present Elgot algebras for a signature that uses undefined
elements and also conditionals. Let $\Sigma$ be a signature, and let $H =
H_\Sigma$ be the associated endofunctor on $\Set$.  Let $(A,a)$ be any
$H_\Sigma$-algebra, and let $\uparrow$ be any element of $A$.  We shall define an
Elgot algebra structure on $A$ related to the natural computation tree
semantics of recursive definitions, where solutions are obtained by rewriting. 
The idea is that $\uparrow$ is our ``scapegoat'' for ungrounded definitions. 
 
We shall assume that the algebra $A$ interprets the function symbols in
$\Sigma$ in a strict fashion: if any argument $a_i$ is $\uparrow$, then the
overall value $g_A(a_1,\ldots, a_n)$ is $\upa$ as well.  Conversely, if
$g_A(a_1,\ldots, a_n) = \upa$, we require that ($n\geq 1$ and) some $a_i$
is $\uparrow$.  We make this assumption for all function symbols $g$
\emph{except for} the \emph{conditional symbol} $\cond\in\Sigma_3$.  We make a
different assumption on $\cond$.  For this, fix an element $0\in A$ other than
$\uparrow$.  We want
\begin{equation}
  \cond_A(x,y,z) = \
  \left\{
    \begin{array}{ll}
      y  & \mbox{if $x = 0$}  \\
      z  & \mbox{if $x\neq 0$ and  $x\neq \upa$}  \\
      \uparrow & \mbox{otherwise} \\
    \end{array}
  \right.
  \label{eq-defcond}
\end{equation}

To summarize, in this section we work with algebras for signature
functors on $\Set$ which come with designated objects $\uparrow$ and
$0$ satisfying the assumptions above.  

We shall work with partial functions and we use some notation which is
standard. For partial functions $p, q: X \pto A$, $p(x)\uar$ means that $p$ is
not defined on $x$, and we write $p(x)\dar$ if $p$ is defined on $x$. Finally,
the Kleene equality
$p(x)\simeq q(y)$ means that if either $p(x)$ or $q(y)$ is defined, then so is
the other; and in this case, the values are the same.

\begin{defn}\label{defn:comptree}
  Let $e: X \to HX + A$ be a flat equation morphism in $A$.
  We define a partial function $\wh{e} : X \pto A$  as follows:
  \begin{enumerate}
  \item If $e(x) = a$ and $a \neq \upa$, then $\wh{e}(x) \simeq a$ .
  \item  If $e(x) = g(x_1,\ldots, x_k)$, $g \neq \cond$, and for each $i$,
    $\wh{e}(x_i) \simeq a_i$, then $\wh{e}(x) \simeq g_A(a_1,\ldots, a_k)$.
  \item   If $e(x)= \cond(y,z,w)$ and
    $\wh{e}(y) \simeq 0$, %and $\wh{e}(z)\dar$,
    then $\wh{e}(x) \simeq \wh{e}(z)$.
  \item   If $e(x)= \cond(y,z,w)$ and
    $\wh{e}(y) \dar $ but $\wh{e}(y) \not\simeq 0$, %and $\wh{e}(w)\dar$,
    then $\wh{e}(x) \simeq \wh{e}(w)$.
  \end{enumerate}
  We call $\wh{e}$ the  {\em computation function corresponding to $e$}.
  \label{definition-Elgot-comp}
\end{defn}

We intend  this to be a definition of a partial function by recursion, so that
we may prove facts about $\wh{e}$ by induction.  Here is a first example, one which
will be important in Proposition~\ref{proposition-Elgot-computation} below: if
$\wh{e}(x)\dar$, then $\wh{e}(x)\neq \upa$.
 
Now that we have $\wh{e}$, we define $\sol{e}(x)$ to be $\wh{e}(x)$ if
$\wh{e}$ is defined; if it is not, we set $\sol{e}(x) = \upa$.  (Note that
$\sol{e}(x) = \upa$ iff $\wh{e}(x)\uar$.)

In the statement of the result below, we also mention the main way that one
obtains structures which satisfy the standing hypotheses of this section.

\begin{prop} 
  Let $A_0 = (A_0, a_0)$ be any $H_\Sigma$-algebra, let $0\in A_0$, let $\upa
  \notin A_0$, and let $A = A_0 \cup\{\,\uparrow\,\}$. Let $A = (A,a)$ be
  defined in terms of this data by extending $a_0$ to the function
  $a:H_\Sigma A \to A$ strictly on all function symbols except $\cond$, and
  with $\cond_A$ given by (\ref{eq-defcond}).  Let $\funsol$ be as above.
  Then $(A, a, \funsol)$ is an Elgot algebra for $H_\Sigma$.
  \label{proposition-Elgot-computation}
\end{prop}

\begin{proof}
  Clearly the assumption of this section hold for the algebra $A$.  These
  assumptions ensure that $A$ is $\CPO$-enrichable. In fact, equip $A$ with the
  flat cpo structure with the least element $\upa$. Then all operations on $A$
  are easily checked to be monotone, whence continuous; thus, $a: H_\Sigma A
  \to A$ is a continuous algebra. By Example~\ref{ex:elgot}(ii), we obtain an
  Elgot algebra $(A, a, \starop)$. We will prove that for any flat equation
  morphism $e: X \to HX + A$ the least solution $e^*$ agrees with the map
  $\sol{e}$ given by the computation function $\wh{e}$. To this end recall
  first that the set $\Par(X,A)$ of partial functions from $X$ to $A$ is a cpo
  with the order given by $f \sqsubseteq g$ if for all $x \in X$, $f(x)\dar$
  implies $g(x)\dar$ and $f(x) = g(x)$. Now observe that the definition of
  $\wh{e}$ by recursion means that $\wh{e}$ is the join of an increasing
  chain in $\Par(X,A)$. In fact, let $\wh{e}_0$ be the everywhere undefined
  function; and given $\wh{e}_n$ define $\wh{e}_{n+1}$ as follows: in the
  clauses~(i)--(iv) in Definition~\ref{defn:comptree} replace the term
  $\wh{e}(x)$ by $\wh{e}_{n+1}(x)$, and replace all other occurrences of $\wh{e}$
  by $\wh{e}_n$. Then clearly,
  $(\wh{e}_n)_{n< \omega}$ is an increasing chain in $\Par(X,A)$ whose join is
  $\wh{e}$.  \comment{
  \begin{displaymath}
    \wh{e}_{n+1} = [a, A] \o (\ol{H} \wh{e}_n + A) \o e\,,
  \end{displaymath}
  where $\ol{H}$ is the extension of $H$ to the category $\Par$ of sets and
  partial functions. 
  \begin{REMARK}
    I think there is a slight problem here concerning $\cond$. If $e(x) =
    \cond(y,z,w)$, $\wh{e}_n(y) \simeq 0$ and $\wh{e}_n(w)\upa$ we need to
    have $\wh{e}_{n+1}(x) \simeq \wh{e}_n(y)$. Otherwise the
    equation~\refeq{eq:solhat} below is false. In the definition of $\wh{e}$
    above I have changed~(iii) and~(iv) accordingly. But for $h = \ol{H}
    \wh{e}_n$ we have $h(g(\vec{x}))\dar$ iff all $\wh{e}(x)\dar$. 
    I think this can be fixed if the above equation is rewritten in a
    non-categorical way. 
  \end{REMARK}
  
  The fact that $(\wh{e}_n)_{n < \omega}$ is an increasing
  chain follows from the fact that $\Par$ is a $\CPO$-enrichable category and
  $\ol{H}$ is locally continuous since $H$ is finitary. 
  }

  Now recall from Example~\ref{ex:elgot}(i) that $e^*$ is the join
  of the chain $e^*_n$ in $\CPO(X, A)$, where $X$ is discretely
  ordered. We shall show by induction that for every $x \in X$ the equation
  \begin{equation}
    \label{eq:solhat}
    e^*_n(x) = \left\{
      \begin{array}{l@{\qquad}p{2cm}}
        \wh{e}_n(x)  & if $\wh{e}_n(x)\dar$ \\
        \upa & else
      \end{array}
      \right.
    \end{equation}
    holds. The base case is obvious. For the induction step we proceed by case
    analysis. If $e(x) = a$, then $e^*_{n+1}(x) = a$ and
    so~\refeq{eq:solhat} holds. The second case is $e(x) = g(x_1, \ldots,
    x_k)$, $g \neq \cond$. We have $e^*_{n+1}(x) = g_A(e^*_n(x_1),
    \ldots, e^*_n(x_k))$. By our assumptions, this equals $\upa$ precisely
    if at least one of the $e^*_n(x_j)$ is $\upa$, which in turn holds if
    and only if $\wh{e}_n(x_j)\upa$ for some $j$; and equivalently,
    $\wh{e}_{n+1}(x)\upa$.  Otherwise all $\wh{e}_n(x_j)$ are defined and by
    induction hypothesis we get
    \begin{displaymath}
      e^*_{n+1}(x) = g_A(e^*_n(x_1), \ldots, e^*_n(x_k))
      = g_A(\wh{e}_n(x_1), \ldots, \wh{e}_n(x_k)) = \wh{e}_{n+1}(x)\,.
    \end{displaymath}
    Thirdly, assume that $e(x) = \cond(y,z,w)$. Then similarly as before we
    have 
    \iftcs $\else\begin{displaymath}\fi
      e^*_{n+1}(x) = \cond(e^*_n(y), e^*_n(z), e^*_n(w))%
      \iftcs\/$.\else\,.\end{displaymath}\fi 
    We obtain $e^*_{n+1}(x) = \upa$ whenever $e^*_n(y) = \upa$. 
    But this happens precisely if $\wh{e}_n(y)\upa$, which implies that
    $\wh{e}_{n+1}(x)\upa$. Now if $e^*_n(y) \neq \upa$, then we have
    equivalently that $\wh{e}_n(y)\dar$. We treat here only the case that
    $\wh{e}_n(y) = 0$; the remaining case is similar. In our present case it
    follows that $e^*_{n+1}(x) = e^*_n(z)$ and $\wh{e}_{n+1}(x) \simeq
    \wh{e}_n(z)$. Therefore, by the induction hypothesis, the desired
    equation~\refeq{eq:solhat} holds. 

    Finally, from~\refeq{eq:solhat} we conclude that for the least fixed points
    $e^*$ and $\wh{e}$ we have
    \begin{displaymath}
      e^*(x) = \left\{
        \begin{array}{l@{\qquad}p{2cm}}
          \wh{e}(x) & if $\wh{e}(x)\dar$ \\
          \upa & else.
        \end{array}
      \right.
    \end{displaymath}
    Thus, we get $e^* = \sol{e}$ which completes the proof.
  \end{proof}

\comment{
\begin{proof} Clearly the assumptions of this section hold for $A_0$.
We first check that if $e: X\to HX +A$ is a flat equation morphism,
then $\sol{e}$ is a solution to $e$.

First, we show by induction on the computation function $\wh{e}$  that
if $\wh{e}(x)\dar$, then
\begin{equation}
([Ha,A]\o  ( H\sol{e} + A))(e(x)) \simeq \sol{e}(x).
\label{eq-inproposition-Elgot-computation}
\end{equation}
This is by induction on $\wh{e}$.  The base case is when $e(x) = a$ for 
some $a\in A$.  In this case, (\ref{eq-inproposition-Elgot-computation})
 is immediate.
There are two induction steps, one for $\cond$ and
one for the other symbols.  We shall only work the details concerning $\cond$.
 Suppose that $e(x)$ is of the form $\cond(y,z,w)$.
Assume first that $ \sol{e}(x)\dar$.
Then $\wh{e}(y)$ must be defined, and let us assume that 
it is $0$; the other case is handled similarly.    
Then $\wh{e}(z)\dar$.  So by induction hypothesis,
$([Ha,A]\o  ( H\wh{e} + A))(e(z)) = \wh{e}(z)$.
By point (iii) in the definition of $\wh{e}$, $\wh{e}(x) = \wh{e}(z)$.
So $\sol{e}(x) = \sol{e}(z)$.
But also,
\begin{displaymath}  ([Ha,A]\o  ( H\sol{e} + A)(e(x)) =  \cond^A(e^\dag(y), e^\dag(z), e^\dag(w))
 =  \sol{e}(z)\ .
\end{displaymath}
(Note that we have used the assumption that $\cond$ is interpreted in $A$ in 
the expected way.)
We  conclude that indeed  $ ([Ha,A]\o  ( H\sol{e} + A)(e(x))$ is defined and
has the same value as $\sol{e}(x)$.   The converse is similar, as is the
induction step for function symbols other than $\cond$ in this part of the proof.
 
 So far, we have established (\ref{eq-inproposition-Elgot-computation}) in
 the case that $\wh{e}(x)\dar$.
We also must consider 
(\ref{eq-inproposition-Elgot-computation}) 
in the case when $\sol{e}(x) = \uparrow$ because 
$\wh{e}(x)\uar$.   There are some subcases here.
First, it might be   that $e(x) = \inr(\uparrow)$.
(That is, $e(x)$ is the element $\uparrow$ of the structure.)
In this subcase,  the desired instance of the solution equation is
easy to check.    Another subcase is when $e(x) = g(x_1,\ldots, x_n)$
and $g\neq \cond$.  Here we must have $\wh{e}(x_i)\uar$ for
some $i$.   And then by the assumption that $g$ is interpreted strictly,
we have 
\begin{displaymath}\wh{e}(x) = \uparrow = g^A(\wh{e}(x_1), \ldots, \wh{e}(x_n))\, , 
\end{displaymath}
so again we have the desired equality.   Another subcase
is when $e(x) = \cond(y,z,w)$ and $\wh{e}(y)\uar$.
In this case, the fact that $0\neq\uparrow$ tells us that $\cond^A(\sol{e}(y),  \sol{e}(z),\sol{e}(w))
= \uparrow$.   The rest of the verification is based on this.
Another subcase is  when $\wh{e}(y)\simeq 0$ but $\wh{e}(z)\uar$,
and the final subcase is when $\wh{e}(y)\dar$, $\wh{e} \not\simeq 0$, and
but  $\wh{e}(w)\uar$.  In both of these, the argument is straightforward.

We turn to the verification of the two Elgot algebra properties.
We check the Functoriality Law.
Let  $e: X\to HX + A$ and 
 $f: Y\to HY + A$ be flat equation morphisms in $A$, and let $h$ be such that 
$(Hh + A)\o e = f\o h$. 

We show first by induction on $\wh{e}$ that
if   $\wh{e}(x) \dar$, then $\hat{f}(hx)\simeq \wh{e}(x)$.  
The base case is when $e(x) = a$ for some $a\in A$ other than $\uparrow$. 
 Then also 
$f(hx) = a$, and also $\hat{f}(hx) = a$.   Turning to one of the two induction steps,
suppose  that $e(x) = g(x_1,\ldots, x_n)$  with  $g\neq\cond$.
So $f(hx) =  g(hx_1,\ldots, hx_n)$.
 By induction hypothesis,
$\hat{f}(hx_i) = \wh{e}(x_i)$ for all $i$.   And then we see easily that
\begin{displaymath}\wh{e}(x) = g^A(\wh{e}(x_1), \ldots, \wh{e}(x_n))
=g^A(\hat{f}(hx_1)), \ldots, \hat{f}(hx_n))  = \hat{f}(hx).\end{displaymath}
A second subcase is when $g$ is $\cond$.  In this case, the argument is similar
and we omit the details.

Continuing with the Functoriality Law, we show the converse:
if   $\hat{f}(hx) \dar$, then $\wh{e}(x)\simeq \hat{f}(hx)$.  
The argument is by induction on  $\hat{f}$,
and the steps are essentially the same as the ones we have seen.

We have one final point on the Functoriality Law in addition to the arguments in
 the previous two paragraphs, similar to what we have seen
 with regard to the  solution condition:  If neither  $\wh{e}(x) \dar$ nor
 $\hat{f}(hx)\dar$, then $\sol{e}(x) = \uparrow = \sol{f}(hx)$.

We conclude this proof by checking the Compositionality.
Let $e: X \to HX + Y$
  and $f: Y \to HY + A$.  
We  must verify that
 $\sol{(\sol{f} \after e)} = \sol{(f \plus e)} \o \inl$.
 To save some notation in this proof, we shall write
 $p$ for $\sol{f}\after e$, and $q$ for $f\plus e$.
 An easy induction on the 
  computation relation corresponding to $f$ shows that
  for all $y\in Y$, $\hat{f}(y) \simeq \hat{q}(\inr\, y)$. 
 We henceforth ignore the injections into $X+Y$.  
 It
 is sufficient to show that the
 following  hold:
  \begin{myenumerate}
 \myitem If $\hat{p}(x)\dar$, then 
 $\hat{q}(x) \simeq \hat{p}(x)$.
 \myitem If $\hat{q}(x)\dar$, then 
 $\hat{p}(x)\simeq \hat{q}(x)$.
 \end{myenumerate}
 Once these are established, we have the Compositionality:
  in all cases besides the ones above, $\sol{p}(x)  = \uparrow = \sol{q}(x)$.
 
 The proofs of (i) and (ii) above are similar, and perhaps the most interesting
 step is the base case of (i).   That is, assume that $\hat{p}(x)\dar$
 because $(\sol{f}\after e)(x) = a\in A$.  In this case, we know that $a \neq \uparrow$,
 and also that $\hat{p}(a) \simeq a$.
 This means that $e(x)$ is some element $y\in Y $ with the property that
 $\hat{f}(y)\simeq a$.       
 From what we have seen above, $\hat{q}(y)\simeq a$. 
  But the definition of $f\plus e$ tells us that 
  $q(x) = q(y)$, and so \begin{displaymath}\hat{q}(x) = \hat{q}(y) = a  = \hat{p}(x)\, \end{displaymath}
  just as desired.
 This concludes our work on the base case of (i), and the rest of the steps are similar.
 Point (ii) is also similar, and indeed it is easier.   (The argument above is the only
 place in the proof that we use the feature of the definition of $\wh{e}(x)$ that
 it is $\uparrow$ when $e(x) = \uparrow$ in $A$.)

This completes our proof.
\end{proof}
}

\begin{defn}
  Let $H_\Sigma : \Set \to \Set$ be a signature functor, let $A_0 = (A_0, a_0)$
  be any $H_\Sigma$-algebra as in the hypothesis of 
  Proposition~\ref{proposition-Elgot-computation}. We call the Elgot algebra 
  $(A, a, \funsol)$ the \emph{computation tree Elgot algebra} induced by $A_0$.
\end{defn}

We shall study the interpreted solutions of recursive program schemes in
computation tree Elgot algebras in Section~\ref{sec:opsemint}.

\comment{
\subsection{old approach below}

Let $e: X\to HX + A$ be a flat equation morphism.    Let $\prec_e$
be defined by $y \prec_e x$ if $e_x$ is of the form $f(\ldots, y, \ldots)$;
that is, $y$ occurs on the right-hand side of $e_x$.  Recall that 
a relation $R$ on a set $S$ is wellfounded below a point $x\in S$
if $R$ has no infinite descending sequences in the relation
 which begin with $x$.
We say that $e$ is wellfounded below $x$ if $\prec_e$ is wellfounded
below $x$.  In this case, let $|x|_e\in \Nat$ be the   height of
the $\prec_e$ below $x$.  

%Let $\Xwf$ be the elements of $X$ for which $\prec_e$ is wellfounded
%below $x$.   Then $(\Xwf, \prec_e)$ is a wellfounded set.

We say that $\wh{e}$ is an \emph{associate} of $e$ if the following
conditions hold:
\begin{myenumerate}
\myitem $\wh{e}$ is a partial function from some subset of $\hat{X}\subseteq X$ to $A$.
\myitem  $\prec_e \cap (\hat{X}\times\hat{X})$ is
  wellfounded  on $\hat{X}$.
\myitem If $e_x = a\in A$, then also $\wh{e}_x = a$.
\myitem If $e_x = g(x_1,\ldots, x_n)$ and $\wh{e}_x\dar$, then
for all $i$, $\wh{e}_{x_i}\dar$, and also 
$\wh{e}_x = a(g,\wh{e}_{x_1},\ldots, \wh{e}_{x_n})$.
(Recall that $a$ is the structure of the algebra $A$.)
\myitem  If $e_x = \cond(x_1,x_2,x_3)$, $\wh{e}_x\dar$, 
and $\wh{e}_{x_1} \simeq 0$, then  $\wh{e}_{x_2}\dar$ and
 $\wh{e}_{x} = \wh{e}_{x_2}$.
\myitem  If $e_x = \cond(x_1,x_2,x_3)$,  
  $\wh{e}_x\dar$ 
but $\wh{e}_{x_1}\not \simeq 0$, 
then   $\wh{e}_{x_3}\dar$ and
$\wh{e}_{x} \simeq \wh{e}_{x_3}$.
\end{myenumerate}

\begin{lem} If $\wh{e}$ and $e^{\#}$ are associates
of $e$, then for all $x$ such that $\wh{e}_x\dar$ and $e^{\#}_x\dar$,
$\wh{e}_x \simeq e^{\#}_x$.
\label{lemma-associates-1}
\end{lem}

\begin{proof} Let $\hat{X}$ be the domain of $\wh{e}$.
We show by induction on the rank of $x$ in 
$\prec_e \cap (\hat{X}\times\hat{X})$ that if $e^{\#}$ is any associate
of $e$ whose domain contains $x$, then 
$\wh{e}_x \simeq e^{\#}_x$.
\end{proof}

\begin{lem} If $\wh{e}_x$ is defined and not $\uparrow$, then there
is some associate $e^{\#}$ such that for all $y\in dom(e^{\#})$, 
$e^{\#}_y$ is not $\uparrow$.
\label{lemma-associates-2}
\end{lem}

\begin{proof}
By induction on the rank of $x$ in $\prec_e \cap (\hat{X}\times\hat{X})$.
The most important case is when $e_x$ is of the form $\cond(y,z,w)$.
Note that the variables $z$ and $w$ might well be the same, but this is 
not a problem.
\end{proof}

As a result of Lemma~\ref{lemma-associates-1}, the  union of any set of associates
of a given flat equation morphism $e$ is again an associate.   So we shall
henceforth write $\wh{e}$ for this union.     If $\wh{e}_x\dar$, then 
we write $|x|$ for the rank of $x$ in the wellfounded $\prec$.

We define $\sol{e}_x$ as follows: first, if  there is an associate 
$\wh{e}$ such that $\wh{e}_x\dar$, then we set $\sol{e}_{x} = \wh{e}_x$.
Otherwise, we set $\sol{e}_x = \uparrow$.
 
Let $e: X \to HX + Y$
  and $f: Y \to HY + A$.  
  We check the Compositionality
 $\sol{(\sol{f} \after e)} = \sol{(f \plus e)} \o \inl$.
 To save some notation in this proof, we shall write
 $p$ for $\sol{f}\after e$, and $q$ for $f\plus e$;
 we also will ignore the injections into $X+Y$.  It
 is sufficient to show that the
 following  hold:
  \begin{myenumerate}
 \myitem If $\hat{p}_x$ is defined and not $\uparrow$, then 
 $\hat{q}_x \simeq \hat{p}_x$.
 \myitem If $\hat{q}_x$ is defined and not $\uparrow$, then 
 $\hat{p}_x \simeq \hat{q}_x$.
 \end{myenumerate}
 Once these are established, we have the Compositionality because
 in all cases besides the ones above, $\sol{p}_x  = \uparrow = \sol{q}_x$.

Let $x\in X$.  Suppose first that $\prec_e$ is not wellfounded
below $x$.  Then  
 $\prec_{\sol{f} \after e}$ is not wellfounded below $x$,
and neither is  $\prec_{f \plus e}$.  We easily get that 
$\sol{(\sol{f} \after e)}(x) = \uparrow = \sol{(f \plus e)} \o \inl(x)$
in this case.     A second case is when there is some $y \prec^*_e x$
such that $\prec_f$ is not wellfounded below $y$.  In this case
$\sol{f}(y) = \uparrow$, and again we have 
$\sol{(\sol{f} \after e)}(x) = \uparrow = \sol{(f \plus e)} \o \inl(x)$.
In the last case,   $\prec_{\sol{f} \after e}$ is wellfounded below $x$.
We argue by induction on $|x|_{\sol{f} \after e} $.
let  $e_x = g(x_1,\ldots, x_n,y_1,\ldots,y_m)$.
(For convenience, we write the elements of $X$ before the elements of $Y$.)
For $1\leq j \leq m$, $\prec_f$ is wellfounded below $y_j$.  An induction on $|y_j|_{f}$ shows that
$ \sol{f} \after e(y_j) = \sol{(f \plus e)} \o \inr(y_j)$.
For all $x_i$ on the right hand side, $|x_i|_{\sol{f} \after e} < |x|_{\sol{f} \after e}$, so we may use 
the induction hypothesis to calculate:
 \begin{displaymath}\begin{array}{lcl}
\sol{(\sol{f} \after e)}(x)  & = & a(g,
\sol{(\sol{f} \after e)}(x_1), \ldots, \sol{(\sol{f} \after e)}(x_n),
\sol{(\sol{f} \after e)}(y_1),
\ldots, \sol{(\sol{f} \after e)}(y_m))  \\
 & = & a(g,  \sol{(f \plus e)} \o \inl(x_1),  \sol{(f \plus e)} \o \inl(x_n), 
 \sol{(f \plus e)} \o \inr(y_1),\ldots,  \sol{(f \plus e)} \o \inr(y_m)
) \\
 & = &  \sol{(f \plus e)} \o \inl(x) \\
\end{array}\end{displaymath}

We claim first that $h[X\setminus X_{wf}] = h[X] \setminus Y_{wf}$.
To see this,  first let $x\in [X\setminus \Xwf]$.  Then there
are  \begin{displaymath} x = x_0 \succ   x_1 \succ  x_2 \cdots \ . \end{displaymath}
The condition on $h$ implies that $h(x)  \succ  h(x_1) \succ h(x_2)\cdots$,
so $h(x) \in h[X] \setminus \Ywf$.
In the other direction, let $h(x) \in h[X] \setminus  \Ywf$.
So we have $y_n \in Y$ so that 
  \begin{displaymath} h(x) = y_0 \succ  y_1 \succ  y_2 \cdots \ . \end{displaymath}
By recursion on $n$,  we define $x_n$ such that $x = x_0$,
$h(x_n) = y_n$,
and the $x_n$ are a descending sequence.    It follows that $x\in \Xwf$,
as desired.

We now check that for all $x\in X$, $\sol{e}_x = \sol{f}_{hx}$. 
If $x\notin \Xwf$, then $\sol{e}_x = \uparrow$.
As we have seen, $h(x) \notin \Ywf$, so we also have $ \sol{f}_{hx} = \uparrow$.
We now check our equation  $\sol{e}_x = \sol{f}_{hx}$
for $x\in \Xwf$, and we argue by induction on $|x|_e$.
There are a number of cases here.
Suppose first that $e_x$ is of the form $\cond(x_1,x_2,x_3)$.
Then we also have $f_{hx} = \cond(h x_1, h x_2, hx_3)$.
Note that $|x_i|_e < |x|_e$ for $i  = 1,2,3$.
Suppose that $\sol{e}_{x_1} = 0$.  Then 
we have $\sol{e}_x = \sol{e}_{x_2}$.    Also, 
By induction hypothesis,  $ \sol{e}_{x_1} = sol{f}_{hx_1}$ and
 $ \sol{e}_{x_2} = sol{f}_{hx_2}$.
Therefore  $\sol{f}_{hx} =  \sol{f}_{hx_2}$.

 Then for $x,y\in X$, 
$y\prec_e x$ iff $hy \prec_f hx$.  It follows
that $\prec_e$ is wellfounded below $x$ iff $\prec_f$ is wellfounded below $hx$.
If both of these hold, then  induction on $|x|_e$ shows that
$\sol{e}_x = \sol{f}_{hx}$.   And if neither holds, then both 
$\sol{e}_x$ and $\sol{f}_{hx}$ are $\uparrow$, hence both are equal.
}  %%% end of \rem of the old proof that wasn't right

\subsection{Dramatis Personae}
\label{sec:dramatis}

As we have already mentioned, the 
 classical theory of recursive program schemes rests
on the fact that for a continuous algebra $A$, 
 there is a canonical map $T_\Sigma A \to A$.
The point is   that all  $\Sigma$-trees
over $A$ can be evaluated in $A$ itself.    In a
suitable category of cpos the structures $T_\Sigma X$ play the r\^ole of
\emph{free algebras}.  The freeness is used to define maps {\em out\/} of those
algebras. In our setting, the $\Sigma$-trees are the final coalgebra.  So in
order to generalize the classical theory, we need a setting in which the final
coalgebras $TY$ for $\HY$ are free algebras.  The following result gives such a
setting.  It is fundamental for the rest of the paper and collects the results
of Theorems~2.8 and~2.10 of~\cite{milius2} and Theorem~5.6 of~\cite{amv3}.  We
sketch a proof for the convenience of the reader.

\begin{thm} Let $H$ be any endofunctor of $\A$. The following are equivalent:
  \begin{enumerate}
    \iffull\else\setlength{\myitemsep}{-3pt}\fi
  \item $TY$ is a final coalgebra for $H(\,\_\,) + Y$,
  \item $TY$ is a free completely iterative algebra for $H$ on $Y$, and
  \item $TY$ is a free (complete) Elgot algebra for $H$ on $Y$.
  \end{enumerate}
  
  In more detail: if $(TY, \alpha_Y)$ is a final coalgebra for $H(\,\_\,) + Y$,
  the inverse of $\alpha_Y: TY \to HTY + Y$ yields an $H$-algebra structure
  $\tau_Y: HTY \to TY$ and a morphism $\eta_Y: Y \to TY$, which turn $TY$ into
  a free cia on $Y$. By Proposition~\ref{prop:cia}, $TY$ is an Elgot algebra
  for $H$, and additionally this Elgot algebra is also free on $Y$. Conversely,
  let $(TY, \tau_Y, \funsol)$ be a free Elgot algebra for $H$ with a universal
  arrow $\eta_Y: Y \to TY$. Then $TY$ is a cia, whence a free cia on $Y$, and
  $[\tau_Y, \eta_Y]$ is an isomorphism whose inverse is the structure map of a
  final coalgebra for $H(\,\_\,) + Y$.
  \label{theorem-finalcoalgebra-Elgot}
\end{thm}
\iffull
\begin{proof}[Sketch of Proof]
  Suppose first that $(TY, \alpha_Y)$ is a final coalgebra for $H(\,\_\,) +
  Y$. Let $[\tau_Y, \eta_Y]$ be the inverse of $\alpha_Y$. We show that
  $\tau_Y: HTY \to TY$ is a completely iterative algebra for $H$. In
  fact, for any flat equation morphism $e: X \to HX + TY$, form the following 
  coalgebra
  \begin{displaymath}
  \smalldiag
  c \equiv
  \xymatrix@1@C+1pc{
    X + TY \ar[r]^-{[e, \inr]} & HX + TY \ar[r]^-{HX + \alpha_Y} & 
    HX + HTY + Y \ar[r]^-{\can + Y} & H(X + TY) + Y
    }
  \end{displaymath}
  and define
  \begin{displaymath}
  \sol{e} \equiv
  \xymatrix@1{
    X \ar[r]^-\inl & X + TY \ar[r]^-h & TY\,,
    }
  \end{displaymath}
  where $h$ is the unique homomorphism from the coalgebra $(X + TY, c)$ to the
  final one. It is not difficult to prove that $\sol{e}$ is the unique solution
  of $e$. 

  By Proposition~\ref{prop:cia}, it follows that $TY$ is an Elgot algebra. To
  establish~(ii) and~(iii) in the statement of the Theorem it suffices to show
  that $(TY, \tau_Y, \funsol)$ is a free Elgot algebra on $Y$.  For any
  Elgot algebra $(A, a, \fundsol)$ and any morphism $m: Y \to A$ form the
  equation morphism
  \begin{displaymath}
  m \after \alpha_Y \equiv
  \xymatrix@1{
    TY \ar[r]^-{\alpha_Y} & HTY + Y \ar[rr]^-{HTY + m} & & HTY + A
    }
  \end{displaymath}
  It is shown in Theorem~5.3 of~\cite{amv3} that the solution $h =
  \dsol{(m\after \alpha_Y)}$ yields the unique homomorphism 
  $h: TY \to A$ of Elgot algebras such that $h \o \eta_Y = m$. In fact, the
  latter equation follows from the definition of a solution in
  Diagram~\refeq{diag:ciasol}, and the 
  proof that $h$ preserves solutions uses the Compositionality and
  Functoriality of $\fundsol$. Finally, the uniqueness of $h$ is proved using
  the Compositionality and Functoriality of $\funsol$. The details can be found
  in loc.~cit.

  Now conversely, assume that $(TY, \tau_Y, \funsol)$ is a free Elgot algebra
  for $H$ on $Y$ with a universal arrow $\eta_Y: Y \to TY$. It can be shown that
  $[\tau_Y, \eta_Y]$ is an isomorphism, see Lemma~5.7 of~\cite{amv3}. Denote by
  $\alpha_Y: TY \to HTY + Y$ its inverse, which is a flat equation morphism. We
  use $\alpha_Y$ to show that $(TY, \tau_Y)$ is a cia; i.\,e., for any flat
  equation morphism $e: X \to HX + TY$ the solution $\sol{e}$ is unique. In
  fact, suppose that $s$ is any solution of $e$. It follows that $s$ is a
  morphism of equations from $e$ to the flat equation morphism
  \begin{displaymath}
  f \equiv
  \xymatrix@1{
    TY \ar[r]^-{\alpha_Y} & HTY + Y \ar[rr]^-{HTY + \eta_Y} & & HTY + TY\,.
    }
  \end{displaymath}
  Thus, $\sol{f} \o s = \sol{e}$ by Functoriality of $\funsol$. Next one
  can show using Compositionality  that $\sol{f}: TY \to TY$ is a
  homomorphism of Elgot algebras satisfying $\sol{f} \o \eta_Y = \eta_Y$. Thus,
  by the freeness of $TY$, $\sol{f} = \id$. This proves $s = \sol{e}$ so that
  $(TY, \tau_Y)$ is 
  a cia, which implies that it is the free one on $Y$. It is not difficult to
  show that this yields a final coalgebra for $H(\,\_\,) + Y$. In fact, for any coalgebra
  $c: C \to HC + Y$ the unique solution of the flat equation morphism $\eta_Y
  \after c$ yields a unique homomorphism $(C, c) \to (TY, \alpha_Y)$ of
  coalgebras. 
\end{proof}
\fi

Theorem~\ref{theorem-finalcoalgebra-Elgot} has an important consequence
for our work.   Recall that we assume $H$ is iteratable, so  $H(\,\_\,) + Y$
{\em does\/} have a final coalgebra for all $Y$. 
The next result gives the {\em dramatis personae\/} for the rest of the paper.

\begin{thm}\label{thm:adjoint}
There is a left adjoint to the forgetful functor from
$\Elgot H$, the category of Elgot algebras and their homomorphisms, to the base
category $\A$
\begin{displaymath}
\xymatrix@1@C+1pc{
  \Elgot H \ar@<-1ex>[r]_-U \ar@{}[r]|-\perp & \A\,. \ar@<-1ex>[l]_-L  
}
\end{displaymath}
\iftcs\sloppypar\fi The left-adjoint $L$ assigns to each object $Y$ of $\A$ a
free Elgot algebra $(TY, \tau_Y, \funsol)$ on $Y$ with a universal arrow
$\eta_Y: Y \to TY$ (equivalently, $(TY, \alpha_Y)$ where $\alpha_Y = [\tau_Y,
\eta_Y]^{-1}$ is a final coalgebra for $H(\,\_\,) + Y$). The unit of the
adjunction is $\eta$ whose components are given by the universal arrows of the
free Elgot algebras. The counit $\eps$ gives for each Elgot algebra $(A, a,
\fundsol)$ the unique homomorphism $\aext{a}: TA \to A$ of Elgot algebras such
that $\aext a \o \eta_A = \id$.  We have
\begin{equation}\label{eq:aext}
  \aext{a} = \dsol{(\alpha_A)}: TA \to A\,,
\end{equation}
where $\alpha_A: TA \to HTA + A$ is considered as a flat equation morphism
with parameters in $A$.

Moreover, we obtain additional  structure:
\begin{enumerate}
\renewcommand{\labelenumi}{(\arabic{enumi})}
%\iffull\else\setlength{\myitemsep}{\myitemsep}\fi
\item A monad $(\T H, \eta^H, \mu^H)$ on $\A$ such that
  for all objects $Y$ of $\A$,
  \begin{enumerate}
%    \iffull\else\setlength{\myitemsep}{\myitemsep}\fi
  \item $\T H Y = TY$ is the 
    carrier of a final coalgebra for $H(\,\_\,) + Y$;
  \item $\mu^H_Y$ is the (unique) solution of $\alpha_{TY}$, considered as a
  flat equation morphism with parameters in $TY$. 
  \end{enumerate}
\item A natural transformation $\alpha^H : T \to HT + \Id $.
\item A natural transformation $\tau^H : HT \to T$ such that $[\tau^H,\eta^H]$ is 
  the inverse of $\alpha^H$.
\item A ``canonical embedding'' $\kappa^H$  of $H$ into $T$:
    $
    \quad
    \xymatrix@1{
      \kappa^H \equiv H \ar[r]^-{H \eta^H} & HT \ar[r]^-{\tau^H} & T.
    }
    $
\end{enumerate}
\end{thm}%
\begin{proof}
  For every object $Y$, the free Elgot algebra $TY$ provides a universal arrow
  $\eta_Y: Y \to TY$ from $Y$ to $U$, cf.~\cite{ml}, III.1. This family of
  universal arrows for every object $Y$ completely determines the left-adjoint
  to $U$ (see~\cite{ml}, Theorem IV.1.2). More concretely, $L$ assigns to a
  morphism  $f: Y \to Z$ the unique homomorphism from $TY$ to $TZ$ extending
  $\eta_Z \o f$. Then functoriality of $L$ and naturality of $\eta$ and $\eps$
  as well as the adjunction equations all follow easily from the freeness of
  the Elgot algebras $TY$.

  The obtained  adjunction $L \dashv U$ gives rise to a monad $(\T H,
  \eta^H, \mu^H)$ on $\A$ which assigns to every object of $\A$ the underlying
  object $TY$ of a free Elgot algebra on $Y$, see
  Section~\ref{sec:Eilenberg-Moore}. Thus item~(1a) follows from
  Theorem~\ref{theorem-finalcoalgebra-Elgot}. The monad multiplication $\mu^H$
  is given by $U\eps L$, i.\,e., $\mu^H_Y: TTY \to TY$ is the unique
  homomorphism of Elgot algebras such that $\mu^H_Y\o\eta^H_{TY} = \id$. It
  follows from the proof of Theorem~\ref{theorem-finalcoalgebra-Elgot} that
  $\aext{a} = \dsol{(m \after \alpha_A)}$, where $m$ is the identity on $A$.
  The special instance of this where $A$ is the free Elgot algebra $TY$
  yields~(1b).  From the freeness of the $TY$ we also infer that the algebra
  structures $\tau_Y$, and the coalgebra structures $\alpha_Y$,
  form natural transformations. Finally, by definition we have that $\alpha^H$
  and $[\tau^H, \eta^H]$ are mutually inverse. 
\end{proof}

We call the monad $(\T H, \eta^H, \mu^H)$ the \emph{completely
  iterative monad} generated by $H$.   (The name comes from an important property
which we discuss in Section~\ref{section-CIM}.)
As always, we just write $\T H$, or even $T$ to denote this monad, and
  we shall frequently drop all  superscripts when dealing with the  
structure coming from a single endofunctor $H$.
(But as the reader will see later, we frequently do need to 
consider the structures coming from two endofunctors.  This is particularly
pertinent in our study since in recursive program schemes we usually have
two signatures, hence two functors, see Section~\ref{section-rpsfirst}).

For any Elgot algebra $(A, a, \solop)$ for $H$ we call the homomorphism $\aext
a: TA \to A$ in~\refeq{eq:aext} above the \emph{evaluation
  morphism}\index{evaluation morphism} of that
Elgot algebra. Theorem~\ref{thm:aext} below shows that Elgot algebras are
equivalently presented by their evaluation morphisms.

\subsection{The Eilenberg-Moore Category of $\T H$}
\label{section-EM-T}

\begin{thm}\label{thm:aext}\hspace{-3pt}{\rm\cite{amv3}} 
  The category $\Elgot H$ of Elgot algebras is isomorphic to the
  Eilenberg-Moore category $\A^T$ of monadic algebras for the completely
  iterative monad $T$ generated by $H$. More precisely, for any Elgot algebra
  $(A, a, \funsol)$,  the evaluation morphism $U\epsilon_{A} =
  \aext{a}: TA \to A$  is an Eilenberg-Moore algebra for $T$.

  Conversely, for any Eilenberg-Moore algebra $a: TA \to A$ we obtain 
   an Elgot algebra by using as structure map
  the composite $a \o \kappa_A: HA
  \to A$, and by defining for a flat equation morphism $e: X \to HX + A$ 
 the solution  $\sol{e}
  = a \o h$, where $h$ is the unique coalgebra homomorphism from $(X, e)$ to
  $(TA, \alpha_A)$\iffull\/:
  \begin{displaymath}
  \xymatrix{
    & X \ar[d]_{h} \ar[r]^-e \ar `l[ld] `[dd]_{\sol{e}} [dd] & 
    HX + A \ar[d]^{Hh + A} \\ 
    & TA \ar[r]_-{\alpha_A} \ar[d]_a & HTA + A \\
    & A
    }
  \end{displaymath}
  \else\/.\fi
  
  These two constructions extend to the level of morphisms, and they yield the
  desired isomorphism between the two categories $\Elgot H$ and $\A^T$.
\end{thm}

\addproof{Corollary of Theorem~\ref{thm:aext}}{%
\begin{cor}\label{corr:diags}
  The diagrams
  \begin{displaymath}
  \vcenter{
    \xymatrix{
      HTT \ar[r]^{\tau{T}} \ar[d]_{H \mu}  &  TT \ar[d]^\mu \\
      HT \ar[r]_{\tau} & T
      }
    }
  \qquad\textrm{and}\qquad
  \vcenter{
    \xymatrix{
      HT \ar[r]^{\tau} \ar[d]_{\kappa T} & T \\
      TT \ar[ru]_\mu
      }
    }
  \end{displaymath}
  commute, and for every Elgot algebra $(A, a, \funsol)$ the triangle
  \begin{displaymath}
  \xymatrix{
    HA \ar[r]^-a \ar[d]_{\kappa_A} & A \\
    TA \ar[ru]_{\aext a}
    }
  \end{displaymath}
  commutes.
\end{cor}
\begin{proof}
  To see that the lower triangle commutes, observe first that $\aext{a}$ is a
  homomorphism of Elgot algebras.  So it is an $H$-algebra homomorphism by
  Proposition~\ref{prop:sol-pres-hom}. Now use this fact together with the
  equations $\kappa_A = \tau_A \o H\eta_A$ (see Theorem~\ref{thm:adjoint}(4)) and
  $\aext{a} \o \eta_A = \id$. The special cases of this
  triangle for $A = TY$ and $a = \tau_Y$ yield the commutativity of the upper
  right-hand triangle since $\mu_Y = U\eps_{TY} = \aext{\tau_Y}$. Finally, the
  upper left-hand square commutes since for each $Y$ in $\A$, $\mu_Y$ is a
  homomorphism of Elgot algebras, whence an $H$-algebra homomorphism by
  Proposition~\ref{prop:sol-pres-hom} again. 
\end{proof}
}

\section{Completely Iterative Monads}
\label{section-completely-iterative-monads}

Before we can state a theorem providing solutions of (generalized) recursive
program schemes, we need
to explain what a solution is. In the classical setting one introduces
\emph{second-order substitution} of all $\Sigma$-trees.
This is  substitution of
trees for operation symbols, see~\cite{courcelle}. We present a generalization
of second-order substitution to the final coalgebras given by $\T{H}$.  

In fact, in~\cite{aamv,milius2} it is proved that the monad $\T{H}$ of
Theorem~\ref{thm:adjoint} is characterized by an important
universal property---it is the \emph{free completely iterative monad} on $H$.
This freeness of $\T{H}$ specializes to second-order substitution of
$\Sigma$-trees, a fact we illustrate at the end of the current section. 

Here we shall quickly recall those results of~\cite{aamv}
which we will need in the current paper.  For a well-motivated and more
detailed exposition of the material presented here we refer the reader to one
of~\cite{aamv,milius2}.

\label{sec:setup}

\iffull

\begin{exmp}\label{ex:eqnmor}
  We have seen in Section~\ref{section-Elgot-algebras} that for a signature
  $\Sigma$, flat systems of formal equations have unique solutions whose
  components are $\Sigma$-trees over a set of parameters.  But it is also
  possible to uniquely solve certain non-flat systems of equations. 
  More precisely, for a given signature $\Sigma$, consider a system of
  equations as in~\refeq{eq:system}, but where each right-hand side $t_x$, $x \in X$, is
  any $\Sigma$-tree from $T_\Sigma(X+Y)$ which is not just a single variable
  from $X$. Such systems are called \emph{guarded}. Guardedness suffices to
  ensure the existence of a unique solution of the given system.
  
  \index{guarded!system of equations}
  For example, let $\Sigma$ consist of binary operations $+$ and $*$ and a
  constant $1$. The following system of equations
  \begin{displaymath}
    x_0  
    \approx   
    \vcenter{  
      \xy  
      \POS (00,00) *+{x_1}   = "x1"  
      , (10,00) *++{}     = "star"  
      , (05,-10) *+{y}     = "y"  
      , (15,-10) *+{1}  = "bot"  
      , (05,10) *++{}     = "plus"  
%%%  
      \POS "y" \ar @{-} "star"  
      \POS "bot" \ar @{-} "star"  
      \POS "star" \ar @{-} "plus"  
      \POS "x1" \ar @{-} "plus"  
%%%  
      \POS "star" *{{*}}  
      \POS "plus" *{{+}}  
      \endxy  
      }  
    \qquad  
    x_1 \approx  
    \vcenter{  
      \xy  
      \POS (00,00) *+{x_0}    = "x0"  
      , (10,00) *+{{1}} = "bot"  
      , (05,10) *++{}      = "star"  
%%%  
      \POS "x0" \ar @{-} "star"  
      \POS "bot" \ar @{-} "star"  
%%%  
      \POS "star" *{{*}}  
      \endxy  
      }  
  \end{displaymath}
  with formal variables $X = \{\,x_0, x_1\,\}$ and parameters $Y = \{\,y\,\}$
  is guarded. The solution is given by the following trees in $T_\Sigma Y$:
  \begin{displaymath}  
    \sol{x_0} =  
    \vcenter{  
      \xy  
      \POS (000,000) *+{+} = "p1"
         , (-10,-5)  *+{*} = "s1"
         , ( 10,-5)  *+{*} = "s2"
         , (-15,-15) *+{+} = "p2"
         , (-5,-15)  *+{1} ="b1"
         , (5,-15)   *+{y} ="y1"
         , (15,-15)  *+{1} ="b2"
         , (-25,-25) *+{*} ="s3"
         , (-5,-25) *+{*} = "s4"
         , (-30,-35) *+{+} = "p3"
         , (-20,-35) *+{1} = "b3"
         , (-10,-35) *+{y} = "y2"
         , (000,-35) *+{1} = "b4"
         , (-30,-40) *+{\vdots}

      %%%
      \ar@{-} "p1";"s1"
      \ar@{-} "p1";"s2"
      \ar@{-} "s1";"p2"
      \ar@{-} "s1";"b1"
      \ar@{-} "s2";"y1"
      \ar@{-} "s2";"b2"
      \ar@{-} "p2";"s3"
      \ar@{-} "p2";"s4"
      \ar@{-} "s3";"p3"
      \ar@{-} "s3";"b3"
      \ar@{-} "s4";"y2"
      \ar@{-} "s4";"b4"
      \endxy  
      }
    \qquad  
    \sol{x_1} =  
    \vcenter{  
      \xy  
      \POS  (000,000) *+{*}
      \ar@{-} '(-5,-10) *+{+}
      \ar@{-} '(5, -10)  *+{1}
      \POS (-5,-10) *+{+}
      \ar@{-} '(-15,-20) *+{*}
      \ar@{-} '(5,-20) *+{*}
      \POS (-15,-20) *+{*}
      \ar@{-} '(-20,-30) *+{+}
      \ar@{-} '(-10,-30) *+{1}
      \POS (5,-20) *+{*}
      \ar@{-} '(0,-30) *+{y}
      \ar@{-} '(10,-30) *+{1}
      \POS (-20,-35) *+{\vdots}
      \endxy  
      }  
  \end{displaymath}   
  A system of equations as considered in this example is precisely a map
  $X \to T_\Sigma(X+Y)$. And a solution is given by a map $\sol{e}: X \to
  T_\Sigma Y$ such that the square below commutes:
  \begin{displaymath}
    \xymatrix@C+1pc{
      X \ar[r]^-{\sol{e}}
      \ar[d]_e 
      & 
      T_\Sigma Y
      \\
      T_\Sigma(X+Y)
      \ar[r]_-{T_\Sigma[\sol{e}, \eta_Y]}
      & T_\Sigma T_\Sigma Y \ar[u]_{\mu_Y}
      }
  \end{displaymath}
\end{exmp}

\begin{rem} 
  In lieu of the monad $T_\Sigma$ in Example~\ref{ex:eqnmor} one can 
  more generally consider \emph{equation morphisms} $X \to T(X+Y)$ and their
  solutions $X \to TY$ for every monad $T = \T{H}$ of
  Theorem~\ref{thm:adjoint} above. Notice also that any flat
  equation morphism $e: X \to HX + Y$ gives rise to an equation morphism 
  \begin{displaymath}
    \xymatrix@1@C+1pc{
      X \ar[r]^-{e} 
      & 
      HX + Y
      \ar[r]^-{\kappa_X + \eta_Y}
      &
      TX + TY
      \ar[r]^-{\can}
      &
      T(X+Y)\,.
      }
  \end{displaymath}
  It is easy to see that solutions of this equation morphism are in one-to-one
  correspondence with solutions of $\eta_Y \after e$ in the cia $TY$
  (see~\cite{aamv}, Lemma~3.4). 
\end{rem}

In~\cite{moss,aamv} it was proved that for every monad $\T{H}$ any guarded
equation morphisms has a unique solution.  
Before we state this result precisely we recall the notion of an ideal
monad. It adds to an arbitrary monad $S$ enough structure  to be able to
speak of guarded equations for $S$. 
\fi

\iffull
\setlength{\mylen}{.5pc}
\else
\setlength{\mylen}{1pc}
\fi
\begin{defn}\label{dfn:ideal-monad}\index{ideal!monad}\index{monad!ideal $\sim$} %\mbox { } \par
By an {\em ideal monad} we mean a sextuple
\begin{displaymath}
(S,\eta,\mu,S',\iota,\mu')
\end{displaymath}
consisting of a monad $(S,\eta,\mu)$, and a (right) \emph{ideal} $(S', \mu')$,
which consists of a subfunctor $\iota:S' \subto S$,
that is a monomorphism $\iota$ in the functor category $[\A, \A]$, 
and a natural transformation $\mu':S'S\to S'$ such that the following two
conditions hold:
\begin{enumerate}
%\iffull\else\setlength{\myitemsep}{\myitemsep}\fi
\item $S=S'+\Id$ with coproduct injections $\iota$ and $\eta$

\item $\mu$ restricts to $\mu'$ along $\iota$ in the sense that 
  \iffull
  the square below commutes:
  \begin{displaymath}
    \xy
    \xymatrix{
      S'S
      \ar[0,2]^-{\mu'}
      \ar[1,0]_{\iota S}
      &
      &
      S'
      \ar[1,0]^{\iota}
      \\
      SS
      \ar[0,2]_-{\mu}
      &
      &
      S
      }
    \endxy
  \end{displaymath} 
  \else 
  the equation $\mu \o \iota S = \iota \o \mu'$ holds. 
  \fi
\end{enumerate}

\sloppypar
\index{ideal!monad morphism}
An {\em ideal monad morphism} from an ideal monad 
$(S,\eta^S,\mu^S,S',\iota,{\mu'}^S)$ to an ideal monad
$(U,\eta^U,\mu^U,U',\omega,{\mu'}^U)$ is a monad morphism
$m:(S,\eta^S,\mu^S)\to (U,\eta^U,\mu^U)$ which has
a domain-codomain restriction to the ideals. More precisely, this means that 
there exists a natural transformation $m':S'\to U'$ such that the square below
commutes:
\begin{displaymath}
  \xymatrix{
    S' 
    \ar[r]^-{m'}
    \ar[d]_\iota
    &
    U'
    \ar[d]^{\omega}
    \\
    S 
    \ar[r]_-m
    &
    U
    }
\end{displaymath}

\index{ideal!natural transformation}
 For any endofunctor $H$ and ideal monad
  $S$, a natural transformation $\sigma: H \to S$ is
 \emph{ideal} if it factors 
  through the ideal $\iota: S' \subto  S$ \iffull
  as follows:
  \begin{displaymath}\label{p:ideal}
  \xymatrix{
    H \ar[r]^{\sigma} \ar[rd]_{\sigma'} & S \\
    & S' \ar[u]_{\iota}
  }
  \end{displaymath}   
  \else\/, That is, $\sigma = \iota \o \sigma'$ for some natural transformation
  $\sigma': H \to S'$.
  \fi
\end{defn}

\begin{exmp}
  \label{ex:T-ideal}
  Recall that the underlying functor of the monad $T$ of
  Theorem~\ref{thm:adjoint} is a coproduct of $HT$ and
  $\Id$. Taking for $\iota$ the left-hand coproduct injection
  $\tau: HT \subto T$ and for $\mu'$ the natural
  transformation $H\mu: HTT \to HT$ we see that $T$ is an ideal monad. 
  Furthermore, since $\kappa: H \to T$ in Theorem~\ref{thm:adjoint}(4) is
  $\tau\o H\eta$, $\kappa$ is an ideal natural transformation. 
\end{exmp}

\begin{defn}\label{dfn:compiter}
  \iffull
  %\mbox{ } \nopagebreak \par
  \begin{myenumerate}
  \myitem 
  \else
  (i) \fi For an ideal monad $S$ on $\A$ an \emph{equation
  morphism}\index{equation morphism!for an ideal monad} is a
    morphism 
    \iffull
    \begin{displaymath}e: X \to S(X+Y).\end{displaymath}
    \else
    $e: X \to S(X+Y)$.
    \fi
    It is called \emph{guarded}\index{guarded!equation morphism} if it factors as follows:
    \iffull 
    \begin{displaymath}
    \xymatrix{
      X \ar[r]^-e \ar@{-->}[rd] & S(X + Y) \\
      & S'(X+Y) + Y \ar[u]_{[\iota_{X+Y}, \eta_{X+Y}\o\inr]}
      }
    \end{displaymath}
    \else 
    $e = [\iota, \eta_{X+Y}\o\inr] \o f$ for a morphism $f: X \to S'(X+Y) + Y$. 
    \fi 

  \iffull
  \myitem 
  \else
  \smallskip\noindent
  \index{solution!of an equation morphism}
  (ii) \fi A \emph{solution} of an equation morphism $e$ is a morphism 
    $\sol{e}: X \to SY$ such that the 
    \iffull
    square below commutes:
    \begin{displaymath}\label{p:sol}
    \xymatrix@C+1pc{
      X \ar[r]^-{\sol{e}} \ar[d]_e & SY \\
      S(X + Y) \ar[r]_-{S[\sol{e}, \eta_Y]} & SSY \ar[u]_{\mu_Y} 
      }
    \end{displaymath}
    \else
    equation $\sol{e} = \mu_Y \o S[\sol{e}, \eta_Y] \o e$ holds.
    \fi 

  \iffull
  \myitem 
  \else
  \smallskip\noindent
  \index{monad|completely iterative}\index{completely iterative monad}
  (iii) \fi An ideal monad is called \emph{completely iterative} provided that
    any guarded equation morphism has a unique solution. 
      
  \iffull
  \end{myenumerate}
  \fi
\end{defn}

\label{section-CIM}

The first item of the following result is called the \emph{Parametric Corecursion
  Theorem}
in~\cite{moss} and the \emph{Solution Theorem} in~\cite{aamv}.  See
also~\cite{milius2} for an extension of this result to all cias. The second
item is the main result of~\cite{aamv,milius2}.

\begin{thm} \label{thm:freecim}
  For any iteratable endofunctor $H$, 
  \begin{enumerate}
%    \iffull\else\setlength{\myitemsep}{\myitemsep}\fi
  \item the ideal monad $\T{H}$ is completely iterative, and
  \item $\T{H}$ is free on $H$. More precisely, for all completely iterative
    monads $S$ and ideal natural transformations $\sigma: H \to S$, there exists a
    unique monad morphism $\ol{\sigma}: T \to S$ such that $\ol{\sigma}
    \cdot \kappa^H = \sigma$\iffull\/:
  \begin{equation}\label{diag:freecim}
    \vcenter{
      \xymatrix{
        H \ar[r]^{\kappa^H} \ar[rd]_{\forall \sigma} & \T H \ar@{-->}[d]^{\exists !
          \ol{\sigma}} \\
        & S\,.
        }
      }
  \end{equation}
  \else\/. \fi
  And the induced morphism $\ol{\sigma}$ is an ideal monad morphism.
  \end{enumerate}
\end{thm}
\begin{proof}[Sketch of Proof]
  Let $(S, \eta^S, \mu^S, S', \iota, \mu')$ be a completely iterative monad. 
  For every object $Y$ of $\A$ consider $SY$ as an $H$-algebra with the
  structure
  \begin{displaymath}
  \xymatrix@1{
    HSY \ar[r]^-{\sigma_{SY}} & SSY \ar[r]^-{\mu_Y} & SY\,.
    }
  \end{displaymath}
  This is a completely iterative algebra. In fact, every flat equation morphism
  \mbox{$e:X\to HX+SY$} yields the following equation morphism w.\,r.\,t.~the
  completely iterative monad $S$: 
  \begin{displaymath}
    \xymatrixcolsep{3pc}
    \xymatrix{
      \ol{e}\equiv X
      \ar[0,1]^-{e}
      &
      HX+SY
      \ar[0,1]^-{\sigma_X+SA}
      &
      SX+SY
      \ar[0,1]^-{\can}
      &
      S(X+Y)\,.
      }
  \end{displaymath}
  It is easy to verify that $\ol e$ is guarded, and that solutions of $\ol e$
  w.\,r.\,t.~the completely iterative monad $S$ are in 1-1-correspondence with
  solutions of $e$ in the algebra $SY$. Thus, since $\ol e$ has a unique
  solution, so does $e$.  

  Now use that $(TY, \tau_Y)$ is a free cia on $Y$ to obtain a unique
  $H$-algebra homomorphism $\ol{\sigma}_Y: TY \to SY$ with $\ol{\sigma}_Y \o
  \eta_Y = \eta^S_Y$.  
  One readily proves that $\ol{\sigma}$ is natural in $Y$, that it is a monad
  morphism from $T$ to $S$, that it is uniquely determined, and that it is
  an ideal monad morphism.  See Theorem~4.4 of~\cite{milius2} for the
  details.  
\end{proof}

\comment{
\begin{rem} 
    \myitem Since the inclusion of the ideal $\xymatrix@1@-\mylen{\iota: S' \ar@{
      >->}[r] & S}$ is a monomorphism, the commutativity
  of~\refeq{diag:freecim} is equivalent to stating that
    \begin{displaymath}
    \xymatrix{
      H \ar[r]^{H\eta} \ar[rd]_{\sigma'} & HT \ar[d]^{\ol{\sigma}'} \\
      & S' 
      }
    \end{displaymath}
    commutes, where $\ol{\sigma}': HT \to S'$ is the restriction of
    $\ol{\sigma}$ to the ideal of $T$. 

  \myitem Categorically, the statement of the theorem says that every iteratable
    functor $H$ in $[\A, \A]$ has a universal arrow with respect to the forgetful
    functor
    \begin{displaymath}U: \CIM(\A) \to [\A,\A]\end{displaymath}
    of the category $\CIM(\A)$ of all completely iterative monads and
    ideal monad morphisms. Beware! The functor $U$ assigns to every
    completely 
    iterative monad $S$ its ideal $S'$, not the underlying functor. This
    choice of $U$ corresponds to the requirement that $\sigma: H \to S$ be an
    ideal transformation. 
    
    %Unfortunately, the result does not imply the existence of a left adjoint
    %for $U$ in general, since for an iteratable endofunctor $H$, we do not know
    %whether $HT$ is iteratable. If we assume that $\A$
    \sloppypar Notice that under our current assumptions $T$ is accessible;
    this follows from Remark~5.4 in~\cite{amv2}. Hence, if we restrict to the
    categories of accessible endofunctors on our locally presentable category
    $\A$ and accessible completely iterative monads (that is, completely
    iterative monads with $S$ and $S'$ accessible), then $U$ has a left
    adjoint.
  \end{myenumerate}
\end{rem}
} % end comment

In our work in the subsequent sections we shall often use the special case of
Theorem~\ref{thm:freecim} where the completely iterative monad $S$ is $\T{K}$
for some iteratable endofunctor $K$. For that special case we need the following
explicit description of the restriction of the monad morphism $\ol{\sigma}$ to
the subfunctors $H\T{H}$ and $S' = K\T{K}$. 

\begin{lem}\label{lem:ideal}
  If $H$ and $K$ are iteratable endofunctors and $\sigma: H \to \T{K}$ is an
  ideal transformation, i.\,e., $\sigma = \tau^K \o \sigma'$, then for the
  unique induced ideal monad morphism $\ol{\sigma}$ the following is a
  commutative diagram:
  \begin{displaymath}
  \xymatrix{
    H\T{H} \ar[d]_{\tau^H} \ar[r]^-{\sigma' * \ol{\sigma}} &  
    K\T{K}\T{K} \ar[r]^-{K\mu^K} & K\T{K} \ar[d]^{\tau^K} \\
    \T{H} \ar[rr]_{\ol{\sigma}} & & \T{K}
    }
  \end{displaymath}
  (Recall that $*$ denotes parallel composition of natural transformations.)
\end{lem}
\begin{proof}
We verify that the diagram
\begin{displaymath}
\xymatrix{
  % 1st line
  & H\T{H} \ar[d]^{\kappa^H \T{H}} \ar `l[d] `[dd]_{\tau^H}  [dd]
  \ar[r]^-{\sigma' * \ol{\sigma}} & 
  K \T{K} \T{K} \ar[r]^-{K \mu^K} \ar[d]^{\tau^K \T{K}} & 
  K \T{K} \ar[dd]^{\tau^K} \\
  % 2nd line
  & \T{H} \T{H} \ar[r]^-{\ol{\sigma} * \ol{\sigma}} \ar[d]^{\mu^H} & 
  \T{K} \T{K} \ar[rd]^{\mu^K} \\
  % 3rd line
  & \T{H} \ar[rr]_{\ol{\sigma}} & & \T{K}
  }
\end{displaymath}
commutes.  The left-hand and 
right-hand parts commute  by Corollary~\ref{corr:diags}.  The upper part
commutes by the double interchange law~\refeq{eq:dil} together with
$\ol{\sigma} \o \kappa^H = \sigma = \tau^K \o \sigma'$, and the lower one since
$\ol{\sigma}$ is a monad morphism. 
\end{proof}

\subsection{Second-Order Substitution}
\index{substitution!second-order}

For signature functors on $\Set$, the freeness of $T = \T{H_\Sigma}$ specializes to
\emph{second-order substitution}---substitution of (finite or infinite)
trees for operation symbols.  Second-order substitution is a key point in the classical
theory of recursive program schemes because the notion of an {\em uninterpreted
  solution\/} rests on it.  We believe that the connection of second-order
substitution and any notion of freeness is new.

\begin{exmp}\label{ex:2nd}
  Let $\Sigma$ and $\Gamma$ be signatures (considered as functors $\Nat \to
  \Set$).  Each symbol $\sigma \in \Sigma_n$ is considered as a flat tree in
  $n$ variables.  A second-order substitution gives an ``implementation'' to
  each such $\sigma$ as a $\Gamma$-tree in the same $n$ variables.  We model
  this by a natural transformation $\ell: \Sigma \to T_\Gamma \o J$, i.\,e., a
  family of maps $\ell_n: \Sigma_n \to T_\Gamma \{\,0, \ldots n-1\,\}$, $n \in
  \Nat$.  By the bijective correspondence~\refeq{eq:bij} in 
  Section~\ref{section-rpsfirst}, this gives rise to
  a natural transformation $\lambda: H_\Sigma \to T_\Gamma$. When infinite
  trees are involved there is usually the restriction to so-called
  \emph{non-erasing} substitutions, those where $\ell$ assigns to each 
  $\Sigma$-symbol a $\Gamma$-tree which is not just single node tree labelled
  by a variable. Translated along~\refeq{eq:bij} that means precisely that
  $\lambda$ is an ideal natural transformation. Thus, from 
  Theorem~\ref{thm:freecim} we get a monad morphism $\ol{\lambda}: T_\Sigma \to
  T_\Gamma$. For any set $X$ of variables, its action is that of second-order
  substitution:  $\ol{\lambda}_X$ replaces every $\Sigma$-symbol in a
  tree $t$ from $T_\Sigma X$ by its implementation according to $\lambda$. More
  precisely, let $t$ be a tree from $T_\Sigma X$. If $t = x$ is a variable from
  $X$, then $\ol{\lambda}_X(t) = x$. Otherwise we have $t = \sigma(t_1, \ldots,
  t_n)$ with $\sigma \in \Sigma_n$ and $t_i \in T_\Sigma X$, $i = 1, \ldots,
  n$. Let $\ell_n(\sigma) = t'(0, \ldots, n-1) \in T_\Gamma \{\,0,\ldots,
  n-1\,\}$. Then the operation of second-order substitution satisfies the
  following equation  
  \begin{displaymath}
    \ol{\lambda}_X(t) = t'(\ol{\lambda}_X(t_1),\ldots,\ol{\lambda}_X(t_n)). 
  \end{displaymath}
  For example, suppose that $\Sigma$ consists of two binary symbols $+$ and $*$
  and a constant $1$, and $\Gamma$ consists of a binary symbol $b$, a unary one
  $u$ and a constant $c$.  Furthermore, let $\lambda$ be given by $\ell: \Sigma
  \to T_\Gamma \o J$ as follows:
  \begin{displaymath}
    \ell_0: 1 \mapsto
    \vcenter{
      \xy
      \POS   (000,000) *+{u} = "u"
         ,   (000,-10) *+{c} = "c"
       %%%
       \ar@{-} "u";"c"
       \endxy
       }
     \qquad
     \ell_2: \mathord{+}
%    \vcenter{
%     \xy
%      \POS   (000,000)  *+{+} = "+"
%         ,   ( -5,-10)  *+[o][F-]{0} = "x"
%         ,   (  5,-10)  *+[o][F-]{1} = "y"
%      %%%
%      \ar@{-} "+";"x"
%      \ar@{-} "+";"y"
%      \endxy
%      }
    \mapsto 
    \vcenter{
      \xy 
      \POS (000,000) *+{b}  = "b"
         , ( -5,-10) *+[o][F-]{0}  = "x"
         , (  5,-10) *+{u}  = "u"
         , (  5,-20) *+[o][F-]{1}  = "y"
      %%%
      \ar@{-} "b";"x"
      \ar@{-} "b";"u"
      \ar@{-} "u";"y"
      \endxy 
      }
  \qquad  
  \mathord{*}
%  \vcenter{
%    \xy
%    \POS   (000,000)  *+{*} = "*"
%       ,   ( -5,-10)  *+[o][F-]{0} = "x"
%       ,   (  5,-10)  *+[o][F-]{1} = "y"
%    %%%
%    \ar@{-} "*";"x"
%    \ar@{-} "*";"y"
%    \endxy
%    }
  \mapsto 
  \vcenter{
    \xy 
    \POS (000,000) *+{b} = "b"
       , ( -5,-10) *+{u} = "u"
       , (  5,-10) *+[o][F-]{1} = "y"
       , ( -5,-20) *+[o][F-]{0} = "x"
    %%%
    \ar@{-} "b";"u"
    \ar@{-} "b";"y"
    \ar@{-} "u";"x"
    \endxy
    }
\end{displaymath}
and else $\ell_n$ is the unique map from the empty set. 
For the set $X = \{\,x,x'\,\}$, the second-order substitution morphism
$\ol{\lambda}_X$ acts for example as follows:
\begin{displaymath}
\vcenter{
  \xy
  \POS  (000,000) *+{*} = "*"
     ,  ( -5,-10) *+{+} = "+"
     ,  (  5,-10) *+{1} = "1"
     ,  (-10,-20) *+{x} = "z"
     ,  (  0,-20) *+{x'}= "zp"
  %%%
  \ar@{-} "*";"+"
  \ar@{-} "*";"1"
  \ar@{-} "+";"z"
  \ar@{-} "+";"zp"
  \endxy
  }
\mapsto
\vcenter{
  \xy
  \POS  (000,000) *+{b} = "b1"
     ,  ( -5,-10) *+{u} = "ul"
     ,  (  5,-10) *+{u} = "ur"
     ,  ( -5,-20) *+{b} = "b2"
     ,  (  5,-20) *+{c} = "1"
     ,  (-10,-30) *+{x} = "z"
     ,  (  0,-30) *+{u} = "u"
     ,  (  0,-40) *+{x'}="zp"
  %%%
  \ar@{-} "b1";"ul"
  \ar@{-} "b1";"ur"
  \ar@{-} "ul";"b2"
  \ar@{-} "ur";"1"
  \ar@{-} "b2";"z"
  \ar@{-} "b2";"u"
  \ar@{-} "u";"zp"
  \endxy
  }
\end{displaymath}
\end{exmp}

\section{$\T{H}$ as Final Coalgebra among Monads}
\label{sec:main}

In this section we will state and prove some technical results which are
essential for the proofs of our results on uninterpreted and interpreted
program schemes.  The culmination of the work comes in
Corollary~\ref{cor:finalcim}.
 
We would like to mention that the results and proofs in
Section~\ref{section-T-final-monad} essentially appear in the work of
Ghani et al.~\cite{glm}. However, that paper does not work in the same category
as we do.  Our setting is, perhaps, conceptually
slightly clearer and we therefore include full proofs.  We do not believe that
any of our subsequent new results in Sections~\ref{sec:sol} and~\ref{section:interpreted} can be obtained by simply applying the
solution theorem of~\cite{glm}.  
 
  \comment{% I have taken this out. 
  The proof of Theorem~\ref{thm:finalmonad} uses similar ideas as the proof of
  Theorem~4.3 (????) in~\cite{glm}. However, our result here is slightly more
  general, and we believe that, perhaps, it is conceptually a bit clearer. To
  see that we have a stronger result recall that in~\cite{glm} one works with
  \emph{$H$-coalgebraic monads}.  These are objects $(S, \sigma)$ of $H/\MA$ such
  that $\xi_{S,\sigma}: HS + \Id \to S$ is an isomorphism. 
  Observe that for any $H$-coalgebraic monad we have a unique homomorphism of
  monads
  \begin{displaymath}h_{(S, \sigma)}: (S,\sigma) \to (\T{H}, \tau^H)\end{displaymath}
  by the above Theorem~\ref{thm:finalmonad}. 

  Furthermore, the main result of~\cite{glm} will need a coproduct 
  \begin{displaymath}
  M \oplus N
  \end{displaymath}
  of two monads $M$ and $N$ in $\MA$. Beware! Coproducts in $\MA$ are
  by no means given on the level of endofunctors.  However, that need not
  concern us here since we are merely interested in the universal property.

  Then one can prove the following
  \begin{thm}\label{thm:glm}
    Let $E$ be a monad and let $S$ be an $H$-coalgebraic monad such that $S
    \oplus E$ exists.  If $e: E \to H(S\oplus E) + \Id$ is a monad morphism,
    then there exists a unique monad morphism $m: E \to \T{H}$ such that the
    following square
    \begin{displaymath}
    \xymatrix{
      E \ar[d]_m \ar[r]^-e & 
      H(S\oplus E) + \Id  \ar[d]^{H [h, m] + \Id}
      \\
      T \ar[r]_-\alpha & HT + \Id
      }
    \end{displaymath}
    commutes.
  \end{thm}
  \proof
  In fact, form the following coalgebra
  \begin{displaymath}
  \xymatrix{
    & H \ar[ld]_{\inl \cdot \sigma} \ar[rd]^{\inl \cdot H\eta^{S\oplus E}} \\
    S\oplus E \ar[rr]_-{[(H\inl + \Id) \cdot \xi_{(S,\sigma)}^{-1}, e]} 
    & & H(S\oplus E) + \Id
    }
  \end{displaymath}
  in $H/\MA$ and apply Theorem~\ref{thm:finalmonad}.
  \endproof
  }  % end of rem   

We still assume that every functor $H$ we consider is an iteratable endofunctor.
Recall that for each object $Y$, $TY$ is a final coalgebra for
the functor $H(\,\_\,) + Y$ on $\A$.

We are going to prove a number of results that strengthen this.  First,
consider the functor category $[\A, \A]$.  $H$ may be regarded as a functor on
this, by $F \mapsto H \o F$.  We also get a related functor $H \o \,\_\,+\Id$.
For the functor $T$, the value of this functor at $T$ is $H\o T + \Id$.  So the
natural transformation $\alpha^H: T \to HT + \Id$ of Theorem~\ref{thm:adjoint}(2)
may be regarded as a coalgebra structure on $T$. The proof of the following
theorem is straightforward and therefore left to the reader. 

\begin{thm} $(T,\alpha)$ is a final coalgebra for $H \o \,\_\,+\Id$.
\label{theorem-tfinal-one}
\end{thm}

% Recall that the
%completely iterative monad $T$ generated by $H$ is a final coalgebra of the
%functor 
%\begin{displaymath}H \o \,\_\,+\Id: [\A, \A] \to [\A, \A].\end{displaymath}

\iffull
\subsection{$\T{H}$ Gives a Final Coalgebra as a Monad}
\label{section-T-final-monad}
\fi

We next consider the subcategory of the comma-category $H/[\A,\A]$ 
whose objects are given by pairs $(S, \sigma: H
\to S)$, where $S$ is a monad on $\A$, and morphisms $h: (S_1, \sigma_1) \to
(S_2, \sigma_2)$ are monad morphisms $h: S_1 \to S_2$ such that $h\o\sigma_1 =
\sigma_2$. We slightly abuse notation and write
\iffull
\begin{displaymath}\label{p:Hmon}\Hmon\end{displaymath}
\else $\Hmon$ \fi
for this category. For example, one object of $\Hmon$ is $(T,\kappa)$, where
$\kappa = \tau\o H\eta$ is the canonical natural transformation of
Theorem~\ref{thm:adjoint}(4).  We show that $H$ determines an
endofunctor $\H$ on this category, and that $(T,\kappa)$ is the underlying
object of a final $\H$-coalgebra.  We then extend this finality result by
considering a subcategory of $\Hmon$.  We keep abusing notation and denote by
$\B$ the category whose objects are the pairs $(S, \sigma)$, where $S$ is {\em
  completely iterative\/} (and therefore an ideal monad, see
Definition~\ref{dfn:compiter}(iii)) and $\sigma$ is an \emph{ideal}
natural transformation; the 
morphisms in $\B$ are given by the ideal monad morphisms, see
Definitions~\ref{dfn:ideal-monad} and~\ref{dfn:compiter}.  Again, $(T,\kappa)$
is an object in this category, and we show that $\H$ restricts to an
endofunctor on $\B$, and that $(T,\kappa)$ is once again a final coalgebra for
$\H$.

Let us begin by defining $\H$ on objects of $\Hmon$ as the assignment
\iffull
\begin{displaymath}\label{p:Hs}
\H: (S, \sigma) \mapsto (HS + \Id, \inl \o H\eta)\,,
\end{displaymath}
\else
$\H: (S, \sigma) \mapsto (HS + \Id, \inl \o H\eta)$
\fi 
and for a morphism $h: (S_1, \sigma_1) \to (S_2, \sigma_2)$ we let $\H(h) = Hh
+ \Id$. 

Furthermore, notice that for any object $(S, \sigma)$ of $\Hmon$,
  there is a
natural transformation 
\begin{displaymath}\label{p:xi}
\xi_{(S, \sigma)} \equiv 
\xymatrix@1@C+1pc{
  HS + \Id \ar[r]^-{[\mu \o \sigma S, \eta]} & S
}
\end{displaymath}
As it turns out, the $\xi_{(S, \sigma)}$ are the components of a natural
transformation $\xi: \H \to \Id$ turning $\H$ into a well-copointed
endofunctor on $H/\MA$. 

\iffull
\else
The ideas for the proofs of the following two results are inspired
by~\cite{glm}. However, that paper works in a different category. Our
results are slightly more general, and perhaps, conceptually a bit clearer. 
\fi

\vbox{
\begin{lem}\label{lem:functor} \iffull\hfill\fi
  \begin{enumerate}
%    \iffull\else\setlength{\myitemsep}{\myitemsep}\fi
  \item $\H$ is an endofunctor of $\Hmon$;
  \item $\xi: \H \to \Id$ is a natural transformation;
  \item  \iffull The functor $\H$ is well-copointed. That is,
    $\xi:\H \to \Id$ is a natural transformation with
    $\H\xi_{(S, \sigma)} = \xi_{\H(S, \sigma)}$
    for all objects $(S, \sigma)$ of $H/\MA$. 
  \end{enumerate}
\end{lem}
}
\begin{proof}
(i)~Given the object $(S, \sigma)$ we define a natural transformation $\nu$
as 
\begin{displaymath}
\xymatrix{
  (HS + \Id)^2 = HS( HS + \Id) + HS + \Id 
  \ar[d]^{HS[\mu \o \sigma S, \eta] + HS + \Id} \\
  HSS + HS + \Id \ar[d]^{[H\mu, \inl] + \Id} \\
  HS + \Id 
}
\end{displaymath}
It is easy to check that $(HS + \Id, \inr, \nu)$ is a
monad, see~\cite{milius}, Lemma~3.4. Together with the natural transformation
\begin{displaymath}
\xymatrix@1{
  H \ar[r]^-{H\eta} & HS \ar[r]^-\inl & HS + \Id
}
\end{displaymath} 
we obtain an object of $H/\MA$. 

Now suppose that $h: (S_1, \sigma_1) \to (S_2, \sigma_2)$ is a morphism in 
$H/\MA$. We establish below that $\H(h) = Hh + \Id$ is a monad
morphism. Together with the commutativity of the diagram
\begin{displaymath}
\xymatrix{
  & & H \ar[ld]_{H\eta_1} \ar[rd]^{H\eta_2} \\
  & HS_1 \ar[ld]_\inl \ar[rr]^{Hh} & & HS_2 \ar[rd]^{\inl} \\
  HS_1 + \Id \ar[rrrr]_{Hh + \Id} & & & & HS_2 + \Id
}
\end{displaymath}
this establishes the action of $\H$ on morphisms. That $\H$ preserves
identities and composition is obvious. Let us check then that $\H(h)$ is a
monad morphism. The unit law is the commutativity of the triangle
\begin{displaymath}
\xymatrix@-1pc{
  HS_1 + \Id \ar[rr]^{Hh + \Id} & & HS_2 + \Id \\
  & \Id \ar[lu]^\inr \ar[ru]_\inr
}
\end{displaymath}
Thus, to complete the proof of~(i) we must check that the following
square commutes:
\begin{displaymath}
\xymatrix@C+2pc{
  (HS_1 + \Id)^2 \ar[d]_{\nu_1}\ar[rr]^{(Hh + \Id) * (Hh + \Id)} & &
  (HS_2 + \Id)^2 \ar[d]^{\nu_2} \\
  HS_1 + \Id \ar[rr]_{Hh + \Id} & & HS_2 + \Id
}
\end{displaymath}
Expanding by using the definition of $\nu_i$, $i=1,2$, this is an essentially
easy chase through some large diagrams using only the monad laws for the $S_i$,
as well as the fact that $h$ is a morphism in $H/\MA$. 
We leave this straightforward task to the reader.

\iftcs\else\medskip\noindent\fi
(ii)~It is easy to prove that each component $\xi_{(S, \sigma)}$ is a monad
morphism, see the proof of Theorem~3.1 in~\cite{milius}. Moreover, by the commutativity
of the following diagram, we obtain a morphism in $H/\MA$:
\begin{displaymath}
\xymatrix@C+1pc{
  H \ar[r]^\sigma \ar[d]_{H\eta} & S \ar[d]^{S\eta} \ar@(r,u)@{=}[rdd] \\
  HS \ar[d]_\inl \ar[r]^{\sigma S} & SS \ar[d]^\inl \ar[rd]^\mu \\
  HS + \Id \ar[r]_-{\sigma S + \Id} & SS + \Id \ar[r]_-{[\mu, \eta]} & S 
}
\end{displaymath}

Finally, we check naturality of $\xi$. Suppose that $h: (S_1, \sigma_1) \to
(S_2, \sigma_2)$ is a morphism in $H/\MA$. Then, we must prove that the
following diagram commutes: 
\begin{displaymath}
\xymatrix@C+2pc{
  HS_1 + \Id \ar[dd]_{Hh + \Id} \ar[r]^{\sigma_1 S_1 + \Id} 
  \ar[rd]_-*+{\labelstyle \sigma_2 S_1 + \Id\hspace{5pt}} & 
  S_1S_1 + \Id \ar[d]^{h S_1 + \Id} \ar[r]^-{[\mu_1, \eta_1]}
  & S_1 \ar[dd]^h \ar@{<-} `u[l] `[ll]_{\xi_{(S_1,\sigma_1)}} [ll] \\
  & S_2S_1 + \Id \ar[d]^{S_2 h + \Id} \\
  HS_2 + \Id \ar[r]_{\sigma_2 S_2 + \Id} &
  S_2 S_2 + \Id \ar[r]_-{[\mu_2, \eta_2]} &
  S_2 \ar@{<-} `d[l] `[ll]^{\xi_{(S_2,\sigma_2)}} [ll]
}
\end{displaymath}

On the left we are using the naturality of $\sigma_2$,  on the
right and in the triangle we use that $h$ is a morphism of $\Hmon$.

\medskip\noindent
(iii)~For any object $(S, \sigma)$ we have by definition that 
\begin{displaymath}
\xi_{\H(S,\sigma)} \equiv [\nu \o \inl \o H\eta(HS+\Id), \inr] : H(HS+\Id)
+ \Id \to HS + \Id.\end{displaymath}
We show that this is the same as $H\xi_{(S,\sigma)} + \Id$.

We consider the components of the coproduct separately. Equality on
the right-hand component is obvious. For the left-hand one we shall verify the
commutativity of the following diagram:
\begin{displaymath}
\smalldiag
\xymatrix@C+2.25pc{
  % 1st line
  H(HS+\Id) \ar[r]^{H\eta (HS+\Id)} \ar[d]_{H\xi} &
  HS(HS+\Id) \ar[d]_{HS\xi} \ar[r]^-\inl & 
  HS(HS+\Id) + HS + \Id = (HS + \Id)^2
  \ar[d]^{HS\xi + HS + \Id} \ar `r[d] `[dd]^\nu [dd] \\
  % 2nd line
  HS \ar[r]^{H\eta S} \ar@{=}[rd] & 
  HSS \ar[r]^-\inl \ar[d]_{H\mu} & 
  HSS + HS + \Id \ar[d]^{[H\mu, \inl] + \Id} & \\
  % 3rd line
  & HS \ar[r]^-\inl & HS + \Id & 
}
\end{displaymath}
The upper and right outer edges compose to yield the left-hand component of
$\xi_{\H(S,\sigma)}$, and the left and lower outer edges yield the left-hand
component of $H\xi_{(S, \sigma)} + \Id$. The desired commutativity of the outer
shape follows, since all inner parts of the diagram trivially commute. 
\end{proof}

\begin{lem}\label{lem:xicoalghom}
Let $((S, \sigma),\beta)$ be an $\H$-coalgebra with the structure 
$%\begin{displaymath}
  \beta: (S, \sigma) \to \H(S,\sigma) 
$ %\end{displaymath}
in $H/\MA$. Then 
\begin{displaymath}
  \xi_{(S,\sigma)}:\H(S,\sigma) \to (S,\sigma)
\end{displaymath} 
is an $\H$-coalgebra homomorphism.
\end{lem}

\begin{proof}
  It is clear that $\H(S,\sigma) = (HS + \Id, \inl \o H\eta)$ is an
  $\H$-coalgebra with the structure $H\beta + \Id$.
Also, 
one readily verifies that
  $\xi_{(S,\sigma)}$ is a morphism in $\Hmon$. From the naturality and
  well-copointedness of $\xi$ (see Lemma~\ref{lem:functor}), the following
  square 
  \begin{displaymath}
  \xymatrix@C+1pc{
    HS + \Id \ar[d]_\xi \ar[r]^-{H\beta + \Id} & 
    H(HS + \Id) + \Id \ar[d]^{H\xi + \Id = \H(\xi) = \xi_{\H(S, \sigma)}} \\
    S \ar[r]_-\beta & HS + \Id
    }
  \end{displaymath}
  commutes.  
\end{proof}

\begin{thm}\label{thm:finalmonad}
  $((T,\kappa),\alpha)$   
  is a final coalgebra for the functor $\H$ on $H/\MA$.
\label{theorem-tmon-final-hma}
\end{thm}

\addproof{Theorem~\ref{thm:finalmonad}}{%
\begin{proof}
Recall that the coalgebra structure 
$%\begin{displaymath}
  \alpha: T \to HT + \Id
$ %\end{displaymath}
is given by the inverse of $[\tau, \eta]: HT + \Id \to T$. We prove that
$\alpha$ is a morphism in $H/\MA$. Indeed, recall from Corollary~\ref{corr:diags} that 
$\tau = \mu \o \kappa T$. Hence,  
$[\tau, \eta] = \xi_{(T, \kappa)}$.  So the natural transformation $\alpha =
[\tau, \eta]^{-1}$ is an inverse of a monad
morphism. Thus, $\alpha$ is itself a monad morphism, and clearly we have
$\alpha \o \kappa = \inl \o H\eta$. 

Now suppose that 
$%\begin{displaymath}
\beta: (S, \sigma) \to \H(S, \sigma)
$ %\end{displaymath} 
is any $\H$-coalgebra. So the natural transformation
$\beta: S \to HS + \Id$ is a monad morphism such that $\beta \o \sigma
= \inl \o H\eta^S$. By Theorem~\ref{theorem-tfinal-one}, $T$ is a
final coalgebra on the level of endofunctors.  Thus there exists a
unique natural transformation $h: S \to T$ such that the following
square commutes:
\begin{equation}\label{diag:h}
\vcenter{
  \xymatrix{
    S \ar[r]^-\beta \ar[d]_h &  HS + \Id \ar[d]^{Hh + \Id} \\
    T \ar[r]_-\alpha & HT + \Id 
    }
  }
\end{equation}
Hence  our only task is to show that $h$ is a morphism in $H/\MA$. 
With an easy computation we establish the unit law: 
\begin{alignat*}{2}
  h \o \eta^S 
  & = \alpha^{-1} \o (Hh + \Id) \o \beta \o \eta^S 
  && \qquad \text{(by~\refeq{diag:h})}
  \\
  & = [\tau, \eta] \o (Hh + \Id) \o \inr 
  && \qquad \text{(since $\alpha^{-1} = [\tau, \eta]$ and $\beta \o \eta^S = \inr$)}
  \\
  & = [\tau, \eta] \o \inr \\
  & = \eta
  && \qquad \text{(computation with $\inr$)}
\end{alignat*}
and from this it follows that 
\begin{alignat*}{2}
  h \o \sigma 
  & = \alpha^{-1} \o (Hh + \Id) \o \beta \o \sigma 
  && \qquad \text{(by~\refeq{diag:h})}
  \\
  & = [\tau, \eta] \o (Hh + \Id) \o \inl \o H\eta^S 
  && \qquad \text{(since $\alpha^{-1} = [\tau, \eta]$} 
  \\
  &  && \qquad \text{and $\beta\o \sigma =
  \inl \o H\eta^S$)}
  \\
  & = \tau \o Hh \o H\eta^S 
  && \qquad \text{(composition with $\inl$)}
  \\
  & = \tau \o H\eta 
  && \qquad \text{(since $h\o \eta^S = \eta$)}
  \\
  & = \kappa && \qquad \text{(by Theorem~\ref{thm:adjoint}(4))}\,.
\end{alignat*}
Finally, we check that the following square commutes:
\begin{equation}\label{diag:monadlaw}
  \vcenter{
    \xymatrix{
      SS \ar[d]_{h*h} \ar[r]^{\mu^S} & S \ar[d]^h \\
      TT \ar[r]_\mu & T
      }
    }
\end{equation}
In order to do this we establish below that the arrows in this square
are coalgebra homomorphisms. Then~\refeq{diag:monadlaw} is immediate because
$T$ is a final coalgebra for the functor $H \o \,\_\, + \Id: [\A, \A] \to [\A,
\A]$. Firstly, we need to specify the coalgebra
structures on the objects in \refeq{diag:monadlaw}. For $S$ and $T$,
we of course use $\beta$ and $\alpha$, respectively.  For $SS$, we use
the following coalgebra structure
\begin{displaymath}
\smalldiag
\xymatrix@1@C+2pc{
  SS \ar[r]^-{\beta * \beta} & 
  (HS + \Id)^2 = HS(HS + \Id) + HS + \Id 
  \ar[rr]^-{[HS\xi, H\eta^S S] + \Id} & &
  HSS + \Id
  }
\end{displaymath}
We shall now establish that $\mu^S: SS \to S$ is a coalgebra
homomorphism. That is, we prove that the following diagram commutes:

\begin{displaymath}
\smalldiag
\xymatrix@C+2pc{
  % 1st line
  SS \ar[r]^-{\beta * \beta} \ar[d]_{\mu^S} & 
  (HS + \Id)^2 = HS(HS + \Id) + HS + \Id 
  \ar[rr]^-{[HS\xi, H\eta^S S] + \Id} 
  \ar[rrd]_\nu & &
  HSS + \Id \ar[d]^{H\mu^S + \Id} \\
  % 2nd line
  S \ar[rrr]_\beta & & & HS + \Id 
  }
\end{displaymath}
The left-hand part commutes since $\beta$ is a monad morphism and the
right-hand one is the definition of $\nu$ (use that $H\mu^S \o H\eta S =
1_{HS}$). Similarly, there is a coalgebra structure on $TT$ such that $\mu: TT
\to T$ is a coalgebra homomorphism. Finally, we show that $h*h: SS \to TT$ is a
coalgebra homomorphism. To this end we shall verify the commutativity of the
diagram below: 

\begin{displaymath}
\smalldiag
\xymatrix@C+1.75pc@R+.5pc{
  % 1st line
  SS \ar[r]^-{\beta * \beta} \ar[d]_{h * h} & 
  (HS + \Id)^2 = HS(HS + \Id) + HS + \Id 
  \ar[rr]^-{[HS\xi_{(S, \sigma)}, H\eta^S S] + \Id} 
  \ar[d]|*+{\labelstyle (Hh + \Id)^2 = Hh(Hh + \Id) + Hh + \Id} & &
  HSS + \Id \ar[d]^{H(h * h) + \Id} \\
  % 2nd line
  TT \ar[r]_-{\alpha * \alpha} & (HT + \Id)^2 = HT(HT + \Id) + HT + \Id
  \ar[rr]_-{[HT\xi_{(T, \kappa)}, H\eta T] + \Id} & & 
  HTT + \Id 
  }
\end{displaymath}
We write $(Hh + \Id)^2$ to abbreviate $(Hh + \Id) * (Hh + \Id)$. So by the
double interchange law~\refeq{eq:dil} and
by~\refeq{diag:h} the left-hand square
commutes. We consider the right-hand square 
componentwise: The right-hand component is obvious. For the middle component we
remove $H$ and obtain the following commutative square:
\begin{displaymath}
\xymatrix{
  S \ar[r]^{\eta^S S} \ar[d]_h & SS \ar[d]^{h * h} \\
  T \ar[r]_{\eta T} & TT 
  }
\end{displaymath}
Finally, for the left-hand component we remove $H$ again to obtain
\begin{displaymath}
\xymatrix@C+1pc{
  S(HS + \Id) \ar[r]^-{S \xi_{(S, \sigma)}} \ar[d]_{h * (Hh + \Id)} & SS
  \ar[d]^{h * h} \\ 
  T (HT + \Id) \ar[r]_-{T\xi_{(T, \kappa)}} & TT
  }
\end{displaymath}
By the double interchange law~\refeq{eq:dil} and the fact that $F\xi = \id_F *
\xi$, it suffices to check that the square  
\begin{displaymath}
\xymatrix@C+1pc{
  HS + \Id \ar[r]^-{\xi_{(S, \sigma)}} \ar[d]_{Hh + \Id} & S \ar[d]^h \\
  HT + \Id \ar[r]_-{\xi_{(T, \kappa)}} & T
  }
\end{displaymath}
commutes. Notice that we are not entitled to use naturality of $\xi$ here,
since we do not yet know that $h$ is a monad morphism. Instead, we invoke the
finality of $T$ and show that all arrows in the above square are coalgebra
homomorphisms for the functor $H \o \,\_\, + \Id$ on $[\A, \A]$. Indeed, 
since $h$ is a coalgebra homomorphism, so is $Hh + \Id$;  the other two
morphisms are coalgebra homomorphisms by Lemma~\ref{lem:xicoalghom}. This
completes the proof.
\end{proof}
}

\iffull
\subsection{$\T{H}$ Gives a Final Coalgebra as a Completely Iterative Monad}
\fi

The next result is the main technical tool for our treatment of recursive
program schemes in Section~\ref{sec:sol} below. Recall that we denote by $\B$
the category whose objects are the pairs $(S, \sigma)$, where $S$ is a completely
iterative monad and $\sigma$ is an ideal natural transformation; the morphisms in
$\B$ are given by the ideal monad morphisms.  So $\B$ is a subcategory of
$H/\MA$.  We show in Corollary~\ref{cor:finalcim} just below that
$((T,\kappa),\alpha)$ is a final coalgebra for the same functor $\H$ that we
have been working with.  In the language of Theorem~\ref{thm:finalmonad}, the
main point of the corollary is that if $\beta: S \to HS + \Id$ is an
\emph{ideal} monad morphism which is a coalgebra for $\H$ on completely
iterative monad $S$, then the morphism into $T$ may be taken to be an ideal monad
morphism as well.

\begin{cor}\label{cor:finalcim}
   $\H$ restricts to an endofunctor on $\B$, and
  $((T,\kappa),\alpha)$  
  is a final $\H$-coalgebra \iffull in $\B$.\else.\fi
\end{cor}

\addproof{Corollary~\ref{cor:finalcim}}{%
\begin{proof}
 Lemma 3.5 of~\cite{milius} gives a result propagating
the complete iterativeness of monads:  for any completely iterative
monad $S$ with ideal $\iota: S' \subto S$, and any
natural transformation $\sigma: H \to S$ such that $\sigma = \iota \o \sigma'$
for some $\sigma': H \to S'$,
the monad $HS + \Id$ is completely iterative, too---its ideal is $\inl: HS
\subto HS + \Id$, of course. 
Moreover, for any monad morphism $h$, $\H(h) = Hh + \Id$ is an ideal monad
morphism. Hence  $\H$ restricts to an endofunctor on $\B$, which by abuse of
notation we denote by $\H$ again.

By Theorem~\ref{thm:freecim}, $(T, \kappa)$ lies in $\B$, and it is clear that the
coalgebra structure $\alpha = [\tau, \eta]^{-1}$ is an ideal monad morphism:
the restriction of $\alpha$ to the ideals is $\id: HT \to HT$. 
Now suppose that $\beta: (S, \sigma) \to (HS + \Id, \inl \o H\eta^S)$ is an
$\H$-coalgebra; that is, $\beta$ is an ideal monad morphism between completely
iterative monads such that the following square 
\begin{displaymath}
\xymatrix{
  H \ar[d]_\sigma \ar[r]^{H\eta^S} & HS \ar[d]^\inl \\
  S \ar[r]_-\beta & HS + \Id 
  }
\end{displaymath}
\iffull\sloppypar\fi
commutes. By Theorem~\ref{thm:finalmonad}, we obtain a unique coalgebra
homomorphism $h: (S, \sigma) \to (T, \kappa)$ in $H/\MA$. Our only task is to
check that  $h$ is an ideal monad morphism, so that it
is a morphism in $\B$.   
Since $\beta$ is an ideal monad morphism,
 there is some  $\beta': S'\to HS$  such that the upper-left
square below commutes:
\begin{displaymath}
\xymatrix{
  S' \ar[d]_\iota \ar[r]^-{\beta'} & 
  HS \ar[d]^\inl \ar `r[rd] [dddr]^{Hh} & \\
  S \ar[d]_h \ar[r]^-\beta & HS + \Id \ar[d]^{Hh + \Id} & \\
  T \ar[r]^-{[\tau, \eta]^{-1}} & HT + Id \\
  & & HT \ar[lu]_\inl \ar `l[ll]_(.65)\tau [llu]
}
\end{displaymath}
We verify that the rest of the  interior regions   commute.
The middle square commutes since $h$ is a coalgebra homomorphism.   The
two other parts are obvious.   Hence the outer shape commutes, proving that $Hh
\o \beta': S' \to HT$ is a restriction of $h$ to the ideal $S'$ of $S$.
\end{proof}
}

\section{Uninterpreted Recursive Program Schemes}
\label{sec:sol}

In the classical treatment of recursive program schemes, one gives an
uninterpreted semantics to systems like \refeq{eq:algex} which are in
\emph{Greibach normal form}\index{Greibach normal form}, i.\,e., every tree on
the right-hand side of the system has as its head symbol a symbol from the
signature $\Sigma$ of givens. The semantics assigns to each of the recursively
defined operation symbols a $\Sigma$-tree. These trees are obtained as the
result of unfolding the recursive specification of the RPS. We illustrated this
with an example in~\refeq{eq:induced} in the Introduction.

We have seen in Section~\ref{section-background-monads} that $\Sigma$-trees
are the carrier of a final
coalgebra for the signature functor 
$H_\Sigma$. It is the universal
property of this final coalgebra which allows one to give a semantics to the given RPS. 
Using the technical tools
\iffull developed \else presented \fi 
in Section~\ref{sec:main} we will
now provide a conceptually easy and general way to give an uninterpreted
semantics to recursive program schemes considered more abstractly as natural
transformations, see our discussion in Section~\ref{section-rpsfirst}. 

But before we do this,
 we need to say what a solution of an RPS is. To do this we
use the universal property of the monads $\T{H}$ as presented in
Section~\ref{section-completely-iterative-monads}. The universal property
provides an abstract version of second-order substitution. 

Here are our central  definitions, generalizing recursive program schemes 
from signatures to completely iterative monads.

\begin{defn}\label{dfn:algeq}\index{recursive program scheme!definition
    of}\index{RPS|see{recursive program scheme}}
  Let $\Var$ and $H$ be endofunctors on $\A$.  A \emph{recursive program
    scheme} (or \emph{RPS}, for short) is a natural transformation
  \begin{displaymath}
    \xymatrix@1{e: \Var \ar[r] & \T{H+\Var}}\,.
  \end{displaymath}
  We sometimes call $\Var$ the {\em variables}, and $H$ the {\em givens}.
  
  \index{guarded!recursive program scheme}
  The RPS $e$ is called \emph{guarded} if there exists a natural transformation
  \begin{displaymath}
    f: \Var \to H\T{H+\Var}
  \end{displaymath}
  such that the following diagram commutes:
  \begin{equation}\label{eq:rpsguarded}
    \vcenter{
      \xymatrix{
        \Var \ar[r]^-e \ar `d[rdd]_(.7)f [rdd] & \T{H+\Var} \\
        & (H+\Var)\T{H+\Var} \ar[u]_-{\tau^{H+\Var}} \\
        & H\T{H+\Var} \ar[u]_{\inl * \T{H+\Var}}
      }
    }
  \end{equation}
  
  \index{solution!uninterpreted $\sim$ of an RPS}
  A \emph{solution} of $e$ is an ideal natural transformation 
  $\sol{e}: \Var \to \T{H}$
  such that the following triangle commutes:
  \begin{equation}\label{diag:algeq}
    \vcenter{
      \xymatrix{
        \Var \ar[r]^{\sol{e}} \ar[d]_e & \T{H} \\
        \T{H+\Var} \ar[ru]_(.6)*+{\labelstyle \ol{[\kappa^H, \sol{e}]}}
        }
      }
  \end{equation}
\end{defn}

\begin{rem}
  Recall that $\ol{[\kappa^H, \sol{e}]}$ is the unique ideal monad morphism
  extending $\sigma = [\kappa^H, \sol{e}]: H+\Var \to \T{H}$ (see
  Theorem~\ref{thm:freecim}). Observe that therefore it is 
  important to require that $\sol{e}$ be an ideal natural transformation,
 since
  otherwise  $\ol{\sigma}$ is not defined. 
\end{rem}

\begin{rem} \label{rem:cl}
  \begin{myenumerate}
  \myitem From Section~\ref{section-rpsfirst},
    our definition is a generalization of the classical
    notion of RPS (to the category theoretic setting), and it extends the
    classical work as well by allowing infinite trees on the right-hand sides of
    equations. 
  
  \myitem The classical notion of Greibach normal form (see~\cite{courcelle})
  requires that a system of formal equations such as~\refeq{eq:crps} has the
  roots of all right-hand sides labelled in $\Sigma$. Equivalently, the
  corresponding natural transformation $\Phi \to \Tsf \o J$ factors as $\Phi
  \to H_\Sigma \Tsf \o J \subto \Tsf \o J$. Translated along the bijective
  correspondence~\refeq{eq:bij} that means that a systems~\refeq{eq:crps} is in
  Greibach normal form iff the corresponding RPS $e: H_\Phi \to \T{H_\Sigma +
  H_\Phi}$ is guarded in the sense of Definition~\ref{dfn:algeq}. 
\comment{
  Our notion of guardedness
    captures precisely the requirement that all right-hand sides of
    \refeq{eq:crps} have their root labelled by a symbol from the givens 
    $\Sigma$. In the classical treatment of RPS this is precisely what is called
    Greibach normal form of an RPS, see~\cite{courcelle}.}
    
  \myitem Suppose that $H = H_\Sigma$ and $V = H_\Phi$ are signature functors 
  of $\Set$, and consider the recursive program scheme $e:
    H_\Phi \to \T{H_\Sigma + H_\Phi}$ as a set of formal equations as
    in~\refeq{eq:crps}. Then for any set $X$ of syntactic variables the
    $X$-component $\sol{e}_X : H_\Phi X \to T_\Sigma X$ of a solution assigns
    to any flat tree
    $(f, x_1, \ldots, x_n) = (f, \vec{x})$ from $H_\Phi X$ a $\Sigma$-tree over
    $X$. The commutativity of the triangle~\refeq{diag:algeq}
    gives the following property of solutions: apply to the right-hand side
    $t^f(\vec x)$ of $f(\vec{x})$ in the given RPS the second-order
    substitution that 
    replaces each operation symbol of $\Phi$ by its solution, and each
    operation symbol of $\Sigma$ by itself---this is the action of
    $\ol{[\kappa^H, \sol{e}]}_X$. The resulting tree in $T_\Sigma X$ 
    is the same tree as $\sol{e}_X(f, \vec{x})$.
    
  \myitem Any guarded equation morphism $e: X \to T(X+Y)$ w.\,r.\,t.~the free
    completely iterative monad $T= \T{H}$ (see Definition~\ref{dfn:compiter})
    can be turned into a guarded recursive program scheme with the variables
    $C_X$ and the givens $H+ C_Y$, where $C_X$ and $C_Y$ denote constant
    functors.  From our result in Theorem~\ref{thm:sol} below, one obtains a
    unique solution of that recursive program scheme, and one can prove that
    this unique solutions yields the unique solution of $e$ in the sense of
    Definition~\ref{dfn:compiter}. The non-trivial proof of this fact is left
    to the future paper~\cite{properties}.
  \end{myenumerate}
\end{rem}

\begin{exmp}\label{ex:natrps}
  Let us now present two classical RPSs as well as an example of RPS which 
  is not captured with the  classical setting.
  \begin{myenumerate}\index{factorial!guarded RPS for $\sim$}
  \myitem Recall from the Introduction the formal equations of~\refeq{eq:algex}
    and the ubiquitous~\refeq{eq:run} defining the factorial function. As
    explained in Example~\ref{ex:rps}, these give rise to recursive program
    schemes. Since both~\refeq{eq:algex} and~\refeq{eq:run} are in Greibach
    normal form, we obtain two guarded RPSs in the sense of
    Definition~\ref{dfn:algeq} above. For example, the RPS obtained
    from~\refeq{eq:run} comes from the natural transformation $\Phi \to
    H_\Sigma \Tsf \o J \subto \Tsf \o J$ with 
    \begin{displaymath}
      f \mapsto (\cond, (0, \one, f(\pred(0)) * 0)) \in \Sigma_3 \times (\Tsf
      \{\,0\,\})^3 \subto H_\Sigma \Tsf \{\,0\,\}\,,
    \end{displaymath}
    and similarly for~\refeq{eq:algex}.
 
  \myitem Sometimes one might wish to recursively define new operations
    from old ones where the new operations should satisfy certain extra
    properties automatically. We demonstrate this with an RPS recursively
    defining a new operation which is commutative.  Suppose the signature
    $\Sigma$ of givens consists of a ternary symbol $F$ and a unary one $G$.
    Let us assume that we want to require that $F$ satisfies the equation
    $F(x,y,z) = F(y,x,z)$ in any interpretation. This is modeled by
    the endofunctor $HX = X^3/\mathord{\sim} + X$, where $\sim$ is the smallest
    equivalence on $X^3$ with $(x,y,z) \sim (y,x,z)$. To be an $H$-algebra is
    equivalent to being an algebra $A$ with a unary operation $G_A$ and a
    ternary one $F_A$ satisfying $F_A(x,y,z) = F_A(y,x,z)$.  Suppose that one
    wants to define a commutative binary operation $\phi$ by the formal
    equation
    \begin{equation}\label{eq:commrps}
      \phi(x,y) \approx F(x,y,\phi(Gx, Gy))\,.
    \end{equation}
    To do it we express $\phi$ by the endofunctor $\Var$ assigning to a set $X$
    the set $\{\,\{\,x,y\,\} \mid x, y \in X \,\}$ of unordered pairs of $X$.
    It is not difficult to see that the formal equation~\refeq{eq:commrps}
    gives rise to a guarded RPS $e: \Var \to \T{H+\Var}$.  In fact, to see the
    naturality use the description of the terminal coalgebra $\T{H+\Var} Y$
    given in~\cite{am}, see Example~\ref{ex:finitary}(i).  Notice that in the
    classical setting we are unable to demand that (the solution of) $\phi$ be
    a commutative operation at this stage: one would use general facts to
    obtain a unique solution, and then one would need to devise a special
    argument to verify commutativity of that solution. Once again, our general
    theory ensures that any solution of our RPS will be commutative.
  \end{myenumerate}
\end{exmp}

The main result of this section is the following theorem. Before we present its
proof below let us illustrate the result with a few examples. 

\begin{thm}\label{thm:sol}
  Every guarded recursive program scheme has a unique solution.
\end{thm}

\begin{examples}\label{ex:rps-solutions}
  We present here the solutions of the RPSs of Example~\ref{ex:natrps}. 
  \begin{myenumerate}
  \myitem The unique solution of the RPS induced by the
    equations~\refeq{eq:algex} is an ideal natural transformation $\sol{e}:
    H_\Phi \to \T{H_\Sigma}$. Equivalently, we have a natural transformation
    $\Phi \to T_\Sigma \o J$, see~\refeq{eq:bij}. That is, the
    solution $\sol{e}$ is essentially given by two $\Sigma$-trees (one for each
    of the operation symbols $\phi$ and $\psi$) over a singleton set, say
    $\{\,x\,\}$. It follows from the proof of Theorem~\ref{thm:sol} that those
    $\Sigma$-trees are the ones given in~\refeq{eq:soltrees}, see
    Example~\ref{ex:extsolrps} below.
    
    \index{factorial!uninterpreted RPS solution}%
    Similarly, the unique solution of the RPS induced by the
    equation~\refeq{eq:run} is essentially given by the $\Sigma$-tree over the
    set $\{\,n\,\}$ below: 
    \begin{equation}\label{eq:factree}
      \vcenter{
        \xy
        \POS   (0,0)    *+{\cond} = "if1"
           ,   (-15,-10) *+{n}     = "n1"
           ,   (0,-10)  *+{\one} = "11"
           ,   (15,-10)  *+{*}    = "*1"
           ,   (0,-20)   *+{\cond} = "if2"
           ,   (30,-20)  *+{n}     = "n2"
           ,   (-15,-30) *+{\pred(n)}     = "pred1"
           ,   (0,-30)  *+{\one} = "12"
           ,   (15,-30)  *+{*}    = "*2"
           ,   (0,-40)   *+{\cond} = "if3"
           ,   (30,-40)  *+{\pred(n)}     = "pred2"
           ,   (-20,-50) *+{\pred(\pred(n))}     = "predpred1"
           ,   (0,-50)  *+{\one} = "13"
           ,   (15,-50)  *+{*}    = "*3"
           ,   (0,-60)   *+{\cond} = "if4"
           ,   (30,-60)  *+{\pred(\pred(n))}     = "predpred2"
           ,   (0,-70)   *{}   = "end"
         %%%
         \ar@{-} "if1";"n1"
         \ar@{-} "if1";"11"
         \ar@{-} "if1";"*1"
         \ar@{-} "*1";"if2"
         \ar@{-} "*1";"n2"
         \ar@{-} "if2";"pred1"
         \ar@{-} "if2";"12"
         \ar@{-} "if2";"*2"
         \ar@{-} "*2";"if3"
         \ar@{-} "*2";"pred2"
         \ar@{-} "if3";"predpred1"
         \ar@{-} "if3";"13"
         \ar@{-} "if3";"*3"
         \ar@{-} "*3";"if4"
         \ar@{-} "*3";"predpred2"
         \ar@{.} "if4";"end"
       \endxy
       }
   \end{equation}
   Notice that the nodes labelled by a term correspond to appropriately labelled
   finite subtrees. 

  \myitem We continue example~\ref{ex:natrps}(ii) and describe the
  uninterpreted solution of the guarded RPS $e$ arising from the formal
  equation~\refeq{eq:commrps} defining a commutative operation. The components
  of $\sol{e}_X: \Var X \to \T{H} X$ assign to an unordered pair $\{\,x,y\,\}$
  in $\Var X$ the tree
  \begin{displaymath}
  \xy
  \POS (0,0)*+{F}
  \ar@{-} '(-10,-10)*+{\{\,x,y\,\}}
  \ar@{-} '( 10,-10)*+{F}
  \POS (10,-10)*+{F}
  \ar@{-} '(0,-20)*+{\{\,Gx, Gy\,\}}
  \ar@{-} '(20,-20)*+{F}
  \POS (20,-20)*+{F} = "F"
  \ar@{-} '(10,-30)*+{\{\,GGx, GGy\,\}}
  \POS (30,-30) *+{\phantom{M}} = "end"
  \ar@{.} "F";"end"
  \endxy
  \end{displaymath}
  where for every node labelled by $F$ the order of the first two children
  cannot be distinguished; we indicate this with set-brackets in the picture
  above.
  \end{myenumerate}
\end{examples}

\begin{rem}
  Notice that in the classical setting not every recursive program scheme
  which has a unique solution needs to be in Greibach normal form. For example,
  consider the system formed by the first equation in~\refeq{eq:algex} and by
  the equation $\psi(x) \approx \phi(\psi(x))$. This system gives rise to an
  unguarded RPS. Thus, Theorem~\ref{thm:sol} does not provide a solution of
  this RPS. However, the system is easily rewritten to an equivalent one in
  Greibach normal form which gives a guarded RPS that we can uniquely solve
  using our Theorem~\ref{thm:sol}.
\end{rem}

The rest of this section is devoted to the proof of
Theorem~\ref{thm:sol}. We illustrate each crucial step with the help of our
examples. Before we turn to the proof of the main theorem we need
to establish some preliminary lemmas.

\begin{lem}
  \label{lem:t1}
  Let $H$ and $K$ be endofunctors on $\A$. Suppose we have objects $(S,
  \sigma)$ of $\Hmon$ and $(R, \rho)$ of $K/\MA$. Let $n: H \to K$ be a natural
  transformation and let $m: S \to R$ be a monad morphism such that the
  following square 
  \begin{displaymath}
  \xymatrix{
    H \ar[r]^-n \ar[d]_\sigma & K \ar[d]^\rho \\
    S \ar[r]_m & R}
  \end{displaymath}
  commutes. Then $n * m + \Id: HS +  \Id \to KR + \Id$ is a monad morphism such
  that the following square
  \begin{displaymath}
  \xymatrix@C+1.5pc{
    H \ar[r]^-n \ar[d]_{\inl \o H\eta^S} & K \ar[d]^{\inl \o K\eta^R} \\
    HS + \Id \ar[r]_-{n*m + \Id} & KR + \Id
    }
  \end{displaymath}
  commutes.
\end{lem}
\begin{proof}
  Naturality and the unit law are clear, and the preservation of the monad
  multiplication is a straightforward diagram chasing argument  which we leave
  to the reader. 
\end{proof}

\begin{lem}
  Consider any guarded RPS $e$, with a factor $f$ as in~\refeq{eq:rpsguarded}. 
  There exists a unique (ideal) monad morphism 
  \begin{displaymath}\ext e : \T{H+\Var} \to \T{H+\Var}\end{displaymath} 
  such that the following triangle commutes:
  \begin{displaymath}
  \xymatrix{
    H+\Var \ar[r]^-{\kappa^{H+\Var}} 
    \ar[rd]_(.4)*+{\labelstyle [\kappa^{H+\Var} \o
      \inl, e]} &  
    \T{H+\Var} \ar[d]^{\ext e} \\
    & \T{H+\Var}
    }
  \end{displaymath}
%%  Furthermore, recall that the monad morphism 
  There is also a unique  (ideal) monad morphism
  \begin{displaymath}\ol e: \T{H+\Var} \to H\T{H+\Var} + \Id\end{displaymath} 
  such that the following diagram commutes:
 \begin{equation}
   \vcenter{
     \xymatrix{
       H+\Var\ar[d]_{[H\eta^{H+\Var}, f]} \ar[r]^{\kappa^{H+\Var}} &
       \T{H+\Var} \ar[d]^{\ol e} \\
       H\T{H+\Var} \ar[r]_-\inl & H\T{H+\Var} + \Id
       }
     }
   \label{eq-ole}
 \end{equation}
 \label{lemma-prelim-uninterpreted-rps}
\end{lem}
\begin{proof}
  We get $\ext e$ from Theorem~\ref{thm:freecim}.  Indeed, it is easily checked
  that 
  \begin{displaymath}
    [\kappa^{H+\Var} \o \inl, e]: H+ \Var \to \T{H+\Var}
  \end{displaymath}
  is an ideal natural transformation using that $e$ is guarded.  As for $\ol
  e$, recall that $H$ ``embeds'' into $\T{H+\Var}$ via the natural
  transformation
  \begin{displaymath}
  \xymatrix@1@+1pc{
    H \ar[r]^-\inl & H + \Var \ar[r]^-{\kappa^{H+\Var}} & \T{H+\Var}\,.
    }
  \end{displaymath}
  Since $\kappa^{H+\Var}$ is an ideal natural transformation, so is this one.
  Hence we have
an object $(\T{H+\Var}, \kappa^{H+ \Var} \o \inl)$   of $\B$.
  It follows from Corollary~\ref{cor:finalcim} that $H\T{H + \Var} + \Id$
  carries the structure of a completely iterative monad.  The natural
  transformation $\inl \o [H\eta^{H+\Var}, f]: H+\Var \to H\T{H+\Var} + \Id$,
  see~\refeq{eq-ole}, is obviously ideal. Thus, we obtain $\ol e$ as desired
  from another application of Theorem~\ref{thm:freecim}.
\end{proof}

\begin{rem}
  \label{rem:extrps}   
  In the leading example of a classical RPS for given signatures, the formation
  of the morphism $\ol e$ corresponds to the formation of a flat system of
  equations, where for every tree there is a formal variable.  More
  precisely, suppose we have signatures $\Sigma$ and $\Phi$, and an RPS as
  in~\refeq{eq:crps} which is in Greibach normal form. The component of
  $\ol{e}$ at some set $X$ of syntactic variables can be seen as a set of
  formal equations.  Here is a description of $\ol{e}_X$: For every tree $t \in
  \Tsf X$, we have a formal variable $\ul t$. And for each formal
  variable we have one formal equation:
  \begin{displaymath}\label{p:ult}
    \begin{array}{rcl@{\quad}p{9cm}}
      \ul t & \approx & x & 
      if $t$ is a single node tree with root label $x \in X$, or 
      \\
      \ul t & \approx & \sigma(\ul{t_1}, \ldots, \ul{t_n}) & 
      for some $n \in \Nat$ and some $\sigma \in \Sigma_n$ otherwise,
    \end{array}
  \end{displaymath}
  where the tree $s = \sigma(t_1, \ldots, t_n)$ is the result of the following
  second-order substitution applied to $t$: every symbol
  of $\Phi$ is substituted by its right-hand side in the given RPS, and every
  symbol of $\Sigma$ by itself. Since the given RPS is guarded, the head symbol
  of $s$ is a symbol of $\Sigma$ for all trees $t$. 
%  Incidentally, the right-hand sides of this system corresponds to the
%  application of one step of Kleene's computation rule, see~\cite{guessarian}.
\end{rem}

\begin{exmp}\label{ex:extrps}\index{factorial!flat system of equations}
  For the guarded RPS of~\refeq{eq:algex} the flat system obtained
  from $\ol{e}$ for $X = \{\,x\,\}$ includes the equations of the
  system~\refeq{eq:induced} from the Introduction.
  
  We also give a fragment of the flat system obtained as the extension of the
  RPS~\refeq{eq:run}, see also Example~\ref{ex:natrps}(i). Here the set of
  syntactic variables is $X = \{n \}$, and the formal equations described by
  $\ol{e}_X$ include the following ones:
  \begin{eqnarray}
    \label{eq:extrps}
    \ul{f(n)} & \approx & \cond(\ul n, \ul{\one}, \ul{f(\pred(n)) *
    n}) \nonumber \\ 
  \ul n & \approx & n \nonumber \\ 
  \ul{\one} & \approx & \one \\ 
  \ul{f(\pred(n))} & \approx & \cond(\ul{\pred(n)}, \ul{\one},
    \ul{f(\pred(\pred(n))) *  \pred(n)}) \nonumber \\ 
  \ul{f(\pred(n)) * n} & \approx & \ul{\cond(\pred(n), \one,
    f(\pred(\pred(n))) * \pred(n))} * \ul{n} \nonumber \\
    & \vdots \nonumber
  \end{eqnarray}
\end{exmp}

\begin{lem}
  \label{lem:t2}
  The following diagram commutes:
  \begin{displaymath}
  \xymatrix{
    \T{H+\Var} \ar `d[ddr]_(.7){\ext e} [rdd] \ar[r]^-{\ol e} &
    H \T{H+\Var} + \Id \ar[d]^{\inl \T{H+\Var} + \Id} \\
    & (H+\Var)\T{H+\Var} + \Id \ar[d]^{[\tau^{H+\Var}, \eta^{H+\Var}]} \\
    & \T{H+\Var}
    }
  \end{displaymath}
\end{lem}
\begin{proof}
  By the first part of 
  Lemma~\ref{lemma-prelim-uninterpreted-rps}, 
  it suffices to show that the composite in the
  above diagram is an ideal monad morphism extending $[\kappa^{H+\Var}\o \inl,
  e]$.  For the extension property, we shall prove that the diagram below
  commutes, where we write $T$ for $\T{H+\Var}$:
  \begin{displaymath}
  \xymatrix@C+1pc{
    H+ \Var \ar[dd]_{\kappa^{H+\Var}} 
    \ar `r[rrrrd]^{[\kappa^{H+\Var} \o \inl, e]} [rrrrdd] 
    \ar[rd]^(.6)*++{\labelstyle [H\eta^{H+\Var}, f]} \\
    & HT \ar[d]_\inl \ar[r]^-{\inl T} & 
    (H+\Var) T \ar[d]_\inl \ar[rrd]^*+{\labelstyle \tau^{H+\Var}} & & \\
    T \ar[r]_-{\ol e} & H T + \Id \ar[r]_-{\inl T + \Id} &
    (H+ \Var)T + \Id \ar[rr]_-{[\tau^{H+\Var}, \eta^{H+\Var}]} & &
    T 
    }
  \end{displaymath}
  Indeed, the left-hand part commutes by Lemma~\ref{lemma-prelim-uninterpreted-rps}.
  For the left-hand component of the upper part notice that, using the
  definitions of $\inl * \eta^{H+\Var}$ and $\kappa^{H+\Var}$, 
  \begin{displaymath}
  \tau^{H+\Var} \o \inl T \o H \eta^{H+\Var}    = 
  \tau^{H+\Var} \o (H+\Var)\eta^{H + \Var} \o\inl = 
  \kappa^{H+ \Var} \o \inl\,.
  \end{displaymath}
  The right-hand component of this part commutes since $e$ is guarded,
  see~\refeq{eq:rpsguarded}, and the remaining parts are trivial.

  We show that all parts of the lower edge in the above diagram are monad
  morphisms. For $\ol e$, see Lemma~\ref{lemma-prelim-uninterpreted-rps}. 
  For $\inl * T + \Id$, apply
  Lemma~\ref{lem:t1} to $n = \inl$ and $m = 1_T$.  And the for the last part,
  $[\tau, \eta] = [\mu \o \kappa T, \eta]$, notice that it is the component at
  $(\T{H+\Var}, \kappa^{H+\Var})$ of the natural transformation $\xi$ of 
  Lemma~\ref{lem:functor} applied to $(H+\Var)/\MA$. 
\end{proof}

\comment{
\begin{lem}
  \label{lem:t3}
  The diagram below commutes:
  \begin{displaymath}
  \xymatrix@C+1pc{
    \T{H+\Var} \ar[r]^-{\alpha^{H+\Var}} \ar `d[ddr]_(.7){\ext{e}} [ddr] & 
    (H+\Var)\T{H+\Var} + \Id \ar[d]^{[\kappa^{H+\Var}\o\inl,e]\T{H+\Var} +
      \Id} \\
    & \T{H+\Var} \T{H+\Var} +\Id \ar[d]^{[\mu^{H+\Var}, \eta^{H+\Var}]} \\
    & \T{H+\Var}
    }
  \end{displaymath}
\end{lem}
\begin{proof}
   By the first part of 
   Lemma~\ref{lemma-prelim-uninterpreted-rps}, 
  it suffices to show that the composite in the
  above diagram is an ideal monad morphism extending $[\kappa^{H+\Var}\o\inl,
  e]$.     Let us write $\lambda$ for
  $[\kappa^{H+\Var}\o\inl,e]$ and $T$ for $\T{H+\Var}$. Now consider the
  following commutative diagram
  \begin{displaymath}
  \xymatrix@C+1pc{
    H+\Var \ar[dd]_{\kappa^{H+\Var}} \ar[rr]^-{\lambda} 
    \ar[rd]_*+{\labelstyle (H+\Var)\eta}
    & & 
    T \ar[d]^{T \eta} \ar `r[rd] [rdd]^\id \\
    & (H+\Var) T \ar[d]^{\inl} \ar[r]^-{\lambda T} & 
    TT \ar[d]^\inl \ar[rd]^\mu & \\
    T \ar[r]_-{\alpha^{H+\Var}} & (H+\Var)T + \Id \ar[r]_-{\lambda T +\Id} &
    TT + \Id \ar[r]_-{[\mu, \eta]} & T
    }
  \end{displaymath}
  which establishes the extension part. To see that the morphism of the lower
  edge is a monad morphism, recall
  from Theorem~\ref{theorem-tmon-final-hma} that $\alpha^{H+\Var}$ is the
  structure map of a final coalgebra for a functor on $(H+\Var)/\MA$, whence 
  a monad morphism, and $[\mu\o\lambda T, \eta]$ is the component at $(T,
  \lambda)$ of the natural transformation $\xi$ (see Lemma~\ref{lem:functor}). 
\end{proof}
}

\begin{proof}[Proof of Theorem~\ref{thm:sol} (see page~\pageref{thm:sol})] 
Consider $\ol e$ from Lemma~\ref{lemma-prelim-uninterpreted-rps}.
It is  a coalgebra structure for the functor 
\iftcs $\else\begin{displaymath}\fi
  \H: \B \to \B%
\iftcs\/$. \else\,.\end{displaymath}\fi 
In fact, $\ol e$
is a morphism in $\B$; it is an ideal monad morphism and by~\refeq{eq-ole} we have
\begin{equation}\label{eq:coalg}
  \ol e \o \kappa^{H+\Var} \o \inl = 
  \inl \o [H\eta^{H+\Var}, f] \o \inl = 
  \inl \o H\eta^{H+\Var}\,.
\end{equation}

Now apply Corollary~\ref{cor:finalcim} to obtain a unique $\H$-coalgebra
homomorphism from the above coalgebra $\ol e$ to the final
one. In more detail, we 
obtain a unique ideal monad morphism $h: \T{H+\Var} \to \T{H}$ such that the
following diagram commutes:
\begin{equation}\label{diag:defsol}
  \vcenter{
  \xymatrix@C+1pc{
    H \ar[r]^-\inl \ar[rrd]_{\kappa^H} & H + \Var \ar[r]^-{\kappa^{H+\Var}} & 
    \T{H+\Var} \ar[r]^-{\ol e} \ar[d]_h & 
    H\T{H+\Var} + \Id \ar[d]^{Hh + \Id} \\
    & & \T{H} \ar[r]_-{[\tau^H, \eta^H]^{-1}} & H \T{H} + \Id
    }
  }
\end{equation}
%Observe that since $\T{H+\Var} = \T{H} \oplus \T\Var$, $h$ is of the form
%$[h_1, h_2]$. It is now easy to see that $h_1$ is an $\H$ coalgebra
%homomorphism from the final coalgebra to itself, whence $h_1 = 1_{\T{H}}$.

Now let
\begin{equation}\label{eq:defsol}
  \sol{e} \equiv 
  \xymatrix@1@C+1pc{\Var \ar[r]^-\inr & H+\Var \ar[r]^-{\kappa^{H+\Var}} & 
    \T{H+\Var} \ar[r]^-h  & \T{H}\,.
    }
\end{equation}
We shall prove that $\sol{e}$ uniquely solves $e$.

\medskip\noindent 
(a)~$\sol{e}$ is a solution of $e$. Since $h$ is an ideal
monad morphism and $\kappa^{H+\Var}$ is an ideal natural transformation, we see
that $\sol{e}$ is an ideal natural transformation. Next observe that by
definition we have
\begin{displaymath}h = \ol{[\kappa^H, \sol{e}]},\end{displaymath}
and we also get
\begin{alignat}{2}
%  \begin{array}{rcl@{\qquad}p{4cm}}
    \sol{e} & = h \o \kappa^{H+\Var} \o \inr 
    && \qquad\text{(by~\refeq{eq:defsol})} \nonumber
    \\ 
    & = [\tau^H, \eta^H] \o (Hh+ \Id) \o \ol e \o \kappa^{H + \Var} \o \inr
    && \qquad \text{(by~\refeq{diag:defsol})} \label{eq:sol} 
    \\ 
    & = [\tau^H, \eta^H] \o (Hh+ \Id) \o \inl \o f 
    && \qquad \text{(by~\refeq{eq-ole})} \nonumber 
    \\ 
    & = \tau^H \o Hh \o f. \nonumber
%  \end{array}
%\end{equation}
  \end{alignat}
  Now we are ready to verify that the following diagram commutes:
\begin{equation}\label{diag:main}
\smalldiag
\vcenter{
\xymatrix@C+1pc{
  % 1st line
  \Var \ar[rrrr]^{\sol{e}} \ar[dddd]_e \ar[rdd]^f & 
  \ar@{}[rd]|{\mathrm{(i)}} & & & \T{H} \\
  % 2nd line
  & & H\T{H} \ar@{=}[r] \ar[d]|*+{\labelstyle H\eta^H \T{H}} & H\T{H} \ar[ru]_{\tau^H} \\
  % 3rd line
  & H\T{H+\Var} \ar[ru]^{Hh} \ar[r]^{H\eta^H * h} \ar[d]|*+{\labelstyle \inl \T{H+\Var}}  
  & H\T{H}\T{H} \ar[ru]_{H\mu^H} \ar@{}[rrdd]|{\mathrm{(ii)}} \\
  % 4th line
  & (H+\Var) \T{H+\Var} \ar[ru]_{\ \ \ \ [H\eta^H, Hh\o f] * h } %*!<10pt,0pt>{\scriptsize [H\eta^H, Hh\o f] * h}
  \ar[ld]^{\tau^{H+\Var}} & & & \\
  % 5th line
  \T{H+\Var} \ar `[rrrru] [rrrruuuu]_{h = \ol{[\kappa^H, \sol{e}]}}
  & & & & 
  }
}
\end{equation}
Indeed, part (i)~commutes by~\refeq{eq:sol}. For part (ii), observe first that
from Theorem~\ref{thm:adjoint}(4) and~\refeq{eq:sol} we get the equation 
\begin{equation}\label{eq:kapsol}
  [\kappa^H, \sol{e}] = 
  \xymatrix@1@C+1pc{
    H+\Var \ar[rr]^-{[H\eta^H, Hh \o f]} & & H\T{H} \ar[r]^{\tau^H} & \T{H}\,.
    }
\end{equation}
Now apply Lemma~\ref{lem:ideal} to $H+\Var$ and $H$ and $\sigma = [\kappa^H,
\sol{e}]$. The other parts of~\refeq{diag:main} are obvious.  

\medskip\noindent 
(b)~Uniqueness of solutions.  Suppose that $s: \Var \to
\T{H}$ is a solution of $e$. Since solutions are ideal natural transformations by
definition, there exists a natural transformation $s': \Var \to H\T{H}$ such
that $s = \tau^H \o s'$. We shall show below that the ideal monad morphism $h$
from~\refeq{diag:defsol} is equal to
\begin{equation}\label{eq:coinduct}
  \ol{[\kappa^H, s]}: \T{H+\Var} \to \T{H}
\end{equation}
using coinduction.  So we show that $\ol{[\kappa^H, s]}$ is a coalgebra
homomorphism. Then, since $(\T{H}, \kappa^H)$ is a final $\H$-coalgebra, we can
conclude that $h = \ol{[\kappa^H, s]}$ and therefore
\begin{displaymath}\sol{e} = \ol{[\kappa^H, \sol{e}]} \o e = h \o e = \ol{[\kappa^H, s]} \o e = s,\end{displaymath}
where the last equality holds since $s$ is a solution of $e$.

For the coinduction argument, we replace $h$ in Diagram~\refeq{diag:defsol} by 
$\ol{[\kappa^H, s]}$ and check that the modified diagram commutes. In fact, for
the left-hand triangle we obtain:
\begin{equation}\label{eq:left}
  \ol{[\kappa^H, s]} \o \kappa^{H+\Var} \o \inl = 
  [\kappa^H, s] \o \inl = \kappa^H\,.
\end{equation}
To verify the modified version of the right-hand square of
\refeq{diag:defsol}, we use the freeness of the completely iterative monad
$\T{H+\Var}$. Thus, it is sufficient that this diagram of ideal monad morphisms
commutes when precomposed with the universal arrow $\kappa^{H+\Var}$.
Furthermore we consider the components of the coproduct $H+\Var$ separately.
Let us write $x$ as a short notation for $\ol{[\kappa^H, s]}$. Then, for the
left-hand coproduct component we obtain the following equations:
%\setlongtables
%\begin{longtable}{>{$}r<{$}>{$}c<{$}>{$}l<{$}@{\hspace{1.5em}}l}
%  \endhead
%  \endfoot
\begin{alignat*}{2}
  \multicolumn{3}{l}{
    $[\tau^H, \eta^H] \o (Hx + \Id) \o \ol e \o \kappa^{H+\Var} \o \inl$
    }
  \\
  & \quad = [\tau^H, \eta^H] \o (Hx + \Id) \o \inl \o H\eta^{H+\Var} 
  && \qquad\text{(see \refeq{eq:coalg})} \\
  & \quad = \tau^H \o Hx \o H\eta^{H+\Var} \\
  & \quad = \tau^H \o H\eta^H 
  && \qquad \text{(since $x \o \eta^{H+\Var} = \eta^H$)} 
  \\
  & \quad = \kappa^H  
  && \qquad \text{(see Theorem~\ref{thm:adjoint}(4))} 
  \\
  & \quad = x \o \kappa^{H +\Var} \o \inl 
  && \qquad \text{(see \refeq{eq:left})}
%\end{longtable}
\end{alignat*}
In order to prove that the right-hand component commutes, we use a diagram
similar to Diagram (\ref{diag:main}). Just replace in (\ref{diag:main}) $Hh \o
f$ by $s'$, all other occurrences of $h$ by $x$, and $\sol{e}$ by s. We prove
that part~(i) in this modified diagram commutes. In fact, this follows from
the fact that all other parts and the outside square commute: the outside
square commutes since $s$ is a solution of $e$; part~(ii) in the modified
diagram commutes by Lemma~\ref{lem:ideal} again; and all other parts are
clear. Thus, we proved that the following equation holds:
\begin{equation}\label{eq:49}
  s = \tau^H \o Hx \o f.
\end{equation}
From this we obtain the equations
%\setlongtables
%\begin{longtable}{>{$}r<{$}>{$}c<{$}>{$}l<{$}@{\hspace{1.5em}}l}
%  \endhead
%  \endfoot
\begin{alignat*}{2}
  \multicolumn{3}{l}{ 
    $[\tau^H, \eta^H] \o (Hx + \Id) \o \ol e \o \kappa^{H+\Var} \o \inr$
    }
  \\
  & \quad =  \tau^H \o Hx \o f 
  && \qquad \text{(similar to last three lines of~\refeq{eq:sol})} 
  \\
  & \quad = s 
  && \qquad \text{(by~\refeq{eq:49})}
  \\
  & \quad = x \o \kappa^{H+\Var} \o \inr 
  && \qquad\text{(since $x = \ol{[\kappa^H, s]}$)}\,.
\end{alignat*}
%\end{longtable}
This establishes that $h$ from~\refeq{eq:defsol} is equal to 
$\ol{[\kappa^H, s]}$ from~\refeq{eq:coinduct}.
\end{proof}

\begin{rem}
  Recall that the formation of $\ol e$
  corresponds, in the leading example of an RPS for given signatures, to the
  formation of a flat system of equations, see Remark~\ref{rem:extrps}. Now the
  map $h_X$ of~\refeq{diag:defsol} assigns to every variable $\ul t \in
  T_{\Sigma + \Phi} X$ of the flat system the $\Sigma$-tree given by unfolding
  the recursive specification given by this flat system, i.\,e., $h_X$ is the
  unique solution of the flat equation morphism $\ol{e}_X$ in the cia $T_\Sigma
  X$. 
\end{rem}

\begin{exmp}\label{ex:extsolrps}  In 
  equations~\refeq{eq:algex} in the beginning of this paper, we introduced a
  guarded RPS as it would be classically presented.  It induces an RPS (in our
  sense), as discussed in Example~\ref{ex:natrps}(i).  The unique solution of
  the flat equation morphism $\ol{e}_X$ corresponding to the flat system
  described in Example~\ref{ex:extrps} is $h_X: T_{\Sigma + \Phi} X \to
  T_\Sigma X$ (see also~\refeq{eq:induced} from the Introduction). The
  definition of $\sol e$ in~\refeq{eq:defsol} means that we only consider the
  solution for the formal variables $\ul{\phi(x)}$ and $\ul{\psi(x)}$ in that
  system. These solutions are precisely the $\Sigma$-trees~\refeq{eq:soltrees} 
  as described in Example~\ref{ex:rps-solutions}(i).

  Similarly, for the guarded RPS induced by~\refeq{eq:run}, the solution is
  obtained by considering only the unique solution for the variable $\ul{f(x)}$
  of the flat system~\refeq{eq:extrps}. Clearly, this yields the desired
  tree~\refeq{eq:factree}. 
\end{exmp}

\section{Interpreted Recursive Program Schemes}
\label{section:interpreted}

We have seen in the previous section that for every guarded recursive program
scheme we can find a unique uninterpreted solution. In practice, however, one
is more interested in finding {\em interpreted\/} solutions. In the classical
treatment of recursive program schemes, this means that a recursive program
scheme defining new operation symbols of a signature $\Phi$ from given ones in
a signature $\Sigma$ comes together with some $\Sigma$-algebra $A$.  An
interpreted solution of the recursive program scheme in question is, then, an
operation on $A$ for each operation symbol in $\Phi$ such that the formal
equations of the RPS become valid identities in $A$.

Of course, in general an algebra $A$ will not admit interpreted solutions.
We shall 
\iffull
prove 
\else
show
\fi
in this section that any Elgot algebra $(A, a, (\,\_\,)^*)$ as defined in 
Section~\ref{section-Elgot-algebras}  admits 
an interpreted solution of any guarded recursive program scheme. Moreover, if
$A$ is a completely iterative algebra, interpreted solutions are unique. 
We also define the notion of a \emph{standard} interpreted solution and
\iffull
prove 
\else
state
\fi
that uninterpreted solutions and  standard interpreted ones are consistent with one
another as expected. This is a fundamental result for algebraic semantics.

We turn to applications after proving our main results.  In
Subsection~\ref{sec:opsemint} we study the computation tree semantics of RPSs
arising from the computation tree Elgot algebras of
Section~\ref{section-EASUEAC}. Then, in Subsection~\ref{sec:cpoint} we prove
that in the category $\CPO$ our interpreted program scheme solutions agree with
the usual denotational semantics obtained by computing least fixed points.
Similarly, we show in Subsection~\ref{sec:cmsint} for the category $\CMS$ of
complete metric spaces that our solutions are the same as the ones computed
using Banach's Fixed Point Theorem. Furthermore, we present new examples of
recursive program scheme solutions pertaining to fractal self-similarity.  We
are not aware of any previous work connecting recursion to implicitly defined
sets.
%Finally, in
%Subsection~\ref{sec:non-wellfounded} we discuss unique solutions of recursive
%equations in non-wellfounded sets as special instances the categorical results
%we are going to state and prove now.

\begin{defn}
\label{dfn:intsol}
Let $(A, a)$ be an $H$-algebra.  Let $e: \Var \to \T{H+\Var}$ be an RPS, and
recall the associated monad morphism $\ext{e}: \T{H+\Var} \to \T{H+\Var}$ from
Lemma~\ref{lemma-prelim-uninterpreted-rps}.  An \emph{interpreted solution} of
$e$ in $A$ is a morphism $\isol e A: \Var A \to A$ such that the following
conditions hold:
\begin{enumerate}
\item  There is an operation $(-)^{+}$ of taking solutions under which
  $(A, [a,\isol e A], (-)^{+})$ is an  Elgot algebra for $H+\Var$.
\item For $\beta = \aext{[a, \isol e A]}$, the diagram below commutes:
  \begin{equation}
    \label{eq:betaintsol}
    \vcenter{
      \xymatrix{
        \T{H + \Var} A \ar[r]^{\ext{e}_A} \ar[rd]_\beta & 
        \T{H+ \Var}A\ar[d]^\beta \\
        & A 
        }
      }
  \end{equation}
  (Recall that this is the solution of $\alpha^{H+\Var}_A : \T{H+\Var}A \to
  (H+\Var)\T{H+\Var}A + A$ in the Elgot algebra $A$ from part (i). Put
  differently, $\beta$ it is the Eilenberg-Moore algebra structure associated
  to that Elgot algebra.)
\end{enumerate}
\end{defn}

\begin{rem} \iftcs\else\mbox{ } \hfill\fi%
  \begin{myenumerate}
  \myitem The subscript $A$ in $\isol e A$ is only present to remind us of the
    codomain $A$. That is, $\isol e A$ is not a component of any natural
    transformation.  
     
    \myitem In our leading example where $H = H_\Sigma$ and $\Var = H_\Phi$ are
    signature functors on $\Set$, the commutativity of~\refeq{eq:betaintsol}
    states precisely that the evaluation map $\beta$ arising from  an
    interpreted solution interprets symbols of $\Phi$ in any tree from $\Tsf
    A$ by operations that satisfy the equations given by the recursive program
    scheme. More precisely, suppose that $e$ is a recursive program scheme
    given by formal equations as in~\refeq{eq:crps}. The interpreted solution
    $\isol e A$ gives for each $n$-ary operation symbol $f$ of the signature
    $\Phi$ an operation $f_A: A^n \to A$, and together with the given
    operations for symbols from $\Sigma$ these new operations give an Elgot
    algebra structure for $\Sigma + \Phi$ on $A$. The commutativity
    of~\refeq{eq:betaintsol} expresses the following property of the associated
    evaluation map $\beta$: take any tree $t$ in $\Tsf A$ and second-order
    substitute every operation symbol from $\Phi$ in $t$ by its right-hand side
    in the given recursive program scheme to obtain the tree $t'$---this is the
    result of applying $\ext{e}_A$ to $t$; then the result of evaluating
    $t'$ with $\beta$ in $A$ is $\beta(t)$. 
    
    \myitem The requirement that $[a, \isol e A]$ be part of the structure of
    an Elgot algebra may seem odd at first. However, we need to assume this in
    order to be able to define $\beta = \aext{[a, \isol e A]}$
    in~\refeq{eq:betaintsol}. Furthermore, the requirement has a clear
    practical advantage: operations defined recursively by means of an
    interpreted solution of an RPS may be used in subsequent recursive
    definitions. For example, for the signature functors on $\Set$ as in~(ii)
    above the Elgot algebra with structure map $[a, \isol e A]$ has operations
    for all operation symbols of $\Sigma + \Phi$. Thus, it can be used as an
    interpretation of givens for any further recursive program scheme with
    signature $\Sigma + \Phi$ of givens.
  \end{myenumerate}
\end{rem}

\begin{lem}\label{lem:intsoleq}
  Let $\isol e A: VA \to A$ be an interpreted solution of an RPS $e$ in an
  Elgot algebra $(A, a, (\,\_\,)^*)$.  Consider also $\beta = \aext{[a, \isol e
    A]}: \T{H+\Var} A \to A$.  Then the following three triangles commute:
  \begin{eqnarray}
    \label{eq:betaext}
    \vcenter{
      \xymatrix{
        (H+\Var)A \ar[r]^-{\kappa^{H+\Var}_A} \ar[rd]_{[a,\isol e A]} 
        &
        \T{H+\Var} A \ar[d]^\beta 
        \\
        & 
        A
        }
      }
    \\
  \label{eq:intsol(i)}
    \vcenter{
      \xymatrix{
        HA \ar[r]^-{\inl_A} \ar[rrd]_a & 
        (H+ \Var ) A \ar[r]^-{\kappa^{H+\Var}_A} &  
        \T{H + \Var}A \ar[d]^{\beta} \\
        & & A
        }
      } 
    \\
    \label{diag:intsol}
    \vcenter{
      \xymatrix{
        \Var A \ar[d]_{e_A} \ar[r]^-{\isol e A} & A \\
        \T{H+\Var}A \ar[ru]_(.6){\beta}
        }
      }
  \end{eqnarray}
\end{lem}
\begin{proof}
  For~\refeq{eq:betaext} see Corollary~\ref{corr:diags}, and~\refeq{eq:intsol(i)} follows
  easily from this.  For~\refeq{diag:intsol}, we verify the commutativity of
  the diagram below, where we write $T$ as a short notation for $\T{H+\Var}$ and
  $\lambda$ for $[\kappa^{H+\Var}\o\inl,e]: H+V \to T$:
  \begin{displaymath}
  \xymatrix{
    \Var A \ar `d[ddr]_(.7){e_A} [ddrr] \ar[rd]^\inr \ar[rrr]^-{\isol e A} 
    & & & A \\
    & (H+\Var)A \ar[r]^-{\kappa^{H+\Var}_A} \ar[rd]_{\lambda_A} & 
    TA \ar[ru]_\beta \ar[d]^{\ext{e}_A} & \\
    & & TA \ar `r[ru] [ruu]_(.3){\beta = \aext{[a,\isol e A]}} & 
    }
  \end{displaymath}
  The inner triangle commutes due to the definition of $\ext{e}$ (see
  Lemma~\ref{lemma-prelim-uninterpreted-rps}), the right-hand one due
  to~\refeq{eq:betaintsol}, and the other two parts are clear. 
\end{proof}

In the previously published version of this paper we incorrectly used~\refeq{diag:intsol} in
the definition of an interpreted solution in lieu of~\refeq{eq:betaintsol}. The
following example shows that the two definitions are not equivalent as we
originally claimed. 

\begin{exmp}
  Consider the signature $\Sigma$ with a unary operation symbol $F$. The
  associated signature functor is $H_\Sigma = \Id$. Take the $\Sigma$-algebra
  $A = \{\,0,1\,\}$ with the unary operation $F_A = \id_A$. This is a
  $\CPO$-enrichable algebra in the sense of Example~\ref{ex:elgot}(ii); thus,
  $A$ is an Elgot algebra for $H_\Sigma$. Notice that $\T{H_\Sigma} X =
  \{\,F\,\}^* \times X + 1$, where the element of $1$ stands for the infinite
  sequence $F^\omega = FFF\cdots$. So the associated evaluation map $T_\Sigma A \to A$
  assigns to each $(F^n, i)$, $i = 0,1$, the element $i$ and to $F^\omega$ the
  least element $0$ of $A$.
  
  Now consider the formal equation $f(x) \approx F(x)$ which defines a guarded
  RPS $e: \Var \to \T{H_\Sigma + \Var}$, where $\Var = \Id$, as explained in
  Section~\ref{section-rpsfirst}. We verify now that the unary operation $f_A =
  \id_A$ is an $\Var$-algebra structure such that condition~(i) in
  Definition~\ref{dfn:intsol} holds and such that~\refeq{diag:intsol} commutes.
  Observe first that
  $$
  \T{\Id + H_\Sigma} X = \{\,f,F\,\}^* \times X + \{\,f,F\,\}^\omega\,.
  $$
  To see that $[F_A, f_A]: A + A \to A$ is part of the structure of an Elgot
  algebra consider the map $\beta: \T{\Id + H_\Sigma} A \to A$ defined by 
  \begin{eqnarray*}
    \beta: (w,i) & \mapsto & i \qquad \textrm{for $w\in \{\,f,F\,\}^*$}, \\
    v & \mapsto & 
    \left\{
    \begin{array}{lp{5cm}}
      1 & if $v = uf^\omega$, $u \in \{\,f,F\,\}^*$, \\
      0 & else.
    \end{array}
    \right.
  \end{eqnarray*}
  It is not difficult to verify that this map $\beta$ is an Eilenberg-Moore
  algebra for $\T{\Id + H_\Sigma}$ (for details see~\cite{abm}). By the first
  line of the definition we also see that $\beta \o \kappa^{\Id + H_\Sigma}_A =
  [f_A, F_A]$. Now the commutativity of~\refeq{diag:intsol} expresses precisely
  that $f_A(i) = F_A(i)$ holds, which is true since $f_A = F_A = \id_A$. 

  However, $f_A$ is not an interpreted solution of $e$ in $A$ in the sense of
  Definition~\ref{dfn:intsol}. The commutativity of~\refeq{eq:betaintsol}
  expresses that the evaluation map $\beta$ is consistent with the second order
  substitution $\ext{e}_A$ which replaces in any infinite tree from $\T{\Id +
  H_\Sigma}A$ any variable symbol from the RPS $e$ by its right-hand
  side, and so in our example $\ext{e}_A$ replaces every symbol $f$ by a symbol
  $F$. Thus, we have 
  $$
  \beta(f^\omega) = 1 \neq 0 = \beta (F^\omega) = \beta \o \ext{e}_A
  (f^\omega)\,.
  $$
\end{exmp}

\begin{thm}
\label{thm:intsol}
Let $(A, a, (\,\_\,)^*)$ be an Elgot algebra for $H$, and let $e: \Var \to
\T{H+\Var}$ be a guarded RPS. Then the following hold:
\begin{enumerate}
%\iffull\else\setlength{\myitemsep}{\myitemsep}\fi
\item there exists an interpreted solution $\isol e A$ of $e$ in $A$, 
\item if $A$ is a completely iterative algebra, then $\isol e A$ is the unique
  interpreted solution of $e$ in $A$. 
\end{enumerate}
\end{thm}

\begin{proof}
(i)~Given an Elgot algebra $(A, a, (\,\_\,)^*)$ and a guarded recursive program
  scheme $e: \Var\to \T{H+\Var}$
 consider $\ol e: \T{H + \Var} \to H \o \T{H + \Var} +
  \Id$ from Lemma~\ref{lemma-prelim-uninterpreted-rps}.
 Its component at $A$ yields a
  flat equation morphism
  \begin{equation}\label{eq:g}
    g \equiv \T{H + \Var} A \to H \T{H + \Var} A + A\,,
  \end{equation}
  with respect to $(A, a, (\,\_\,)^*)$ and we take its solution
  \begin{equation}\label{eq:defbeta}
    \beta \equiv
    \xymatrix@1{\T{H+\Var}A \ar[r]^-{g^*} & A} 
  \end{equation}

  We will prove that the morphism
  \begin{equation}\label{eq:defisol}
    \isol e A \equiv
    \xymatrix@1@C+1pc{
      \Var A \ar[r]^-{\inr} &
      (H + \Var) A \ar[r]^-{\kappa^{H+\Var}_A} &
      \T{H+\Var} A \ar[r]^-{g^*} &
      A\,
    }
  \end{equation}
  is an interpreted solution of $e$ in $A$. We proceed in four steps (a)--(d).

  (a)~We first check that $\beta$ an evaluation morphism; more precisely,
  $\beta$ is the structure of an Eilenberg-Moore algebra for $\T{H+\Var}$.
  To this end we first establish the equation
  \begin{equation}\label{eq:beta}
    \beta = \aext{a} \o h_A\,,
  \end{equation}
  where $h: \T{H + \Var} \to \T{H}$ is the monad morphism that we obtain from
  the recursive program scheme $e$, see~\refeq{diag:defsol}. (It is also
  useful to recall that $h = \ol{[\kappa^H,\sol{e}]}$.)  Recall that $\aext{a}
  = (\alpha_A)^*$, see Theorem~\ref{thm:adjoint}, and that $h_A$ is a
  homomorphism of coalgebras for $H(\,\_\,) + A$, see
  Diagram~\refeq{diag:defsol}. Thus, by the Functoriality of $(\,\_\,)^*$, we
  obtain $g^* = (\alpha_A)^* \o h_A$.  By~\refeq{eq:g} and~\refeq{eq:aext},
  this is the desired Equation~\refeq{eq:beta}.  Now since $h: \T{H + \Var}
  \to \T H$ is a monad morphism, and $\aext{a}$ is the structure morphism of
  an Eilenberg-Moore algebra for $\T H$, $\beta = \aext{a} \o h_A$ is an
  Eilenberg-Moore algebra for $\T{H + \Var}$. In fact, this follows from a
  general fact from category theory, see e.\,g.~Proposition~4.5.9
  in~\cite{borceux}.
  
  (b)~We check that the diagram~\refeq{eq:betaintsol} commutes for $\beta$
  from~\refeq{eq:defbeta}. Let us use $T$ as a short notation for $\T{H+\Var}$
  and $T'$ for $\T H$; we also use $\eta'$, $\tau'$, and $\mu'$ to denote the
  natural transformations associated with $T'$.  Consider the following diagram
  \begin{equation}\label{diag:hfix}
    \vcenter{
    \xymatrix@C+1.5pc{
      T \ar[ddd]_h \ar[r]^-{\ol e}  &
      HT + \Id \ar[ddd]^{Hh + \Id} \ar[r]^-{\inl T + \Id} 
      \ar[rdd]_{\kappa^H * h + \Id} \ar@{}[rd]|(.6){(*)} & 
      (H+ \Var) T + \Id \ar[d]_{\kappa T + \Id} \ar[r]^-{[\tau,\eta]} & 
      T 
      \ar[ddd]^h 
      \ar@{<-} `u[l] `[lll]_{\ext e} [lll]
      \\
      & & TT + \Id \ar[ru]_{[\mu, \eta]} \ar[d]^{h * h + \Id} \\
      & & T'T' + \Id \ar[rd]^{[\mu', \eta']} \\
      T' 
      \ar@{<-} `d[r] `[rrr]_{\id} [rrr]
      & HT' + \Id \ar[l]^-{[\tau', \eta']} \ar[rr]_-{[\tau', \eta']} & &
      T' 
      }
    }
  \end{equation}
  All of its inner parts commute. The lowest part trivially commutes, the
  upper part is Lemma~\ref{lem:t2}, for the left-hand square,
  see~\refeq{diag:defsol}, and for the right-hand part use that $h$ is a monad
  morphism. That part~($*$) commutes follows from commutativity of the
  left-hand triangle of~\refeq{diag:defsol} and the double interchange
  law~\refeq{eq:dil}. The right-hand coproduct component of the remaining inner
  part trivially commutes, and the left-hand component commutes due to
  Corollary~\ref{corr:diags}: $\mu \o \kappa T = \tau$. 
  Thus, the outer shape of diagram~\refeq{diag:hfix} commutes.  We obtain the
  equations
  \begin{alignat*}{2}
    \beta \o \ext{e}_A & =  \aext{a} \o h_A \o \ext{e}_A 
    && \qquad\text{(see~\refeq{eq:beta})}
    \\ 
    & = \aext{a} \o h_A 
    && \qquad \text{(see~\refeq{diag:hfix})}
    \\
    & = \beta 
    && \qquad \text{(see~\refeq{eq:beta})}
  \end{alignat*}

  (c)~We argue that the diagram below commutes:
  \begin{displaymath}
    \xymatrix{
      HA
      \ar[r]^-\inl
      \ar[rrd]^{\kappa^H_A}
      \ar `d[ddrr] [ddrr]_a
      &
      (H+\Var)A
      \ar[r]^-{\kappa^{H+\Var}_A}
      &
      TA
      \ar[d]^{h_A}
      \ar `r[d] `[dd]^{\beta} [dd]
      \\
      & &
      T'A
      \ar[d]^{\aext{a}}
      \\
      & &
      A
    }
  \end{displaymath}
  Indeed, the upper triangle commutes by~\refeq{diag:defsol}, the right-hand part commutes
  by~\refeq{eq:beta} and for the commutativity of the lower part see
  Corollary~\ref{corr:diags}. Thus, the outward shape of the diagram commutes.

  (d)~We are ready to prove that $\isol e A$ as defined in~\refeq{eq:defisol}
  is an interpreted solution of $e$ in $A$. From the diagram in~(c) and the
  definition of $\isol e A$ we see that the equation
  \begin{displaymath}
    \beta \o \kappa^{H+\Var}_A = [a, \isol e A]
  \end{displaymath}
  holds.  Recall that by~(a) above, $\beta$ is an evaluation morphism.
  This fact, together with the equation just above, imply
  that $[a,\isol e A]$ is the structure morphism of an Elgot
  algebra on $A$ (cf.~Theorem~\ref{thm:aext}). This establishes item~(i) in
  Definition~\ref{dfn:intsol}, and item~(ii) was established in (b) above.
  This completes the proof of part~(i).
  
  \medskip\noindent (ii) Let $(A, a, (\,\_\,)^*)$ be a cia. We show
  that the solution $\isol e A$ defined in~(i) is unique. 
  We will now verify that any
  evaluation morphism $\beta: \T{H+\Var} A \to A$ for which
  diagram~\refeq{eq:betaintsol} commutes is a solution
  of $g$ as defined in~\refeq{eq:g}. To this end consider the following diagram, where
  we write $T = \T{H+\Var}$ for short once more:

  \begin{displaymath}
    \xymatrix@C+.5pc{
      % 1st
      TA \ar[ddd]_\beta \ar[r]^-{\ol{e}_A = g} & 
      HTA + A \ar[ddd]^{H\beta + A}  \ar[r]^-{\inl_{TA} + A} & 
      (H+\Var)TA + A \ar[dd]_{(H+\Var)\beta + A} \ar[rd]_*+{\labelstyle \kappa_{TA} + A} 
      \ar[rr]^-{[\tau_A, \eta_A]} & & 
      TA 
      \ar[ddd]^\beta 
      \ar@{<-} `u[l] `[llll]_{\ext{e}_A} [llll]
      \\
      % 2nd
      & & & TTA + A \ar[ru]_*+{\labelstyle [\mu_A, \eta_A]} \ar[d]^{T\beta + A} \\
      & & (H+\Var)A + A \ar[r]_-{\kappa_A + A} & TA + A \ar[rd]_{[\beta, A]} \\
      % 3rd
      A
      \ar@{<-} `d[r] `[rrrr]_\id [rrrr]
      & HA + A \ar[l]^-{[a, A]} \ar[rrr]_-{[a, A]} \ar[ru]_(.6)*+{\labelstyle
      \inl_A + A} & & & 
      A 
      }
  \end{displaymath}
  Its outer shape commutes due to~\refeq{eq:betaintsol}, the right-hand part
  since $\beta$ is an Eilenberg-Moore algebra structure, the upper-right
  triangle follows from Corollary~\ref{corr:diags}, and the lower
  right-hand part follows from~\refeq{eq:intsol(i)}. Thus, since all other
  parts are obviously commutative, the left-hand inner square commutes. But
  this shows that $\beta$ is a solution of $g$.   By the
  uniqueness of solutions, we have $\beta = g^*$. 
  
  To complete our proof, we appeal to the first part of Lemma~\ref{lem:intsoleq}
  to see that the interpreted solutions are always determined by their evaluation morphisms $\beta$.
This means that interpreted solutions in cias are unique.
\end{proof}

\index{solution!standard interpreted $\sim$} 
\begin{defn}\label{dfn:standard}
  For any guarded RPS $e$ and any Elgot algebra $(A,a, \starop)$, let $\isol e
  A$ be the interpreted solution obtained from the proof of
  Theorem~\ref{thm:intsol} as stated below
  \begin{displaymath}
  \isol e A \equiv 
  \xymatrix@1@C+1pc{
    \Var A \ar[r]^-{\inr} & 
    (H + \Var) A \ar[r]^-{\kappa^{H+\Var}_A} &
    \T{H+\Var} A \ar[r]^-{g^*} &
    A\,,
    }
  \end{displaymath}
  where $g$ is the flat equation morphism of~\refeq{eq:g}. We call this the
  \emph{standard interpreted solution} of $e$ in $A$. 
\end{defn}

It follows from the proof of Theorem~\ref{thm:intsol} that uninterpreted 
solutions correspond to certain interpreted ones
 in a canonical way. We now make this precise
 and thereby prove what could be called ``Fundamental
Theorem of Algebraic Semantics''. 

\begin{thm}\label{thm:fundamental}
  Let $(A, a, (\,\_\,)^*)$ be an Elgot algebra and consider its evaluation
  morphism $\aext{a}: \T H A \to A$. Let $e$ be any guarded recursive 
  program scheme, let $\isol e A: \Var A \to A$ be the standard interpreted
  solution of $e$ in $A$, and let $\sol{e}: \Var \to
  \T{H}$ be the (uninterpreted) solution of Theorem~\ref{thm:sol}. Then the 
  \iffull
  triangle
  \begin{equation}\label{diag:fundamental}
    \vcenter{
      \xymatrix{
        \Var A \ar[r]^-{(\sol{e})_A} \ar[rd]_{\isol e A} & 
        \T{H} A \ar[d]^{\aext a}
        \\
        & A 
        }
      }
  \end{equation}
  commutes.
  \else 
  equation $\aext a \o (\sol e)_A = \isol e A$ holds.
  \fi

  Furthermore, the standard interpreted solution $\isol e A$ is uniquely
  determined by the commutativity of the above triangle. 
\end{thm}

\begin{rem}   
  Notice that $(\sol e)_A$ is the component at $A$ of the natural
  transformation $\sol e$. And once again, $\isol e A$ is not the component
  at $A$ of any natural transformation but merely a morphism from $\Var A$ to
  $A$.  
\end{rem}

\addproof{Theorem~\ref{thm:fundamental}}{%
\begin{proof}
  In fact, we have the commutative diagram
  \begin{displaymath}
  \xymatrix@C+1pc{
    VA 
    \ar[rr]^{\sol{e}_A} 
    \ar[dr]_{e_A}
    \ar `d[ddr]_(.7){\isol{e}{A}} [ddrr]
    & 
    & 
    \T{H}A  
    \ar[dd]^{\aext{a}}
    \\
    & 
    \T{H+V}A 
    \ar[ru]_*+{\labelstyle \ol{[\kappa^H, \sol{e}]}_A} 
    \ar[dr]_(.4){\aext{[a,\isol{e}{A}]}} 
    & 
    \\
    & & 
    A
    }
  \end{displaymath}
  The lower left part commutes by Lemma~\ref{lem:intsoleq},
  see~\refeq{diag:intsol}; the top triangle is the definition of an
  uninterpreted solution, see~\refeq{diag:algeq}; and the triangle on the right
  commutes by~\refeq{eq:beta} since $h = \ol{[\kappa^H, \sol{e}]}$ and $\beta =
  \aext{[a, \isol e A]}$.  Hence~\refeq{diag:fundamental} follows immediately.

  It is obvious that $\isol e A$ is uniquely determined by the commutativity of
  the triangle~\refeq{diag:fundamental}, as neither $\aext a$ nor $\sol{e}$
  depend on $\isol e A$.  
\end{proof}
}

\subsection{Interpreted Solutions in Computation Tree Elgot Algebras}
\label{sec:opsemint}
\index{factorial!computation tree semantics}

In this section, we work through some details concerning the factorial example
of Equation~\refeq{eq:run}, namely
\begin{displaymath}\label{p:Natupa}f(n) \approx \cond(n, \one, f(\pred(n) * n))\end{displaymath}
and studied further in
Example~\ref{ex:natrps}(i). Recall that we work with the signature $\Sigma$
containing a constant $\one$, one unary symbol $\pred$, a
binary symbol $*$ and a ternary one $\cond$.  Let $H_{\Sigma}$ be the
associated endofunctor on $\Set$.  Let $\upa$ be an object which is not a
natural number, and let $\Nat_{\upa}$ be the $H_{\Sigma}$-algebra with carrier
$\{0,1,2, \ldots\}\cup \{\upa\}$ and whose operations are the strict extensions
of the standard operations.  (For example $\upa * 3 = \upa$ in this algebra.)

We shall use the computation tree Elgot algebra structure 
 $(\Nat_{\upa}, a, (\,\_\,)^*)$ from Section~\ref{section-EASUEAC}.
That structure used a particular element of the carrier set
in connection with the conditional $\cond$, and in our
structure we take $0$.
Before looking at the interpreted solution to the factorial RPS, it
might be useful to spell out the associated evaluation morphism
${\aext a}: \T{H_\Sigma} \Nat_{\upa} \to \Nat_{\upa}$.
Let $t$ be a finite or infinite $\Sigma$-tree over $\Nat_{\upa}$;
so the leaves of $t$ might be labelled with natural number or $\upa$,
but not with formal variables. 
Here is a pertinent example:
 \begin{displaymath}
  \vcenter{
    \xy
    \POS   (0,0)    *+{\cond} = "if1"
       ,   (-15,-10) *+{1}     = "n1"
       ,   (0,-10)  *+{\one} = "11"
       ,   (15,-10)  *+{*}    = "*1"
       ,   (0,-20)   *+{\cond} = "if2"
       ,   (30,-20)  *+{1}     = "n2"
       ,   (-15,-30) *+{\pred(1)}     = "pred1"
       ,   (0,-30)  *+{\one} = "12"
       ,   (15,-30)  *+{*}    = "*2"
       ,   (0,-40)   *+{\cond} = "if3"
       ,   (30,-40)  *+{\pred(1)}     = "pred2"
       ,   (-20,-50) *+{\pred(\pred(1))}     = "predpred1"
       ,   (0,-50)  *+{\one} = "13"
       ,   (15,-50)  *+{*}    = "*3"
       ,   (0,-60)   *+{\cond} = "if4"
       ,   (30,-60)  *+{\pred(\pred(1))}     = "predpred2"
       ,   (0,-70)   *{}   = "end"
       %%%
       \ar@{-} "if1";"n1"
       \ar@{-} "if1";"11"
       \ar@{-} "if1";"*1"
       \ar@{-} "*1";"if2"
       \ar@{-} "*1";"n2"
       \ar@{-} "if2";"pred1"
       \ar@{-} "if2";"12"
       \ar@{-} "if2";"*2"
       \ar@{-} "*2";"if3"
       \ar@{-} "*2";"pred2"
       \ar@{-} "if3";"predpred1"
       \ar@{-} "if3";"13"
       \ar@{-} "if3";"*3"
       \ar@{-} "*3";"if4"
       \ar@{-} "*3";"predpred2"
       \ar@{.} "if4";"end"
    \endxy
    }
  \end{displaymath}
  We got this by taking the uninterpreted solution of our RPS, as depicted
  in~\refeq{eq:factree}, and then substituting the number $1$ for the formal
  variable $n$.  Note that the nodes labelled $\cond$ have three children.
   Let $t$ again be a $\Sigma$-tree over   $\Nat_{\upa}$.
Here is how we define $\aext a(t)$.  We look for a finite subtree $u$ of $t$
  with the property that if a node belongs to $u$ and is labelled by a function
  symbol other than $\cond$, then all its children belong to $u$ as well; and
  if the node is labeled by $\cond$, then either the first and second children
  belong to $u$, or else the first and third children do. For such a finite
  subtree $u$, we can evaluate the nodes in a bottom-up fashion using the
  $H_{\Sigma}$-algebra structure. We require that for a conditional node $x$,
  the first child evaluates to $0$ (from our Elgot algebra structure) iff the
  second child is in $u$. If such a finite $u$ exists, then we can read off
  an element of $\Nat_{\upa}$. This element is $\aext a(t)$.  If no finite $u$
  exists, we set $\aext a(t) = \upa$.  Returning to our example above, the
  finite subtree would be
 \begin{displaymath}
  \vcenter{
    \xy
    \POS   (0,0)    *+{\cond} = "if1"
       ,   (-15,-10) *+{1}     = "n1"
 %      ,   (0,-10)  *+{\one} = "11"
       ,   (15,-10)  *+{*}    = "*1"
       ,   (0,-20)   *+{\cond} = "if2"
       ,   (30,-20)  *+{1}     = "n2"
       ,   (-15,-30) *+{\pred(1)}     = "pred1"
       ,   (0,-30)  *+{\one} = "12"
       %%%
       \ar@{-} "if1";"n1"
%       \ar@{-} "if1";"11"
       \ar@{-} "if1";"*1"
       \ar@{-} "*1";"if2"
       \ar@{-} "*1";"n2"
       \ar@{-} "if2";"pred1"
       \ar@{-} "if2";"12"
    \endxy
    }
  \end{displaymath}
And for our example tree $t$, $\aext a(t) =  1$.

We are now in a position to discuss the interpreted solution of our RPS.
Recall that the signature $\Phi$ of recursively defined symbols contains only
the unary symbol $f$.  The corresponding signature functor is $H_\Phi$, and
$H_\Phi(\Nat_{\upa})$ is the set $\{f(0), f(1),
\ldots\}\cup\{f(\upa)\}$.  The RPS itself is a natural transformation $e :
H_\Phi \to \T{H_{\Sigma} + H_\Phi}$.  The uninterpreted solution is the natural
transformation $\sol e : H_\Phi \to \T{H_{\Sigma}}$ corresponding to the tree
shown in~\refeq{eq:factree}.  We are concerned here with the interpreted
solution $ \isol{e}{\Nat_{\upa}}: H_\Phi(\Nat_{\upa}) \to \Nat_{\upa}$ of our
RPS.  In light of the Fundamental Theorem~\ref{thm:fundamental}, this is $\aext
a \o (\sol e)_{\Nat_{\upa}}$.  We show by an easy induction on $n\in N$ that
this interpreted solution takes $f(n)$ to $n!$, and that it takes $f(\upa)$ to
$\upa$.

We could also establish this same result directly, without
Theorem~\ref{thm:fundamental}.  To do this, we follow the proof of
Theorem~\ref{thm:intsol}.  We turn our RPS $e$ into a related natural
transformation $ \ol e: \T{H_\Sigma + H_\Phi} \to H_\Sigma \T{H_\Sigma +
  H_\Phi} + \Id$.  Then ${\ol e}_{\Nat_{\upa}}$ is a flat equation morphism in
the Elgot algebra $\Nat_{\upa}$, and its solution is the interpreted solution
of our RPS.  Here is a fragment of ${\ol e}_{\Nat_{\upa}}$:
\begin{eqnarray*}
  \ul{f(0)} & \approx & \cond(\ul{0}, \ul{\one}, \ul{f(\pred(0)) *
    0})   \\ 
%  \ul{f(\pred(0)) * 0} & \approx & \ul{f(\pred(0))} * 0\\ 
   \ul{f(1)} & \approx & \cond(\ul{1}, \ul{\one}, \ul{f(\pred(1)) *
    1})   \\ 
  \ul{f(\pred(1))} & \approx & \cond(\ul{\pred(1)}, \ul{\one},
    \ul{f(\pred(\pred(1))) *  \pred(1)})\\ 
  \ul{f(\pred(1)) * 1} & \approx & \ul{\cond(\pred(1), \one,
    f(\pred(\pred(1))) * \pred(1))} * \ul{1}\\ 
  \ul{\pred(1)} & \approx & \pred(\ul{1})\\ 
  \ul{\one} & \approx & 1\\ 
\end{eqnarray*}
One can see that for each natural number $n$, the solution to this 
flat equation morphism assigns to $\ul{f(n)}$ the number $n!$.

\subsection{Interpreted Solutions in $\CPO$}
\label{sec:cpoint}

We shall show in this subsection that if we have $\A = \CPO$ as our base
category, then interpreted
solutions of   guarded RPSs $e$ in an Elgot algebra $(A, a, (\,\_\,)^*)$ are given as
least fixed points of a continuous operator on a function space. In this way
we recover denotational semantics from our categorical interpreted semantics
of recursive program schemes. 

\begin{exmp}
  \label{ex:runelgot}\index{factorial!classical semantics}
%As in Section~\ref{sec:opsemint}, 
We study the RPS of Equation~\refeq{eq:run}
as formalized in
Example~\ref{ex:natrps}(i).
 As we know, the intended interpreted solution is the factorial function on the 
  natural numbers $\Nat$. 
  
  This time we turn the natural numbers into an object of $\CPO$ so as to
  obtain a suitable Elgot algebra in which we can find an interpreted solution
  of~\refeq{eq:run}.  Let $\Natb$ be the flat cpo obtained from the discretely
  ordered $\Nat$ by adding a bottom element $\bot$\iffull\/, so $x \leq y$ iff
  $x = \bot$ or $x=y$\fi\/.  We equip $\Natb$ with the strict operations $
  \one_{\Natb}$, $ \pred_{\Natb}$, and $ *_{\Natb}$.  Being strict, they are
  hence continuous.  In addition, we use the continuous function
  \begin{eqnarray*}
  \cond_{\Natb}(n, x, y) & = & 
    \left\{ 
      \begin{array}{cl}
        \bot & \textrm{if $n=\bot$} \\
        x    & \textrm{if $n=0$} \\
        y    & \textrm{else}
      \end{array}
    \right.
  \end{eqnarray*}
  Indeed, this is what we saw in \refeq{eq-defcond} for the computation tree
  semantics, except we write $\bot$ for $\uparrow$.  Hence we have a continuous
  $\Sigma$-algebra with $\bot$.  Therefore $\Natb$ is an Elgot algebra for
  $H_\Sigma: \Set \to \Set$, see Example~\ref{ex:elgot}(ii).

  The standard interpreted solution $\isol e \Natb: H_\Phi \Natb \to \Natb$
  will certainly be \emph{some} function or other on $\Natb$. But how do we
  know that this function is the desired factorial function?  Usually one would
  simply regard the RPS~\refeq{eq:run} itself as a continuous function $R$ on
  $\CPO(\Natb, \Natb)$ acting as
  \begin{displaymath}
   f(\,\_\,) \mapsto \cond_{\Natb}(\,\_\,, 1, f(\pred_{\Natb}(\,\_\,) *_{\Natb}
   \,\_\,)\,, \,\_\,)\, ;
   \end{displaymath}
    Hence $R$ is the operator described in~\refeq{eq:R} in the
   introduction.  That means that we interpret all the operation symbols of
   $\Sigma$ in the algebra $\Natb$. The usual denotational semantics assigns to
   the formal equation of~\refeq{eq:run} with the interpretation in $\Natb$ the
   least fixed point of $R$.  Clearly this yields the desired factorial
   function.  And it is not difficult to work out that the least fixed point of
   $R$ coincides with the standard interpreted solution $\isol e \Natb$
   obtained from Theorem~\ref{thm:intsol}.  We shall do this shortly in greater
   generality.
\end{exmp}

In general, any recursive program scheme can be turned into a
continuous operator $R$ on the function space $\CPO(\Var A, A)$.
Theorem~\ref{thm:cpointsol} below shows that the
least fixed point of $R$ is the same as the interpreted solution obtained from
Theorem~\ref{thm:intsol}.

We assume throughout this subsection that $H$, $\Var$ and $H+\Var$ are locally
continuous (and, as always, iteratable) endofunctors of $\CPO$.  We also assume
that the free completely iterative monad $\T{H+\Var}$ is locally continuous.  
We consider a fixed guarded RPS $e: \Var \to \T{H + \Var}$, and an $H$-algebra
$(A, a)$ with a least element $\bot$. By Example~\ref{ex:elgot}(i), we know
that this carries the structure of an Elgot algebra $(A, a, (\,\_\,)^*)$,
where $(\,\_\,)^*$ assigns to every flat equation morphism a least solution.
As before, we will use the notation $\aext{a}: \T{H}A \to A$ for the induced
evaluation morphism. 

For any continuous map $t: \Var A \to A$, we have an Elgot algebra on $A$ for
the functor $H+V$; the structure is $[a, t]$, and again the solution operation
takes least solutions to flat equation morphisms. Due to
Corollary~\ref{corr:diags}, we have
\begin{equation}\label{eq:iso}
  \aext{[a, t]} \o \kappa^{H + \Var}_A = [a, t]\,.
\end{equation}

Before we prove the main result of this subsection we need to recall a result
from~\cite{abm} which states that the operation of taking least solutions in any
continuous algebra $[a,t]: (H+\Var)A \to A$ extends to non-flat equation
morphisms as in Section~\ref{section-completely-iterative-monads}. Moreover,
those least solutions are properly connected to the unique solutions in the
free completely iterative monad on $H+\Var$. 

Let us write $T = \T{H+\Var}$ and $\gamma = \aext{[a,t]}$ for short. Now
consider any (non-flat) guarded equation morphism $d: X \to T(X+A)$ (see
Definition~\ref{dfn:compiter}).  The least solution of $d$ in the algebra $A$
is the join $d^*: X \to A$ of the following $\omega$-chain $(d_i^*)$ in
$\CPO(X,A)$: $d_0^* = \op{const}_{\bot}$ and given $d_i^*: X \to A$, we define
$d_{i+1}^*$ by the commutativity of the following square
\begin{equation}\label{diag:sold}
  \vcenter{
    \xymatrix@C+1pc{
      X \ar[r]^-{d^*_{i+1}}
      \ar[d]_d
      &
      A
      \\
      T(X+A)
      \ar[r]_-{T[d_i^*, A]}
      &
      TA
      \ar[u]_\gamma
      }
    }
\end{equation}
Notice that we use here that $T$ is locally continuous in order to see that
$(d^*_i)$ is an $\omega$-chain. Furthermore, observe that $d^*$ is a solution
of $d$ in $A$ in the sense that the equation $d^* = \gamma \o T[d^*, A] \o d$
holds.  

\begin{thm}\label{thm:bloom} {\rm (see~\cite{abm})}
  The least solution of $d$ in $A$ is the morphism $\gamma \o \sol{d}$, where
  $\sol{d}: X \to TA$ is the unique solution of $d$ w.\,r.\,t.~the completely
  iterative monad $T$; in symbols:
  $$
  d^* \equiv
  \xymatrix@1{
    X \ar[r]^-{\sol{d}} & TA \ar[r]^-{\gamma} & A\,.
    }
  $$
\end{thm}

We are now prepared to prove the main result of the current subsection.

\begin{thm}\label{thm:cpointsol}
  The following function $R$ on $\CPO(\Var A, A)$
  \begin{equation}\label{eq:functional}
    f \mapsto 
    \xymatrix@1{
      \Var A \ar[r]^-{e_A} & \T{H+\Var} A \ar[r]^-{\aext{[a, f]}} & A
      }
  \end{equation} 
  is continuous. Its least fixed point is the standard interpreted solution
  $\isol e A: \Var A \to A$ of Theorem~\ref{thm:intsol}.
\end{thm}
\begin{proof}
  (i)~To see the continuity of $R$ is suffices to prove that the function
  \iftcs $\else\begin{displaymath}\fi
    \aext{(\,\_\,)}: \CPO(HA, A) \to \CPO(\T{H}A, A)%
    \iftcs$ \else\end{displaymath}\fi
  is continuous. Let
  us write $T$ for $\T{H}$.  Recall that for any continuous map $a: HA \to A$,
  the evaluation morphism $\aext{a}$ is the least solution of the flat equation morphism
  $\alpha_A: TA \to HTA + A$.   This means that
   $\aext{a}$ is the least fixed point of the continuous function 
  \begin{displaymath}
    F(a, -): \CPO(TA, A) \to \CPO(TA, A) \qquad f \mapsto [a, A] \o (Hf + A) \o
    \alpha_A\,. 
  \end{displaymath}
  Observe that $F$ is continuous in the first argument $a$, and so $F$ is a
  continuous function on the product $\CPO(HA, A) \times \CPO(TA, A)$. It
  follows from standard arguments that taking the least fixed point in the
  second argument yields a continuous  map $\CPO(HA, A) \to \CPO(TA, A)$. But
  this is precisely the desired one $\aext{(\,\_\,)}$.

  (ii)~We prove that $\isol e A$ is the least fixed point of $R$.
  Notice that the least fixed point of $R$ is the join $t$ of the following
  increasing chain in $\CPO(\Var A, A)$:
  \begin{displaymath}
  t_0 = \op{const}_\bot: \Var A \to A, \iftcs\quad\else\qquad\fi
  t_{i+1} \equiv 
    \xymatrix@1{
      \Var A \ar[r]^-{e_A} & \T{H+\Var} A \ar[r]^-{\aext{[a, t_i]}} & A}, 
    \iftcs\quad\else\qquad\fi \textrm{for $i \geq 0$}.
  \end{displaymath}
 
  Recall that the standard interpreted solution $\isol e A$ is defined by the
  equation $\isol e A = \beta \o \kappa^{H+\Var}_A \o \inr$ (see
  Definition~\ref{dfn:standard}), where $\beta = g^*$ is the least solution of
  the flat equation morphism $g$ from~\refeq{eq:g}, which is obtained from the
  component at $A$ of the $\H$-coalgebra $\ol e$, see
  Lemma~\ref{lemma-prelim-uninterpreted-rps} and Theorem~\ref{thm:intsol}.
  Observe that by Lemma~\ref{lem:intsoleq}, $\isol e A$ is a fixed point of $R$
  (see~\refeq{diag:intsol}). Thus, we have
  $%\begin{displaymath}
  t \less \isol e A %\,.\end{displaymath}
  $.
  To show the reverse inequality we will prove in~(iib) below that $\beta \less
  \aext{[a,t]}$ holds. Then we have 
  \begin{displaymath}
  \isol e A = \beta \o \kappa^{H+\Var}_A \o \inr 
  \less \aext{[a,t]} \o \kappa^{H+\Var}_A \o \inr 
  = t\,,
  \end{displaymath}
  where the last equality follows from~\refeq{eq:iso}.

  Let us write $T$ for $\T{H+\Var}$, $\gamma$ as a short notation for
  $\aext{[a,t]}$, and we also write $\lambda = [\kappa^{H+\Var} \o \inl, e]$.  

  (iia)~We will now prove that the following inequality
  \begin{equation}\label{eq:gammaall}
    \gamma \o \ext{e}_A \sqsubseteq \gamma \qquad
    \textrm{holds in $\CPO(TA, A)$,}
  \end{equation}
  where $\ext{e}: T \to T$ is the monad morphism from
  Lemma~\ref{lemma-prelim-uninterpreted-rps}. We use that the domain $TA$
  is a coproduct of $(H+\Var)TA$ and $A$, and we verify~\refeq{eq:gammaall}
  componentwise.  For the right-hand component we extend both sides of the
  inequality by the colimit injection $\eta^{H+\Var}_A: A\to TA$, and then we
  use the unit law of the monad morphism $\ext{e}: T \to T$:
  $$
  \gamma \o \ext{e}_A \o \eta^{H+\Var}_A = \gamma \o \eta^{H+V}_A\,.
  $$
  For the left-hand component of the desired inequality we extend by the
  coproduct injection $\tau^{H+\Var}_A: (H+\Var)TA \to TA$, and so we need to
  establish the inequality  
  \begin{equation}\label{eq:gamma}
    \gamma \o \ext{e}_A \o \tau^{H+\Var}_A \sqsubseteq \gamma \o
    \tau^{H+\Var}_A \qquad \textrm{in $\CPO((H+\Var)TA, A)$.}
  \end{equation}
  
  Recall from~\cite{aamv} (see the proof of Theorem~4.14) that the monad
  morphism $\ext{e}: T \to T$ can be obtained as follows. The component
  $\ext{e}_Y$ for any object $Y$ is $$[\sol{(\lambda*\alpha^{H+\Var})}_Y,
  \eta^{H+\Var}_Y]: TY = (H+\Var)TY + Y \to TY\,,$$
  where we consider the
  component at $Y$ of the natural transformation 
  $$
  \lambda * \alpha^{H+\Var}: (H+\Var) T \to T((H+\Var)T + \Id)
  $$ 
  to be a guarded equation morphism. Let us
  write $d = (\lambda * \alpha^{H+\Var})_A$ for short.  Note that $d: X \to
  T(X+A)$, with $X = (H+V)TA$.

  Then, the left-hand side of our desired inequality~\refeq{eq:gamma} is
  $\gamma \o \sol{d}$, and, by Theorem~\ref{thm:bloom}, we know that this is
  the least solution $d^*$ of the equation morphism $d$ in $A$. So in order to
  establish inequality~\refeq{eq:gamma} above we prove by induction on $i$ that
  $$
  d_i^* \sqsubseteq \gamma \o \tau^{H+\Var}_A \qquad 
  \textrm{holds in $\CPO((H+\Var) TA, A)$, for all $i \in \Nat$.} 
  $$
  The base case is clear. For the induction step we consider the diagram
  below (we write $X = (H+\Var) TA$ for short): 
  $$
  \xymatrix@+1pc{
    &
    X
    \ar[r]^-{\tau^{H+\Var}_A}
    \ar[d]_{\lambda_{TA}}
    \ar@/_1.5pc/[rrr]_-{d_{i+1}^*}
    \ar@{}[rrr]_*+{\begin{turn}{90}$\labelstyle\sqsubseteq$\end{turn}}
    \ar `l[ld] `[ddd]_{(H+\Var)\gamma} [ddd]
    \ar[rd]_d
    &
    TA
    \ar[rr]^-{\gamma}
    &&
    A
    \\
    &
    TTA
    \ar[r]_-{T\alpha^{H+\Var}_A}
    \ar `d[rd] [rd]_(.3){\id}
    &
    T(X + A)
    \ar[d]_{T[\tau^{H+\Var}_A, \eta^{H+\Var}_A]}
    \ar[r]^-{T[d_i^*, A]}
    \ar[rd]|{T[\gamma \o \tau^{H+\Var}_A, A]}
    & 
    TA
    \ar[ru]^{\gamma}
    \ar@{}[d]|{\begin{turn}{-90}$\labelstyle\sqsubseteq$\end{turn}}
    \\
    &
    &
    TTA
    \ar[r]^-{T\gamma}
    &
    TA 
    \ar[ruu]_{\gamma}
    &
    \\
    &
    (H+\Var)A
    \ar `r[rrru] [rrruuu]_{[a,t]}
    \ar[rru]_{\lambda_A}
    }
  $$
  We will verify that the upper inequality holds as indicated. In fact, the
  part below that inequality commutes by the definition of $d_{i+1}^*$ (see
  Diagram~\refeq{diag:sold}); the left-most part commutes by the naturality of
  $\lambda$; the upper left triangle commutes by the definition of $d$; the
  part below that triangle commutes since $[\tau, \eta]$ is the inverse of
  $\alpha$ (see Theorem~\ref{thm:adjoint}(iii)). For the inner triangle remove
  $T$ and consider the coproduct components separately: the left-hand component
  trivially commutes, and for the right-hand one use the unit law $\gamma \o
  \eta^{H+\Var}_A = \id$. The inner inequality holds by the induction
  hypothesis $d_i^* \sqsubseteq \gamma \o \tau^{H+\Var}_A$ and by the fact that
  copairing, application of $T$ and composition with $\gamma$ on the left are
  all continuous whence monotone operations.  (Note that we use the assumption
  that $T$ be locally continuous here.)  Recall that $\gamma = \aext{[a,t]}: TA
  \to A$ is a homomorphism of Elgot algebras, see Theorem~\ref{thm:adjoint}.
  Hence, $\gamma$ is an homomorphism of algebras for $H+\Var$ by
  Proposition~\ref{prop:sol-pres-hom}. Thus, the outer shape commutes.  Finally
  we verify the lower right-hand part componentwise: that the left-hand component
  with domain $HA$ commutes is seen by the following computation
  $$
  \begin{array}{rcl@{\hspace{1cm}}p{8cm}}
    \gamma \o \lambda_A \o \inl & = & \gamma \o \kappa^{H+\Var}_A \o \inl &
    (by the definition of $\lambda$) \\
    & = & \aext{[a,t]} \o \kappa^{H+\Var}_A \o \inl &
    (by the definition of $\gamma$) \\
    & = & [a,t] \o \inl & 
    (by Equation~\refeq{eq:iso}) \\
    & =& a\,,
  \end{array}
  $$
  and the right-hand component with domain $\Var A$ is the equation $\gamma \o
  e_A = t$ which holds since $t$ is a fixed point of $R$. This completes the
  proof of the inequality~\refeq{eq:gammaall}. 

  (iib)~We are ready to establish the inequality $\beta \less
  \aext{[a,t]}$. To this end observe that by Example~\ref{ex:elgot}(i), the
  least solution $\beta$ of $g$ is the join of the chain
  \begin{displaymath}
    \beta_0 = \op{const}_\bot, \qquad 
    \beta_{i+1} = [a, A] \o H(\beta_i + A) \o g, \quad \textrm{for $i \geq 0$}.
  \end{displaymath}
  So in order to establish the desired inequality we prove by induction on $i$ the inequalities
  \begin{equation}\label{eq:induct}
    \beta_i \less \aext{[a, t]}\,,\quad i \in \Nat\,.
  \end{equation}
  The base case is clear. For the induction step we consider the diagram below:
  \begin{displaymath}
  \xymatrix@C+1pc{
    TA \ar[dd]_{\beta_{i+1}} \ar[r]^-{g = \ol{e}_A} 
    &
    HTA + A \ar[r]^-{\inl_{TA} + A} \ar@/_1pc/[dd]_{H\beta_i + A} 
    \ar@/^1pc/[dd]^{H\gamma + A} \ar@{}[dd]|{\less} & 
    (H+\Var)TA + A \ar[r]^-{[\tau, \eta]} \ar[d]_{(H+\Var)\gamma +A} & 
    TA \ar[dd]^\gamma 
    \ar@{<-} `u[l] `[lll]_{\ext{e}_A} [lll] 
    \\
    & & (H+\Var)A + A \ar[rd]_(.4)*+{\labelstyle [a,t,A]} \\
    A \ar@{<-} `d[r] `[rrr]_{\id} [rrr] 
    & 
    HA + A \ar[l]^-{[a, A]} \ar[ru]_-*+{\labelstyle \inl_A + A} \ar[rr]_-{[a,A]} & & 
    A 
    }
  \end{displaymath}
  Its upper part commutes due to Lemma~\ref{lem:t2}, the left-hand part is the
  definition of $\beta_{i+1}$, and the inequality follows from the induction
  hypothesis. For the right-hand part use that $\gamma$ is an algebra
  homomorphism. Finally, the remaining parts of the above diagram clearly
  commute. Thus we obtain the inequality $\beta_{i+1} \less \gamma \o
  \ext{e}_A$, and the desired inequality follows from this by~\refeq{eq:gammaall}.
  This completes the proof. 
\end{proof}

\begin{rem}
  The result of Theorem~\ref{thm:cpointsol} furnishes a concrete formula
  \begin{displaymath}
    \isol e A = \bigsqcup\limits_{n < \omega} \dsol{e}_n
  \end{displaymath}
  for the interpreted solution of the guarded RPS $e$ in the continuous
  algebra $A$. In fact, the least fixed point of $R$ is the join of the
  ascending chain
  \begin{displaymath}
    \bot \sqsubseteq R(\bot) \sqsubseteq R^2(\bot) \sqsubseteq \cdots
  \end{displaymath}
  where $\bot = \const \bot$ is the least element of $\CPO(\Var\! A, A)$. Thus,
  with $\dsol{e}_0 = \const \bot$ and
  \begin{displaymath}
    \dsol{e}_{n+1} \equiv 
    \xymatrix@1@+1pc{
      \Var\! A
      \ar[r]^-{e_A}
      &
      \T{H+\Var} A 
      \ar[r]^-{\aext{[a, \dsol{e}_n]}}
      &
      A
      }
  \end{displaymath}
  we obtain the above formula for $\isol e A$. 
\end{rem}

\begin{rem}
  \label{rem:set}
  Suppose that $H$,  $\Var$ and $H + \Var$ are iteratable endofunctors of $\Set$, which have
  locally continuous liftings $H'$ and $\Var'$ to $\CPO$. Then from
  Example~\ref{ex:lift_cpo} we have a commutative square
  \begin{displaymath}
  \xymatrix@C+2pc{
    \CPO
    \ar[d]_U
    \ar[r]^-{\T{H'+\Var'}}
    &
    \CPO
    \ar[d]^U 
    \\
    \Set
    \ar[r]_-{\T{H+\Var}}
    &
    \Set
    }
  \end{displaymath}
  
  In addition, we now show that $\T{H'+\Var'}$ is locally continuous. In fact,
  for each CPO $X$, $\T{H'+\Var'} X$ is obtained as a limit of an ordinal
  indexed op-chain of the cpos $T_\beta$ (see Example~\ref{ex:lift_cpo}).
  Hence, one readily proves that for any continuous map $f: X \to Y$ the
  continuous map $\T{H+\Var} f$ is obtained by using the universal property of
  that limit. It follows from standard arguments that the assignment $f \mapsto
  \T{H+\Var} f$ is a continous map on $\CPO(X,Y)$ (see~\cite{abm} for details).
  
  Assume that the guarded RPS $e: \Var \to \T{H+\Var}$ has a lifting $e':\Var'
  \to \T{H'+\Var'}$.  This is a natural transformation $e'$ such that $U e' = e
  U$.  Now consider any $\CPO$-enrichable $H$-algebra $(A, a)$ as an Elgot
  algebra, see Example~\ref{ex:elgot}(ii).  Then we can apply
  Theorem~\ref{thm:cpointsol} to obtain the standard interpreted solution
  $\isol e A$ of $e$ in the algebra $A$ as a least fixed point of the above
  function $R$ of~\refeq{eq:functional}.
\end{rem}

\begin{exmp}\label{ex:intsolcpo}\iftcs\else\mbox{ } \hfill\fi
  \iffull
  \begin{myenumerate}
  \myitem 
  \else
  (i) \fi Suppose we have signatures $\Sigma$ and $\Phi$. Then the signature
    functors $H_\Sigma$ and $H_\Phi$ have locally continuous liftings $H_\Sigma'$ and
    $H_\Phi'$. Since the lifting of $H_\Sigma + H_\Phi$ is a lifting of
    $H_{\Sigma + \Phi}$ we know that $\T{H_\Sigma' + H_\Phi'}$ assigns to any
    cpo $X$ the algebra $T_{\Sigma + \Phi} X$ with the cpo structure induced
    by $X$, see~Example~\ref{ex:lift_cpo}. More precisely, to compare a tree $t$ to a tree
    $s$ replace all leaves labelled by a variable from $X$ by a leaf labelled
    by some extra symbol $\star$ to obtain relabelled trees $t'$ and $s'$. Then
    $t \less s$ holds in $\Tsf X$ iff $t'$ and $s'$ are isomorphic as labelled trees,
    and for any leaf of $t$ labelled by a variable $x$ the corresponding leaf
    in $s$ is labelled by a variable $y$ with $x \less y$ in $X$.
    
    Now consider any system as in~\refeq{eq:crps} which is in Greibach normal
    form, and form the associated guarded RPS $e: H_\Phi \to T_{\Sigma +
      \Phi}$. Then $e$ has a lifting $e': H_\Phi' \to \T{H_\Sigma' + H_\Phi'}$.
    In fact, for any cpo $X$ the component $e'_X = e_X: H_\Phi X \to T_{\Sigma
      + \Phi} X$ is a continuous map since the order in $H_\Phi X$ is given
    similarly as for $T_{\Sigma + \Phi} X$ on the level of variables only: $(f,
    \vec x) \sqsubseteq (g, \vec y)$ holds for elements of $H_\Phi X$ if $f = g
    \in \Phi_n$ and $x_i \sqsubseteq y_i$, $i = 1, \ldots, n$, hold in $X$.

    \sloppypar
    Let $(A, a)$ be a $\CPO$-enrichable $H_\Sigma$-algebra; i.\,e., a
    continuous $\Sigma$-algebra with a least element $\bot$. We wish to
    consider the continuous function $R$ on $\CPO(H_\Phi A, A)$ which assigns
    to any continuous algebra structure $\varphi: H_\Phi A \to A$ the algebra
    structure $R(\varphi) = \aext{[a, \varphi]} \o e_A'$.  The structure
    $R(\varphi): H_\Phi A \to A$ gives to each $n$-ary operation symbol $f$ of
    $\Phi$ the operation $t^f_A: A^n \to A$ which is obtained as follows: take
    the term $t^f$ provided by the right-hand side of $f$ in our given RPS,
    then interpret all operation symbols of $\Sigma$ in $t^f$ according to 
    the given algebraic structure $a$ and all operation symbols of $\Phi$
    according to $\varphi$; the action of $t^f_A$ is evaluation of that
    interpreted term.
  
    Theorem~\ref{thm:cpointsol} states that the standard interpreted solution
    $\isol e A$ of $e$ in the algebra $A$ can be obtained by taking the least
    fixed point of $R$; in other words the standard interpreted solution $\isol
    e A$ gives the usual denotational semantics.
  
  \iffull
  \myitem 
  \else 
  \smallskip\noindent (ii) \fi \index{factorial!standard interpreted solution}%
  Apply the previous example to the RPS
  of Example~\ref{ex:natrps}(i).  Then Theorem~\ref{thm:cpointsol} states that
  the standard interpreted solution of the RPS~\refeq{eq:run} in the Elgot
  algebra  $\Natb$ is obtained as the least fixed point of the function $R$ of
  Example~\ref{ex:runelgot}. That is, the standard
  interpreted solution gives the desired factorial function. 
  
  \myitem Recall the guarded RPS $e: \Var \to \T{H + \Var}$ from
  Example~\ref{ex:natrps}(ii) whose uninterpreted solution we have described in
  Example~\ref{ex:rps-solutions}(ii). Consider again the algebra $\Natb$
  together with the following two operations:
  \begin{equation}
    F_{\Natb}(x,y,z) = 
    \left\{
      \begin{array}{l@{\quad}l}
        \bot & \textrm{if $x = \bot$ or $y = \bot$ or $z = \bot$,} \\
        x & \textrm{if $x = y \neq \bot$, and $z \neq \bot$,} \\ 
        z & \textrm{else,}
      \end{array}
    \right.
    \qquad
    G_{\Natb}(x) = 
    \left\{
      \begin{array}{l@{\quad}l}
        \lfloor\frac{x}{2}\rfloor & \textrm{if $x \in \Nat$,} \\
        \bot & x = \bot.
      \end{array}
    \right.
  \end{equation}
  Since the first operation obviously satisfies $F_{\Natb}(x,y,z) =
  F_{\Natb}(y,x,z)$ we have defined an $H$-algebra. It is not difficult to
  check that the set functor $H$ has a locally continuous lifting $H'$ to $\CPO$
  and that $\Natb$ is a continuous $H'$-algebra. In fact, the existence of the
  lifting $H'$ follows from the fact that the unordered pair functor $V: \Set
  \to \Set$ can be lifted to $\CPO$; the lifting assigns to a cpo $(X,\leq)$ the
  set of unordered pairs with the following order: $\{\,x, y\,\} \sqsubseteq
  \{\,x',y'\,\}$ iff either $x \leq x'$ and $y \leq y'$, or $x\leq y'$ and $y\leq
  x'$. Thus, we have defined an Elgot
  algebra for $H: \Set \to \Set$, see Example~\ref{ex:elgot}(ii). The standard
  interpreted solution $\isol{e}{\Natb}: \Var \Natb \to \Natb$ is given by one
  commutative binary operation $\phi_{\Natb}$ on $\Natb$. We leave it to the
  reader to verify that for natural numbers $n$ and $m$, $\phi_{\Natb}(n,m)$ is
  the natural number represented by the greatest common prefix in the binary
  representation of $n$ and $m$, e.\,g., $\phi_{\Natb}(12,13) = 6$.  Notice
  that we do not have to prove separately that $\phi_{\Natb}$ is commutative.
  In Example~\ref{ex:natrps}(ii) we have encoded that extra property directly
  into the RPS $e$ so that any solution must be commutative. 

  \iffull
\myitem \else \smallskip\noindent (iii) \fi Least fixed points are RPS
  solutions. Let $A$ be a poset with joins of all subsets which are at most
  countable, and let $f: A \to A$ be a function preserving joins of ascending
  chains. Take $f$ and binary joins to obtain an algebra structure on $A$ for
  the signature functor $H_\Sigma X = X + X \times X$ expressing a binary
  operation symbol $F$ and a unary one $G$. Obviously, this functor has a
  lifting $H': \CPO \to \CPO$ and $A$ is a $\CPO$-enrichable algebra.  So
  $A$ is an Elgot algebra.  Turn the formal equations from~\refeq{eq:algex} into a
  recursive program scheme $e: H_\Phi \to \T{H_\Sigma + H_\Phi}$ as
  demonstrated in Section~\ref{section-rpsfirst}.
  The RPS $e$ has a lifting $e': \Var' \to \T{H' + \Var'}$, where $\Var'$
  denotes the lifting of $H_\Phi$. The standard interpreted solution $\isol e A: \Var'
  A \to A$ gives two continuous functions $\phi_A$ and $\psi_A$ on
  $A$. Clearly, we have $\phi_A(a) = \bigvee_{n \in \Nat} f^n(a)$, and in
  particular $\phi_A(\bot)$ is the least fixed point of $f$.  \iffull
  \end{myenumerate}
  \fi
\end{exmp}

\subsection{Interpreted Solutions in $\CMS$}
\label{sec:cmsint}

Recall the category $\CMS$ of complete metric spaces from
Example~\ref{ex:cats_iii}, and also the facts that  the
contracting endofunctors are iteratable and closed under finite coproducts.
Let $H, \Var: \CMS \to \CMS$ be such contracting
endofunctors. We shall show in this subsection that for any guarded RPS $e:
\Var \to \T{H +\Var}$,
 we can find a unique interpreted solution in any
non-empty $H$-algebra $A$.  To this end, assume that we have such a
guarded RPS
$e$, and let $(A, a)$ be a non-empty $H$-algebra. Then
$A$ is a cia, and, in particular it carries the structure of an Elgot
algebra. Notice that for any non-expanding map $t: \Var A \to A$ we obtain an algebra
structure $[a, t]: (H+\Var) A \to A$. This turns $A$ into a cia for $H+\Var$,
and thus we have the evaluation morphism  
\begin{displaymath}\aext{[a, t]}: \T{H+\Var} A \to A\,.\end{displaymath}
As in $\CPO$, the RPS $e$ induces a
function $R$ on $\CMS(\Var A, A)$, see~\refeq{eq:functional}.  The standard
procedure for obtaining an interpreted solution would be to 
recall that $\CMS(\Var A, A)$ is a complete metric space
and then to prove that $R$ is
a contracting map on it.
We then invoke Banach's Fixed Point theorem to obtain a
unique fixed point of $R$.  Here we simply apply Theorem~\ref{thm:intsol}.
Notice, however, that we cannot completely avoid Banach's Fixed Point theorem:
it is used in the proof that final coalgebras exist for contracting functors,
see~\cite{are}.

\begin{thm}\label{thm:cmsintsol}
  The unique interpreted solution $\isol e A: \Var A \to A$ of $e$ in $A$ as obtained
  in Theorem~\ref{thm:intsol} is the unique fixed point of the function $R$ on
  $\CMS(\Var A, A)$ defined by~\refeq{eq:functional}. 
\end{thm}
\addproof{Corollary~\ref{thm:cmsintsol}}{%
\begin{proof}
  \comment{%
    In fact, being a fixed point of $R$ is equivalent to being an interpreted
    solution of $e$ in the cia $A$, whose unique existence we have by
    Theorem~\ref{thm:intsol}. }%
  We will prove that fixed points of $R$ are in one-to-one correspondence to
  interpreted solutions. In fact, any interpreted solution $\isol e A: \Var A
  \to A$ is a fixed point of $R$ by Lemma~\ref{lem:intsoleq}
  (see~\refeq{diag:intsol}). 
  Conversely, we will now prove that any fixed point
  $t$ of $R$ is an interpreted solution of $e$ in $A$. Let $t$ be a fixed point
  of $R$ and write $\beta = \aext{[a,t]}$. We are finished if we prove the
  commutativity of~\refeq{eq:betaintsol}. We write $T = \T{H+\Var}$ and
  $\lambda = [\kappa^{H+\Var}_A \o \inl, e_A]$ for short once again. Recall
  that $TA$ is the free cia on $A$ and that $\beta: TA \to A$ is the unique
  homomorphism of Elgot algebras from $TA$ to $A$ extending $\id_A$,
  i.\,e.~such that the equation $\beta \o \eta^{H+\Var}_A = \id_A$ holds.  From
  Example~\ref{ex:cia}(iv) we see that $(A, [a,t])$ is a cia for $H+\Var$.
  Thus, by Proposition~\ref{prop:sol-pres-hom}, $\beta$ is the unique
  homomorphism of algebras for $H+\Var$ extending $\id_A$.  So in order to
  verify the desired equation $\beta \o \ext{e}_A = \beta$ we verify below that
  the following diagram commutes:
  $$
    \xymatrix{
    A 
    \ar[r]^-{\eta^{H+\Var}_A}
    \ar[rd]|*+{\labelstyle \eta^{H+\Var}_A}
    \ar `d[rdd]_{\id} [rdd]
    &
    TA 
    \ar[d]^{\ext{e}_A}
    \ar@{<-} `u[r] `[rr]^{\tau^{H+\Var}_A} [rr]
    &
    TTA
    \ar[l]_{\mu^{H+\Var}_A}
    \ar[d]^{(\ext{e}*\ext{e})_A}
    &
    (H+\Var)TA
    \ar[l]_{\kappa^{H+\Var}_{TA}}
    \ar[d]^{(H+\Var)\ext{e}_A}
    \\
    &
    TA
    \ar[d]^\beta
    &
    TTA
    \ar[d]^{T\beta}
    \ar[l]^{\mu^{H+\Var}_A}
    &
    (H+\Var)TA
    \ar[d]^{(H+\Var)\beta}
    \ar[l]^{\lambda_{TA}}
    \\
    &
    A
    \ar@{<-} `d[r] `[rr]_{[a,t]} [rr]
    &
    TA
    \ar[l]^{\beta}
    &
    (H+\Var)A
    \ar[l]^{\lambda_A}
    }
  $$
  In fact, the upper part commutes by Corrollary~\ref{corr:diags}; the upper triangle
  and the upper left-hand square commute since $\ext{e}$ is a monad morphism;
  the upper right-hand square commutes by the double interchange law and the
  definition of $\ext{e}$ (see Lemma~\ref{lemma-prelim-uninterpreted-rps}); the
  lower triangle commutes and the lower left-hand square both commute since the
  evaluation morphism $\beta$ is an Eilenberg-Moore algebra for the monad $T$;
  the lower right-hand square commutes by the naturality of $\lambda$. Finally,
  we verify the commutativity of the lowest part componentwise: the left-hand
  component commutes by Equation~\refeq{eq:intsol(i)}, and the right-hand part
  $\beta \o e_A = t$ holds since $t$ is a fixed point of $R$.
\end{proof}
}

\begin{rem}\label{rem:treemetric}
  Let $H_\Sigma$ be a signature functor on $\Set$ and denote by $H'$ a lifting
  to $\CMS$ as described in Example~\ref{ex:cia}(vii). For a complete metric
  space $Y$ the final coalgebra $\T{H'}Y$ for $H'(\,\_\,) + Y$ is the set
  $T_\Sigma Y$ of all $\Sigma$-trees over $Y$ equipped with a suitable complete
  metric. This metric can be described as follows. Recall from~\cite{are} that
  $\T{H'}Y$ is obtained as $T_\omega$ after $\omega$ steps of the final
  coalgebra chain for $H'(\,\_\,) + Y$ (see Construction~\ref{constr:coalg}).
  That means the metric on $T_\Sigma Y$ is the smallest metric such that all
  projections $t_{\omega, i}: T_\Sigma Y = T_\omega \to T_i$ are non-expanding.
  We illustrate this with an example adapted from~\cite{are}.  Let $H_\Sigma X
  = X \times X$ be the functor expressing one binary operation symbol $*$. Then
  we can represent $T_0 = 1$ by a single node tree labelled with $\bot$ and
  $T_{i+1} = T_i \times T_i + Y$ by trees which are either single node trees
  labelled in $Y$, or which are composed by joining two trees from $T_i$ with a
  root labelled by $*$:
  \begin{displaymath}
  \begin{array}{rl}
    T_0: & \bot \\
    T_1: & y ,
    \vcenter{
    \xy
    \POS    (000,000)   *+{*}     = "r"
       ,    ( -5, -10)   *+{\bot} = "b1"
       ,    (  5, -10)   *+{\bot} = "b2"
    %%%
    \ar@{-} "r";"b1"
    \ar@{-} "r";"b2"
    \endxy
    }
    \\
    T_2: & y, 
    \vcenter{
      \xy
      \POS    (000,000)   *+{*} = "r"
         ,    ( -5,-10)   *+{y}  = "b1"
         ,    (  5,-10)   *+{y'} = "b2"
      %%%
      \ar@{-} "r";"b1"
      \ar@{-} "r";"b2"
      \endxy
      },\quad
    \vcenter{
      \xy
      \POS    (000,000)   *+{*}   = "r"
         ,    ( -5,-10)   *+{y}   = "y"
         ,    (  5,-10)   *+{*}   = "r2"
         ,    (000,-20)   *+{\bot}= "b1"
         ,    ( 10,-20)   *+{\bot}= "b2"
      %%%
      \ar@{-} "r";"y"
      \ar@{-} "r";"r2"
      \ar@{-} "r2";"b1"
      \ar@{-} "r2";"b2"
      \endxy
      },\quad
    \vcenter{
      \xy
      \POS    (000,000)   *+{*} = "r"
         ,    ( -5,-10)   *+{*} = "r2"
         ,    (  5,-10)   *+{y} = "y"
         ,    (-10,-20)   *+{\bot}="b1"
         ,    (000,-20)   *+{\bot}="b2"
      %%%
      \ar@{-} "r";"y"
      \ar@{-} "r";"r2"
      \ar@{-} "r2";"b1"
      \ar@{-} "r2";"b2"
      \endxy
      },\quad
    \vcenter{
      \xy
      \POS    (000,000)   *+{*} = "r"
         ,    ( -5,-10)   *+{*} = "r2"
         ,    (  5,-10)   *+{*} = "r3"
         ,    ( -8,-20)   *+{\bot}="b1"
         ,    ( -2,-20)   *+{\bot}="b2"
         ,    (  2,-20)   *+{\bot}="b3"
         ,    (  8,-20)   *+{\bot}="b4"
      %%%
      \ar@{-} "r";"r3"
      \ar@{-} "r";"r2"
      \ar@{-} "r2";"b1"
      \ar@{-} "r2";"b2"
      \ar@{-} "r3";"b3"
      \ar@{-} "r3";"b4"
      \endxy
      }
    \\
    \vdots
  \end{array}
  \end{displaymath}
  The distance on $T_1$ is that of $Y$ for single node trees and 1 otherwise. The
  distance on $T_2$ is again that of $Y$ between single node trees, and 1
  between single node trees and all other trees. Furthermore, the distance
  between trees of different shapes is $\frac{1}{2}$, and finally,
  $d_{T_2}(y*y', z*z') = \frac{1}{2}\max\{\,d_Y(y,z), d_Y(y',z')\,\}$ as
  well as $d_{T_2}(y*t, y'*t) = d_{T_2}(t*y,t*y') =
  \frac{1}{2}d_Y(y,y')$, where $t = \bot * \bot$, etc. In general, the
  distance on $T_{i+1}$ is that of $Y$ between single node trees, it is $1$
  between single node trees and trees of height at least $1$, and 
  otherwise we have
  $d_{T_{i+1}}(s*t, s'*t') = \frac{1}{2}\max\{\,d_{T_i} (s,s'),
  d_{T_i}(t,t')\,\}$.   For the metric on  $T_\Sigma Y$,
  we have 
  \begin{displaymath} 
    d_{T_{\Sigma}Y}(s_1,s_2) 
    = 
    \sup_{i < \omega} d_{T_i}(  t_{\omega, i}(s_1), t_{\omega, i} (s_2)).
  \end{displaymath}
  This is the smallest metric for which the projections are non-expanding.
  (One may also verify directly that this definition gives a complete 
  metric space structure and that $H'(\,\_\,) + Y$
  preserves the limit, so that we indeed have a final coalgebra.)
   Finally notice that the metric of
  $T_\Sigma Y$ depends on the choice of the lifting $H'$. For example, if we
  lift the functor $H_\Sigma$ as $H'(X,d) = (X^2, \frac{1}{3}d_{\max})$, the
  factor $\frac{1}{2}$ 
would have to be replaced by $\frac{1}{3}$ systematically.

  \comment{
  with the following metric: For distinct trees $t_1$ and $t_2$
  that differ only by leaves labelled in $Y$ we have 
  \begin{displaymath}d(t_1, t_2) = \sup 2^{-n} d(y,y')\,,\end{displaymath}
  where the $\sup$ ranges over all corresponding pairs $y, y'$ of leaves of
  $t_1$ and $t_2$, respectively.  So $y$ and $y'$ are reached by the same
  path from 
  the roots of $t_1$ and $t_2$, and where $n$ is the depth of $y$ and
  $y'$. Otherwise let $n$ be the smallest number such that $t_1$ and $t_2$
  differ at depth $m$ by a node labelled in $\Sigma$. Then 
  \begin{displaymath}d(t_1, t_2)  = \max\{\,2^{-m}, 2^{-n} d(y,y')\,\}\,,\end{displaymath}
  where the second component of the $\max$ ranges over all corresponding pairs
  of leaves of $t_1$ and $t_2$ labelled in $Y$ with depths $n < m$. 
  Example: Let $\Sigma$ be the signature with a binary operation symbol $F$
  and a unary one $G$. Then for distinct elements $y, y', z \in Y$ we have
  \begin{displaymath}
  d(F(y, Gz), F(y', GGz)) = 
  \max\left\{\,\frac{1}{2}d(y,y'), \frac{1}{2}\,\right\} =
  \frac{1}{2}\,.
  \end{displaymath}
  }
\end{rem}

\begin{exmp} \iftcs\else\mbox{ }\fi%
  \label{examples-rps-interpreted}
\begin{myenumerate}
  \myitem Consider the endofunctor $H': \CMS \to \CMS$ obtained by lifting the
  signature functor $H_\Sigma X = X \times X + X$ expressing a binary operation
  $F$ and a unary one $G$ as described in Example~\ref{ex:lift_cms}.
   We use a contraction factor $\eps = \frac{1}{2}$.  The
  Euclidean interval $I = [0,1]$ together with the operations $F(x,y) =
  \frac{x+y}{4}$ and $G(x) = \frac{\sin(x)}{2}$ is an $H'$-algebra, whence a
  cia. Use only the first equation in~\refeq{eq:algex} to obtain a guarded RPS
  $e: \Id \to \T{H_\Sigma + \Id}$ where $\Id$ expresses the unary operation
  symbol $\phi$. Let $\Var'$ be contracting lifting of $\Id$, again with a contraction
  factor of $\eps = \frac{1}{2}$. Then $e$ gives rise to a guarded RPS $e':
  \Var' \to \T{H'+\Var'}$ in $\CMS$. The unique interpreted solution of $e'$ in
  $I$ consists of a function $\phi_I: I \to I$ satisfying $\phi_I(x) =
  \frac{1}{4}(x + \phi_I(\frac{1}{2}\sin x))$, that is, $\phi_I$ is the unique
  function $f$ satisfying~\refeq{eq-sin}.
Please note that at long last the theory of recursive program schemes
has provided new results: it is not obvious that there is \emph{any}
$\phi_I$ satisfying this equation.
    
  \myitem Self-similar sets are solutions of interpreted program schemes.
    Recall from Example~\ref{ex:cia}(v) that for any complete metric space $(X,
    d)$, we obtain the complete metric space $(C(X), h)$ of all non-empty
    compact subspaces of $X$ with the Hausdorff metric. Furthermore,
    contractive mappings of $X$ yield structures of cias on $C(X)$.  Now
    consider the functor $H'$ on $\CMS$ with $H'(X,d) = (X^3,
    \frac{1}{3}d_{\max})$, where $d_{\max}$ is the maximum metric. It is a
    lifting of the signature functor $H_\Sigma$ on $\Set$ expressing one
    ternary operation $\alpha$.  Let $A = [0,1] \times [0,1]$, be equipped with
    the usual Euclidean metric. Consider the contracting maps $f(x,y) =
    (\frac{1}{3}x,\frac{1}{3}y)$, $g(x,y) = (\frac{1}{3}x + \frac{1}{3},
    \frac{1}{3}y)$, and $h(x,y) = (\frac{1}{3}x + \frac{2}{3}, \frac{1}{3}y)$
    of $A$. Then it follows that $\alpha_A: C(A)^3 \to C(A)$ with
    $\alpha(D,E,F) = f[D] \cup g[E] \cup h[F]$ is a $\frac{1}{3}$-contracting
    map, whence a structure of a cia for $H'$. The formal equation
    \begin{displaymath}
      \phi(x) \approx \alpha(\phi(x), x, \phi(x))
    \end{displaymath}
    gives rise to a guarded RPS $e: Id \to \T{H_\Sigma + \Id}$, where the
    identity functor expresses the operation $\phi$. If we take the lifting of
    $\Id$ to $\CMS$ which is given by $\Var' (X,d) = (X, \frac{1}{3}d)$, then
    $e$ gives rise to a natural transformation $e': \Var' \to \T{H'+\Var'}$.
    Its interpreted solution in the cia $C(A)$ is a
    $\frac{1}{3}$-contracting map $\phi_A: C(A) \to C(A)$ which maps a
    non-empty compact subspace $U$ of $A$ to a space of the following form:
    $\phi_A(U)$ has three parts, the middle one is a copy of $U$ scaled by
    $\frac{1}{3}$, and the left-hand and right-hand one look like copies of the
    whole space $\phi_A(U)$ scaled by $\frac{1}{3}$. For example, we have the
    assignment
    \begin{displaymath}
      \vcenter{
        \hbox{
          \includegraphics[width=2.1cm]{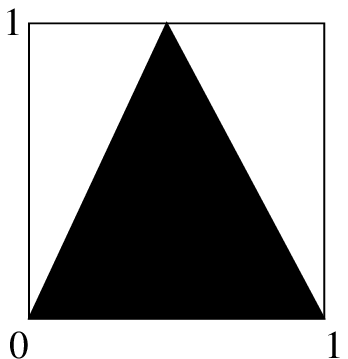}
          }
        }
      \qquad
      \stackrel{\phi_A}{\longmapsto}
      \qquad
      \vcenter{
        \hbox{
          \includegraphics[width=6cm]{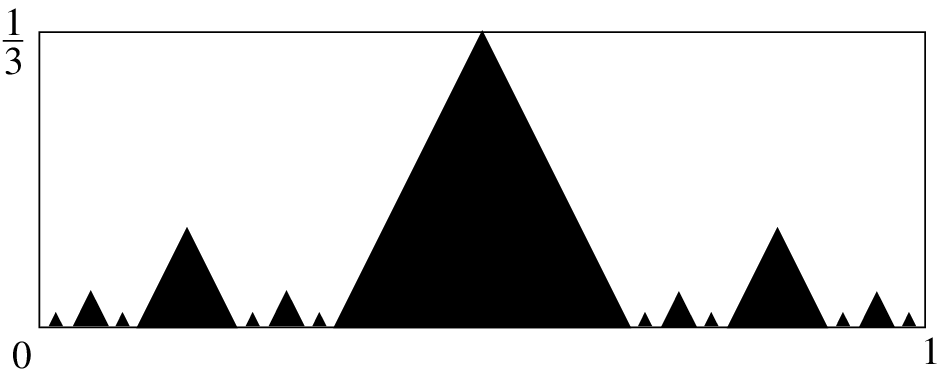}
          }
        }
    \end{displaymath}

    \index{Cantor set}%
  \myitem Coming back to Example~\ref{ex:cia}(vi), let us consider $(C(I),
    \alpha_I)$, where $I = [0,1]$ is the Euclidean interval, $C(I)$ is the set
    of all non-empty closed subsets of $I$, and $\alpha_I$ is
    the structure of a cia arising from $f(x) = \frac{1}{3} x$ and $g(x) =
    \frac{1}{3}x + \frac{2}{3}$ as described is Example~\ref{ex:cia}(v). 
    The formal equation
    \begin{displaymath}
    \phi(x) \approx \alpha(\phi(x) , x)
    \end{displaymath}
    gives similarly as in~(i) above a guarded RPS $e: \Id \to \T{H_\Sigma +
      \Id}$, where $H_\Sigma X = X \times X$ now expresses the binary operation
    $\alpha$. Again, we have liftings $V'(X, d) = (X, \frac{1}{3}d)$ and $H'(X
    ,d) = (X^2, \frac{1}{3} d_\mathrm{max})$ of $\Id$ and $H_\Sigma$,
    respectively. So the RPS $e$ lifts to the guarded RPS $e': V' \to \T{H' +
      V'}$ in $\CMS$. Its unique interpreted solution is given by the
    $\frac{1}{3}$-contracting map $\phi_I: C(I) \to C(I)$ satisfying $\phi_I(t)
    = \alpha_I(\phi_I(t), t) = f[\phi_I(t)] \cup g[t]$ for every non-empty
    closed subset $t$ of the interval $I$.
  \end{myenumerate}
\end{exmp}

\iffull
%\section{Applications}

%\begin{REMARK}
%  To be done.  Here we want to mention work on equivalence of RPSs,
%fold/unfold transformations, etc.  Also Mersch's logical calculus for
%RPSs, since it can be interpreted in an Elgot algebra and has sound
%rules of equivalence.
%\end{REMARK}
\fi

\iffull
\section{Conclusions and Future Work}
We have presented a general and conceptually clear way of treating the
uninterpreted and the interpreted semantics of recursive program schemes in a
category theoretic setting. For this we have used recent results on complete
Elgot algebras and results from the theory of coalgebras. We have shown that
our theory readily specializes to the classical setting yielding denotational
semantics using complete partial orders or complete metric spaces. We have
presented new applications of recursive program scheme solutions including
fractal self-similarity and also applications which cannot be handled by
the classical methods; defining operations satisfying equations like
commutativity. Another new application, recursively defined functions on
non-wellfounded sets, will be treated in a future paper. 

Now one must go forward in reinventing algebraic semantics with category
theoretic methods.  We strongly suspect that there is much to be said about the
relation of our work to operational semantics.  We have not investigated
higher-order recursive program schemes using our tools, and it would be good to
know whether our approach applies in that area as well.  The paper~\cite{mu}
addresses variable binding and infinite terms coalgebraically, and this may
well be relevant.  Back to the classical theory, one of the main goals of the
original theory is to serve as a foundation for program equivalence.  It is not
difficult to prove the soundness of fold/unfold transformations in an algebraic
way using our semantics; this was done in~\cite{moss_sec} for uninterpreted
schemes.  We study the equational properties of our very general formulation of
recursion in~\cite{properties}.  One would like more results of this type.  The
equivalence of interpreted schemes in the natural numbers is undecidable, and
so one naturally wants to study the equivalence of interpreted schemes in {\em
  classes of interpretations}.  The classical theory proposes classes of
interpretations, many of which are defined on ordered algebras,
see~\cite{guessarian}.  It would be good to revisit this part of the classical
theory to see whether Elgot algebras suggest tractable classes of
interpretations.

Another path of future research is the study of algebraic trees with
categorical methods.  In the setting of trees over a signature $\Sigma$ the
solutions of recursive program schemes form the theory of algebraic trees, a
subtheory of the theory of all trees on $\Sigma$. Moreover, algebraic trees are
closed under second-order substitution and they form an iterative theory in the
sense of Elgot~\cite{elgot}. Similar results should be possible to obtain in
our generalized categorical setting. 
\comment{First promising steps in the direction of
categorically studying algebraic trees have been taken in~\cite{amv4}.  }
\enlargethispage{.5cm}%
\fi%
\iftcs%
\begin{ack}
\else
\section*{Acknowledgments}
\fi%
  We are grateful to Ji\v{r}\'\i\ Ad\'{a}mek for  many conversations
  on this and related matters, and for his enthusiasm for our project. We would
  also like to thank the anonymous referees for their suggestions to improve
  this text.
\iftcs%
\end{ack}
\fi%

%
% Appendix for proofs in LNCS version
%
\comment{
\iffull\else
\clearpage
\begin{appendix}
  \section{Proofs}

  \doproofs
\end{appendix}
\fi
}

%
% List of notations
%
\clearpage
\section*{Notations}
%\addcontentsline{toc}{section}{Notations}

\setlongtables
\noindent
\begin{longtable}{lp{12.75cm}}
  \endhead
  \endfoot
  % format: notation & explanation \\
  $\T{H}X$ & a final coalgebra for $\HX$, p.~\pageref{p:thx}; a free cia or a free
  Elgot algebra on $X$, Theorem~\ref{theorem-finalcoalgebra-Elgot}; object map
  of a free completely iterative monad on $H$, Theorems~\ref{thm:adjoint}
  and~\ref{thm:freecim}.
  \\
  $\alpha^H_X$ 
  & 
  $\alpha^H_X = [\tau^H_X, \eta^H_X]^{-1}: \T{H} X \cong H\T{H}X
  + X$, a final coalgebra, p.~\pageref{eq:alpha} and
  Theorems~\ref{theorem-finalcoalgebra-Elgot} and~\ref{thm:adjoint}. 
  \\
  $\tau^H_X$ 
  & 
  $H$-algebra structure of the free cia $\T{H} X$, p.~\pageref{p:taueta} and 
  Theorem~\ref{theorem-finalcoalgebra-Elgot}.
  \\
  $\eta^H_X$
  &
  universal arrow of the free cia $\T{H}X$, p.~\pageref{p:taueta} and
  Theorems~\ref{theorem-finalcoalgebra-Elgot} and~\ref{thm:adjoint}; 
  unit of the free completely iterative monad $\T{H}$, Theorem~\ref{thm:freecim}. 
  \\
  $\mu^H$
  &
  $\mu^H: \T{H}\T{H} \to \T{H}$ is the multiplication of the the free
  completely iterative monad $\T{H}$, Theorem~\ref{thm:adjoint}(1b),
  p.~\pageref{thm:adjoint}, and Theorem~\ref{thm:freecim},
  p.~\pageref{thm:freecim}. 
  \\
  $\kappa^H$ & $\kappa^H: H \to \T{H}$ is the universal arrow of the free
  completely iterative monad $\T{H}$, Theorem~\ref{thm:adjoint}(4),
  p.~\pageref{thm:adjoint}, and Theorem~\ref{thm:freecim},
  p.~\pageref{thm:freecim}.
  \\
  $\Sigma$ 
  & 
  a signature of function symbols, also regarded as a functor $\Sigma: \Nat \to
  \Set$, Example~\ref{ex:signature}, p.~\pageref{ex:signature}.
  \\
  $H_\Sigma$ 
  & 
  a signature functor, see~\refeq{eq:poly}, p.~\pageref{eq:poly}.
  \\
  $(f, \vec{x})$ & generic element of $H_\Sigma X$, $f \in \Sigma_n$, $x \in
  X^n$, p.~\pageref{ex:signature}.
  \\
  $T_\Sigma X$ & all $\Sigma$-trees over $X$, p.~\pageref{p:trees}.
  \\
  $\Set$ & the category of sets and maps, p.~\pageref{ex:signature}f. 
  \\
  $\CPO$ & the category of complete partial orders (not necessarily with a
  least element) and continuous maps, Example~\ref{ex:cats_ii},
  p.~\pageref{ex:cats_ii}. 
  \\
  $\CMS$ & the category of complete metric spaces with distances in $[0,1]$ and
  non-expanding maps, Example~\ref{ex:cats_iii}, p.~\ref{ex:cats_iii}.
  \\
  $T_\lambda$ & the $\lambda$th step in the final coalgebra construction,
  Construction~\ref{constr:coalg}, p.~\pageref{constr:coalg}. 
  \\
  $\Pfin$ & finitary powerset functor $X \mapsto \{\,A \mid A \subseteq 
X\
  \textrm{finite}\,\}$, Example~\ref{ex:finitary}(ii),
  p.~\pageref{ex:finitary}. 
  \\
  $H'$ & lifting of a set endofunctor $H$ to $\CPO$ or $\CMS$, respectively,
  Examples~\ref{ex:lift_cpo}, p.~\pageref{ex:lift_cpo}, and~\ref{ex:lift_cms},
  p.~\pageref{ex:lift_cms}. 
  \\
  $\alpha * \beta$ & parallel composition of natural transformations $\alpha$
  and $\beta$, p.~\pageref{p:parallel}. 
  \\
  $\MA$ & the category of monads and their homomorphisms on the category $\A$,
  p.~\pageref{p:MA}.
  \\
  $J$ & the canonical inclusion $J: \Nat \to \Set$, $n \mapsto \{\,0, \ldots,
  n-1\,\}$, p.~\pageref{eq:bij}.
  \\
  $\fun{\A}{\cat{B}}$ & the category of endofunctors from $\A$ to $\cat{B}$ and
  natural transformations between them, p.~\pageref{p:fun}. 
  \\
  $f(\vec{x}) \approx t^f(\vec{x})$ & classical form of a recursive program
  scheme as system of mutually recursive formal function definitions,
  \refeq{eq:crps}, p.~\pageref{eq:crps}.
  \\
  $e:\Var \to \T{H+\Var}$ & a recursive program scheme, p.~\pageref{p:rpsfirst}
  and Definition~\ref{dfn:algeq}, p.~\pageref{dfn:algeq}. 
  \\
%  $\A^T$ & the Eilenberg-Moore category of the monad $T: \A \to \A$,
%  p.~\pageref{sec:Eilenberg-Moore}. 
%  \\
  $e: X \to HX + A$ & a flat equation morphism, Definition~\ref{dfn:cia},
  p.~\pageref{dfn:cia}. 
  \\
  $e: X \to S(X+Y)$ & an equation morphism w.\,r.\,t.~the ideal monad $S$, 
  Definition~\ref{dfn:compiter}, p.~\pageref{dfn:compiter}. 
  \\
  $\sol{e}$ & a solution of a (flat) equation morphism $e$ or of a recursive
  program scheme $e$, \refeq{diag:ciasol},
  p.~\pageref{diag:ciasol}, p.~\pageref{p:sol}, and~\refeq{diag:algeq},
  p.~\pageref{diag:algeq}, respectively. 
  \\
  $C(X)$ & the non-empty compact subspaces of a complete metric space $X$, 
  Example~\ref{ex:cia}(v), p.~\pageref{p:C}.
  \\
  $h\after e$ & the ``renaming of parameters'' of a flat equation morphism $e:
  X \to HX + A$ by a morphism $h: A \to B$, Remark~\ref{rem:after},
  p.~\pageref{rem:after}. 
  \\
  $f \plus e$ & ``simultaneous'' flat equation morphism obtained from $e: X \to
  HX + Y$ and $f: Y \to HY + A$, Remark~\ref{rem:after},
  p.~\pageref{rem:after}. 
  \\
  $(A, a, \funsol)$ & a (complete) Elgot algebra,
  Definition~\ref{dfn:elgotalg}, p.~\pageref{dfn:elgotalg}. 
  \\
  $p(x)\dar$ ($p(x)\uar$) & partial function $p$ is (not) defined,
  p.~\pageref{defn:comptree}. 
  \\
  $p(x) \simeq q(y)$ & Kleene equality: $p(x)$ is defined iff $q(y)$ is defined and
  if both are defined, they are equal, p.~\pageref{dfn:elgotalg}. 
  \\
  $\Elgot H$ & the category of Elgot algebras and their homomorphisms,
  Theorem~\ref{thm:adjoint}, p.~\pageref{thm:adjoint}.
  \\
  $\aext{a}: TA \to A$ & the evaluation morphism of an Elgot algebra $(A, a,
  \funsol)$, \refeq{eq:aext}, p.~\pageref{eq:aext}, and Theorem~\ref{thm:aext},
  p.~\pageref{thm:aext}. 
  \\
  $(S, \eta, \mu, S', \iota, \mu')$ & an ideal monad,
  Definition~\ref{dfn:ideal-monad}, p.~\pageref{dfn:ideal-monad}. 
  \\
  $\sigma: H \to S$ & an ideal natural transformation from the functor $H$ to
  the ideal monad $S$, $\sigma = \xymatrix@1{H \ar[r]^-{\sigma'} & S'
  \ar[r]^-{\iota} & S}$ for some $\sigma'$, Definition~\ref{dfn:ideal-monad},
  p.~\pageref{p:ideal}. 
  \\
  $\ol{\sigma}$ & unique extension of an ideal natural transformation $\sigma:
  H \to S$ to a monad morphism $\ol{\sigma}: \T{H} \to S$, 
  Theorem~\ref{thm:freecim}, p.~\pageref{thm:freecim}.
  \\
  $\Hmon$ & the subcategory of the comma-category $H/[\A,\A]$
  formed by natural transformation from $H$ to monads $S$ and 
  monad morphisms, p.~\pageref{p:Hmon}. 
  \\
  $\B$ & the subcategory of $\Hmon$ given by ideal natural transformations from
  $H$ into completely iterative monads $S$ and ideal monad morphisms,
  p.~\pageref{p:Hmon}. 
  \\
  $\H$ & the endofunctor on $\Hmon$ and $\B$, respectively, given by 
  $(S, \sigma) \mapsto (HS + \Id, \inl \o H\eta)$, p.~\pageref{p:Hs},
  Lemma~\ref{lem:functor}, p.~\pageref{lem:functor},
  Corollary~\ref{cor:finalcim}, p.~\pageref{cor:finalcim}.
  \\
%  $\xi_{(S, \sigma)}$ & a natural transformation from $\H$ to $\Id$ exhibiting
%  $\H$ as a well-copointed endofunctor, p.~\pageref{p:xi} and
%  Lemma~\ref{lem:functor}, p.~\pageref{lem:functor}. 
%  \\
  $f: \Var \to H\T{H+\Var}$ & a natural transformation exhibiting a recursive
  program scheme $e: \Var \to \T{H+\Var}$ as guarded,
  Definition~\ref{dfn:algeq}, p.~\pageref{dfn:algeq}.
  \\
  $\ul{t}$ & a formal variable arising from a
  recursive program scheme.  Here $t$ is a tree on $\Sigma + \Phi$ 
  allowing some additional variables as well, 
  \refeq{eq:induced}, p.~\pageref{eq:induced}, Remark~\ref{rem:extrps},
  p.~\pageref{p:ult}. 
  \\
  $\ol{e}$ & the ideal monad morphism $\ol e: \T{H+\Var} \to H\T{H+\Var} +\Id$
  is induced by the guarded recursive program scheme $e$, and it yields a
  coalgebra for the endofunctor $\H$ of $\B$,
  Lemma~\ref{lemma-prelim-uninterpreted-rps},
  p.~\pageref{lemma-prelim-uninterpreted-rps}.  
  \\
  $\ext{e}$ & the ideal monad (endo)morphism $\ext{e}: \T{H+\Var} \to
  \T{H+\Var}$ is induced by the guarded recursive program scheme $e$,
  Lemma~\ref{lemma-prelim-uninterpreted-rps},
  p.~\pageref{lemma-prelim-uninterpreted-rps}. 
  \\
  $\isol e A$ & an interpreted solution of a recursive program scheme $e: \Var
  \to \T{H+\Var}$ in an Elgot algebra $(A, a, \starop)$,
  Definition~\ref{dfn:intsol}, p.~\pageref{dfn:intsol}.
  \\
  $\Nat_{\upa}$ & the natural numbers with the computation tree Elgot algebra
  structure, p.~\pageref{p:Natupa} and
  Proposition~\ref{proposition-Elgot-computation},
  p.~\pageref{proposition-Elgot-computation}. 
  \\
  $\Natb$ & the flat cpo obtained from the natural numbers $\Nat$ by adding a
  least element $\bot$, Example~\ref{ex:runelgot}, p.~\pageref{ex:runelgot}.
\end{longtable}

%
% Index 
%
%\addcontentsline{toc}{section}{Index}
%\addtocontents{toc}{\protect{\vspace*{\fill}}} 
\clearpage
\printindex

%
%
% The bibliography
%
%
\iffull
\iftcs  % TCS version
\let\oldbibitem=\bibitem
\renewcommand{\bibitem}[2][\empty]{\oldbibitem{#2}}
\bibliographystyle{plain}

\end{document}